\pdfoutput=1 
\documentclass[usenatbib]{mnras}

\usepackage[T1]{fontenc}
\usepackage{ae,aecompl}
\usepackage[utf8]{inputenc}
\usepackage{multicol}

\usepackage{ulem}

\usepackage{graphicx}	
\usepackage{amsmath}	
\usepackage{amssymb}	
\usepackage{longtable}
\usepackage{pdflscape}
\usepackage{soul}
\usepackage{natbib}

\graphicspath{ {./Images/} }
\usepackage{float}
\usepackage{comment}

\usepackage{etoolbox}

\usepackage{newtxtext,newtxmath}


\usepackage{hyperref}
\newcommand{\orcid}[1]{\href{https://orcid.org/#1}{\includegraphics[width=10pt]{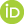}}}

\newcommand{\kms}{\,km~s$^{-1}$}	
\newcommand{\msun}{M$_\odot$}	
\newcommand{\rsun}{R$_\odot$}

\newcommand{\Civ}{C {\sc iv}}	
\newcommand{\Ciii}{C {\sc iii}}	
	
\newcommand{\Niii}{N {\sc iii}}

\defcitealias{Zhan20}{Z20}
\defcitealias{Hiramatsu2020_18zd}{H20}
\defcitealias{Boian_2020_18ZD}{B20}

\title[SN~2018zd]{How low can you go? SN~2018zd as a low-mass Fe core-collapse supernova}

\author[E. Callis et al.]{E.~Callis\orcid{0000-0002-1178-2859},$^{1}$\thanks{E-mail: emma.callis@ucdconnect.ie (EC), morgan.fraser@ucd.ie (MF)}
M.~Fraser\orcid{0000-0003-2191-1674},$^{1}$
A.~Pastorello\orcid{0000-0002-7259-4624},$^{2}$
Subo~Dong,$^{3}$
S.J.~Brennan\orcid{0000-0003-1325-6235},$^{1}$\newauthor
P.~Chen\orcid{0000-0003-0853-6427},$^{3}$
S.~Bose,$^{4,5}$
T.~Reynolds,$^{6}$
L.~Salmon,$^{1}$
P.~Jonker,$^{7,8}$
S.~Benetti\orcid{0000-0002-3256-0016},$^{2}$\newauthor
M.~Berton\orcid{0000-0002-1058-9109},$^{9,10}$
G.~Cannizzaro,$^{7}$
E.~Cappellaro,$^{2}$
E.~Congiu,$^{11}$
S.~Dyrbye,$^{12}$\newauthor
D.~Eappachen,$^{7,8}$
N.~Elias-Rosa\orcid{0000-0002-1381-9125},$^{2,13}$ 
M.~Gromadzki\orcid{0000-0002-1650-1518},$^{14}$
C.P.~Guti\'errez\orcid{0000-0003-2375-2064},$^{9,6}$\newauthor
S.~Holmbo,$^{12}$
T.W.S.~Holoien,$^{15}$
K.~Itagaki,$^{16}$
E.~Kankare\orcid{0000-0001-8257-3512},$^{6}$
S.~Mattila\orcid{0000-0001-7497-2994},$^{6}$\newauthor
R.~Mutel\orcid{0000-0003-1511-6279},$^{17}$
P.~Ochner, $^{2,18}$
R.S.~Post,$^{19}$
J.~Prieto,$^{20,21}$
A.~Reguitti\orcid{0000-0003-4254-2724},$^{22,21,2}$\newauthor
T.~Roth,$^{17}$ 
J.~Ryon,$^{23}$
A.~Sagués-Carracedo\orcid{0000-0002-3498-2167},$^{24}$
B.J.~Shappee,$^{25}$
A.~Siviero,$^{18}$\newauthor
K.G.~Stassun\orcid{0000-0002-3481-9052},$^{26}$
M.~Stritzinger\orcid{0000-0002-5571-1833},$^{12}$
L.~Tomasella,$^{2}$
S.~Villanueva Jr.,$^{4}$\newauthor
T.~Wevers,$^{27,28}$
P.~Wiggins,$^{29}$\\
\\
Affiliations at end of paper
}
\date{Accepted XXX. Received YYY; in original form ZZZ}

\pubyear{2021}

\begin{document}
\label{firstpage}
\pagerange{\pageref{firstpage}--\pageref{lastpage}}
\maketitle

\begin{abstract}
We present spectroscopy and photometry of SN~2018zd, a Type IIP core-collapse supernova with signatures of interaction with circumstantial material in its earliest spectra.
High ionization lines, the earmark of shock breakout, are not seen in the earliest spectral epoch, and are only seen in a single spectrum at 4.9~d after explosion. The strength and brevity of these features imply a confined circumstellar material shell in the immediate vicinity of the progenitor. Once the narrow emission lines disappear, SN~2018zd evolves similarly to a Type IIP SN, although the blue colour and enhanced plateau magnitude of SN~2018zd suggests an additional source of luminosity throughout the plateau phase. 
While SN~2018zd has previously been proposed as an electron-capture SN, we suggest that it is an Fe core-collapse from a low mass red supergiant progenitor.
Differences in interpretation for SN~2018zd arise in part due to the large uncertainty on the distance to the host-galaxy NGC 2146, which we re-derive here to be $15.6^{+6.1}_{-3.0}$~Mpc.
We find the ejected $^{56}$Ni mass for SN~2018zd to be 0.017~\msun, significantly higher than models of ECSNe predict. We also find the Ni/Fe ratio in SN~2018zd to be much lower that would be expected for an ECSN.

\end{abstract}


%

\section{Introduction}
\label{sec:Intro}

Core-collapse supernovae (CCSNe) are one of the most frequently observed classes of optical transients, with the current generation of surveys allowing us to study large samples of these objects in ways not before possible. They are important drivers of the chemical enrichment of the Universe, and in some cases can shed light on the recent mass loss history of their progenitors, helping to improve our understanding of the final stages in the evolution of massive stars. 
CCSNe are divided into classes based on their observational properties \citep{Filippenko_1997,2017hsn..book..195G}, with the hydrogen-rich Type II SNe being the most commonly found and one of the best studied classes. While the bulk of the population of Type IIP SNe, where ``P" refers to the plateau observed in the optical light curves, can be understood as the Fe core-collapse of low to moderate mass red supergiants \citep{Smartt2009_progenitors}, it has become increasingly apparent that some Type IIP SNe do not fit neatly into the standard paradigm  (e.g. \citealp{Botticella10,Inserra12,Takats15_2009ib,PolshawLSQ13fn,Hosseinzadeh16bkv,Ouchi21_SN2009kf}). The study of these unique events broadens the understanding of the class as a whole.

Type IIP SNe are defined by a number of characteristics, most notably their long-lasting optical light curve plateau, and their hydrogen dominated spectra with strong P-Cygni line profiles. Variations in these characteristics between events are relatively well-understood within the framework of current models \citep{Valenti2016}. However, there is a small but growing number of events which evolve similarly to ``normal" Type IIP SNe, but display features which challenge our current understanding.

One of the most common causes of Type IIP SNe differing significantly from the norm is interaction with circumstellar material (CSM). There is a huge variety in the observed characteristics of Type II SNe affected by interaction, with many possible subclasses being the subject of ongoing debate in the literature \citep{Fras20}. 
There is also a growing number of events which only display signatures of interaction in their earliest epochs and evolve similarly to ``normal'' Type IIP SNe afterwards \citep[e.g.][]{Rodriguez2020_LLEVs}. Such events offer an opportunity to better understand the very final stages in the evolution of their progenitors, as the emission features originate from material lost by the star in the final years prior to its collapse \citep{Yaron2017_SN2013fs,Groh_2014_13cu}. 

One of the most direct ways of studying CSM close to the SN is through the so-called ``flash'' lines that may be briefly present immediately after the supernova explosion \citep[e.g.][]{Gal-Yam_2014_13cu_FlashSpec,Khazov_2016_flashspec_SNe_II, Bose_2021_2018gk}. As the SN shock breaks through the surface of the progenitor star, it releases a burst of ionizing photons, which ionizes the nearby CSM. By analysing these flash ionization lines, before the CSM either recombines or is swept up by the ejecta, we can probe the density and composition of the CSM. An in depth example of this method of analysis is SN 2013fs \citep{Yaron2017_SN2013fs}, from which the authors conclude that the wind mass-loss rate required ($\sim10^{-3}$ \msun\ yr$^{-1}$) to explain the short-lived signatures of interaction was orders of magnitudes higher than the $\rm{10^{-6} - 10^{-4} M_{\odot}}$ yr$^{-1}$ that is expected from Red Super-giant stars (RSGs). They infer that the progenitor must have endured a brief, high mass-loss phase shortly before the explosion, and that precursor events such as these might be common.
More recently, however, \cite{Kochanek19} has suggested that these observations can be explained by an RSG with a normal, Thomson optically thin wind in a binary without any need for a pre-SN eruption.

Peculiarity can help delimit the extremes of the CCSN population. Moreover, understanding peculiar events is necessary if we are to use them in cosmology. In recent decades, Type II SNe have become recognised for their value as distance indicators for use in cosmology \citep{Nugent17}. Numerous methods exist in the literature for determining distances to Type IIP SNe \citep{deJaeger_2020_Hubble}. These have been used to create Hubble diagrams out to similar redshifts and with uncertainties comparable to Type Ia Hubble diagrams \citep{Gall_2018_IIP_HubbleDiagram}. Most of these methods make the assumption that the supernova explosion is spherically symmetric, and that the RSG progenitor is exploding into a near-vacuum environment. This latter assumption, in particular, has come under scrutiny in recent years, with a growing number of Type II SNe having been observed to display signatures of interaction in their earliest epochs \citep{Yaron2017_SN2013fs, PolshawLSQ13fn, Hosseinzadeh16bkv}. This affects the photometric distance methods by increasing the plateau luminosity of the events, biasing the inferred distances to be larger. This has major implications for the use of Type II SNe in cosmology, but thorough investigations of objects like SN~2018zd may help us to understand and mitigate these effects.

\begin{figure}
\includegraphics[width=\columnwidth]{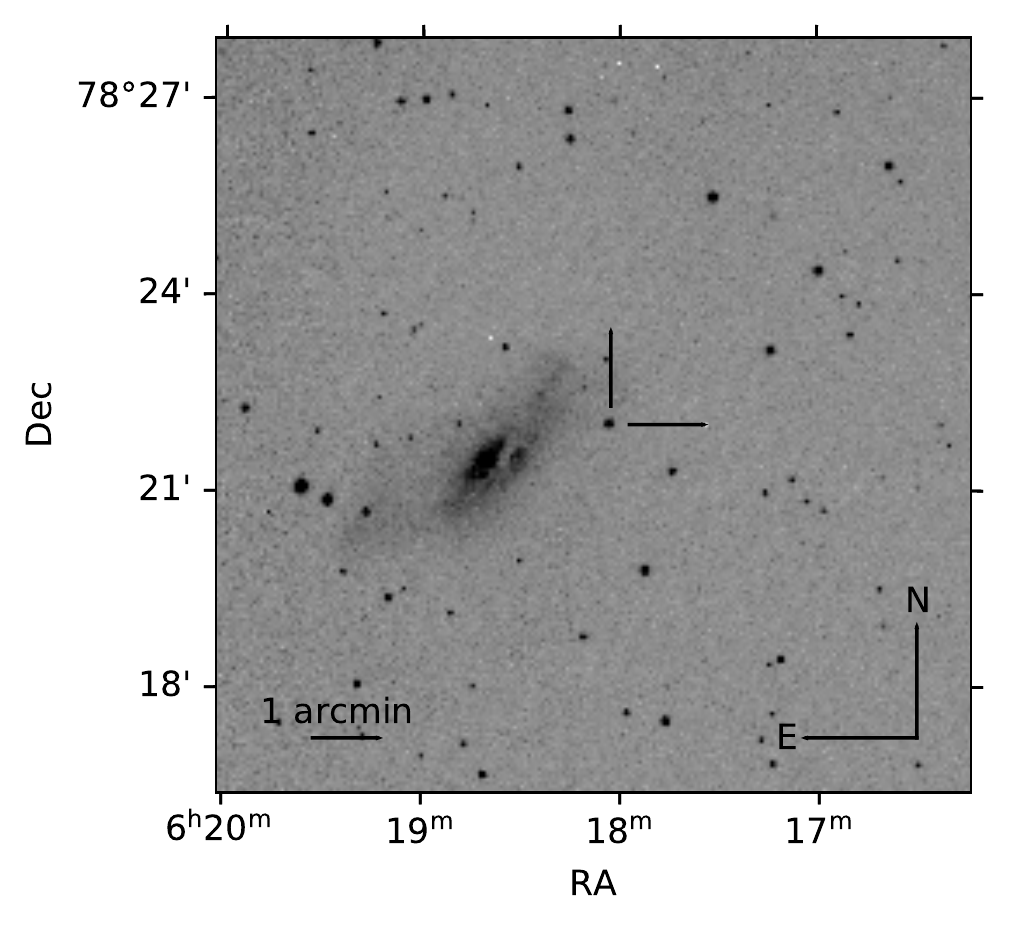}
\caption{Finder chart showing the location of SN~2018zd, based on a $V$-band image from the Post SRO Observatory taken on 2018 April 17, close to the $V$-band maximum.}
\label{fig:finders_chart}
\end{figure}

Motivated by these questions, we present here the results of a comprehensive photometric and spectroscopic observational campaign for SN~2018zd\footnote{The SN was also independently recovered by the Gaia and ATLAS surveys as Gaia18anr (\url{http://gsaweb.ast.cam.ac.uk/alerts/alert/Gaia18anr/}) and ATLAS18mix (\url{https://www.wis-tns.org/object/2018zd}) respectively.} (Fig. \ref{fig:finders_chart}) situated in the nearby starburst galaxy NGC~2146, an unusual Type IIP SN that displayed high ionization lines at early times. It was discovered in NGC~2146 by K. Itagaki on 2018 March 2 \citep{ATel_18zd_11379}, and classified initially as a Type IIn SN because of the narrow emission lines in the spectra \citep{ATel_Zhang_18zd_IIn}. \cite{Zhan20}, \cite{Boian_2020_18ZD}, and \cite{Hiramatsu2020_18zd} (hereafter \citetalias{Zhan20}, \citetalias{Boian_2020_18ZD} and \citetalias{Hiramatsu2020_18zd}, respectively), have published analysis of SN~2018zd, and we will compare our results to theirs throughout this paper.
\citetalias{Boian_2020_18ZD} include SN~2018zd in their sample of SNe with brief, early-time interaction signatures in their spectra, for which they constrain the progenitor and explosion properties by computing early-time spectra using the radiative transfer code { \sc cmfgen} \citep{Hillier_1998_CMFGEN}. \citetalias{Zhan20} published the results of their observational campaign on SN~2018zd, concluding that SN~2018zd could be classified as a transitional event morphologically linking Type II~P and Type II~L SNe, that the progenitor had an initial mass of $\sim12$~$\rm{M_{\odot}}$ and produced a relatively small amount of $^{56}$Ni, 0.013--0.035~$\rm{M_{\odot}}$. \citetalias{Hiramatsu2020_18zd} argue that SN~2018zd is an electron capture SN (ECSN). This result is in part due to their low distance estimate of 9.6 Mpc based on the Standard Candle Method (SCM), which we explore further in Section \ref{sec:SCM}. 
\cite{Paras20} also obtained high cadence imaging data of SN~2018zd, and they reported no evidence of undulations or short timescale variability in the light curve around the time of peak magnitude. 

Section \ref{sec:Observations} details the discovery, pre-explosion observations, observational parameters, follow-up observations and reduction of the data. We adopt a distance of $15.6_{-3.0}^{+6.2}$~Mpc and a reddening of $E(B - V)$=0.19 mag, and a detailed discussion of these parameters can be found in Section \ref{sec:Dist_and Extinction}. Section \ref{sec:Phot-Analysis} discusses the estimated explosion epoch (MJD 58178.5), photometric evolution, the bolometric light curve, and the $^{56}$Ni mass. Section \ref{sec:Spec-evo} details the spectroscopic and velocity evolution. Section \ref{sec:Discussion} discusses the progenitor parameters, mass loss history and SN~2018zd in relation to the standard candle method (SCM) for Type IIP SNe. Finally, our conclusions are presented in Section \ref{sec:Conclusions}.

\section{Observations}
\label{sec:Observations}

\subsection{Pre-explosion observations}
\label{sec:Pre-explosion}

\subsubsection{HST and progenitor imaging}
\label{sec:HST}
\begin{figure*}
\includegraphics[width=\textwidth]{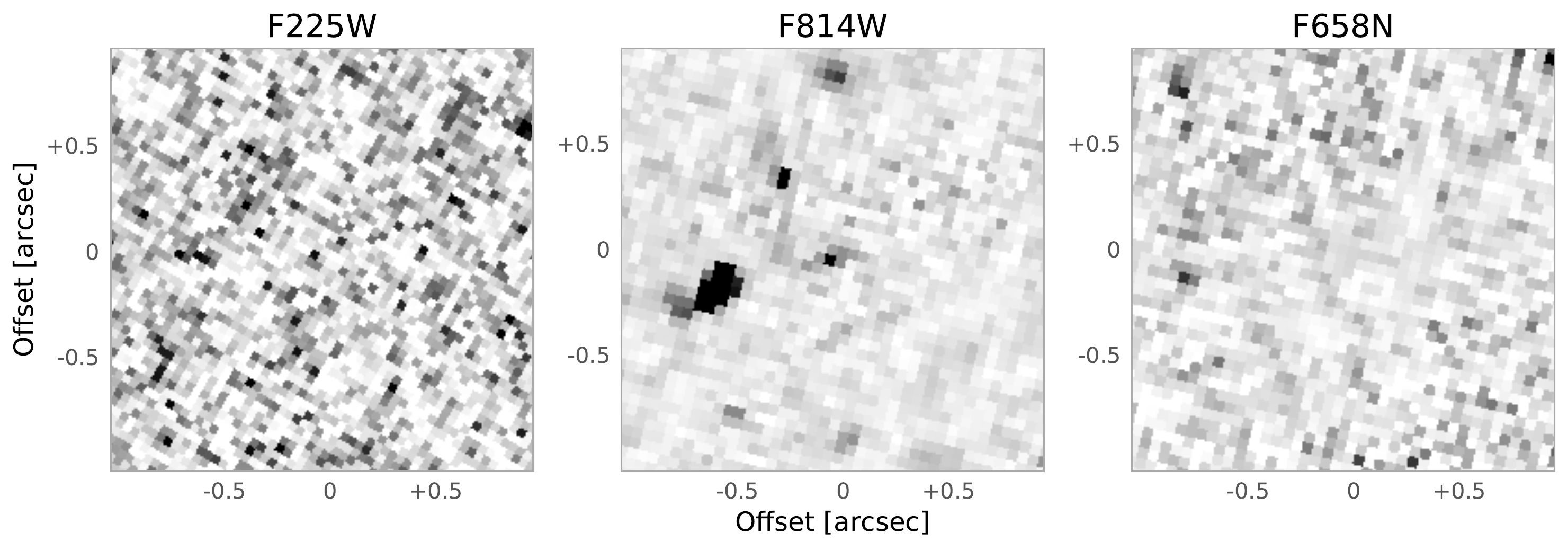}
\includegraphics[width=0.66\textwidth]{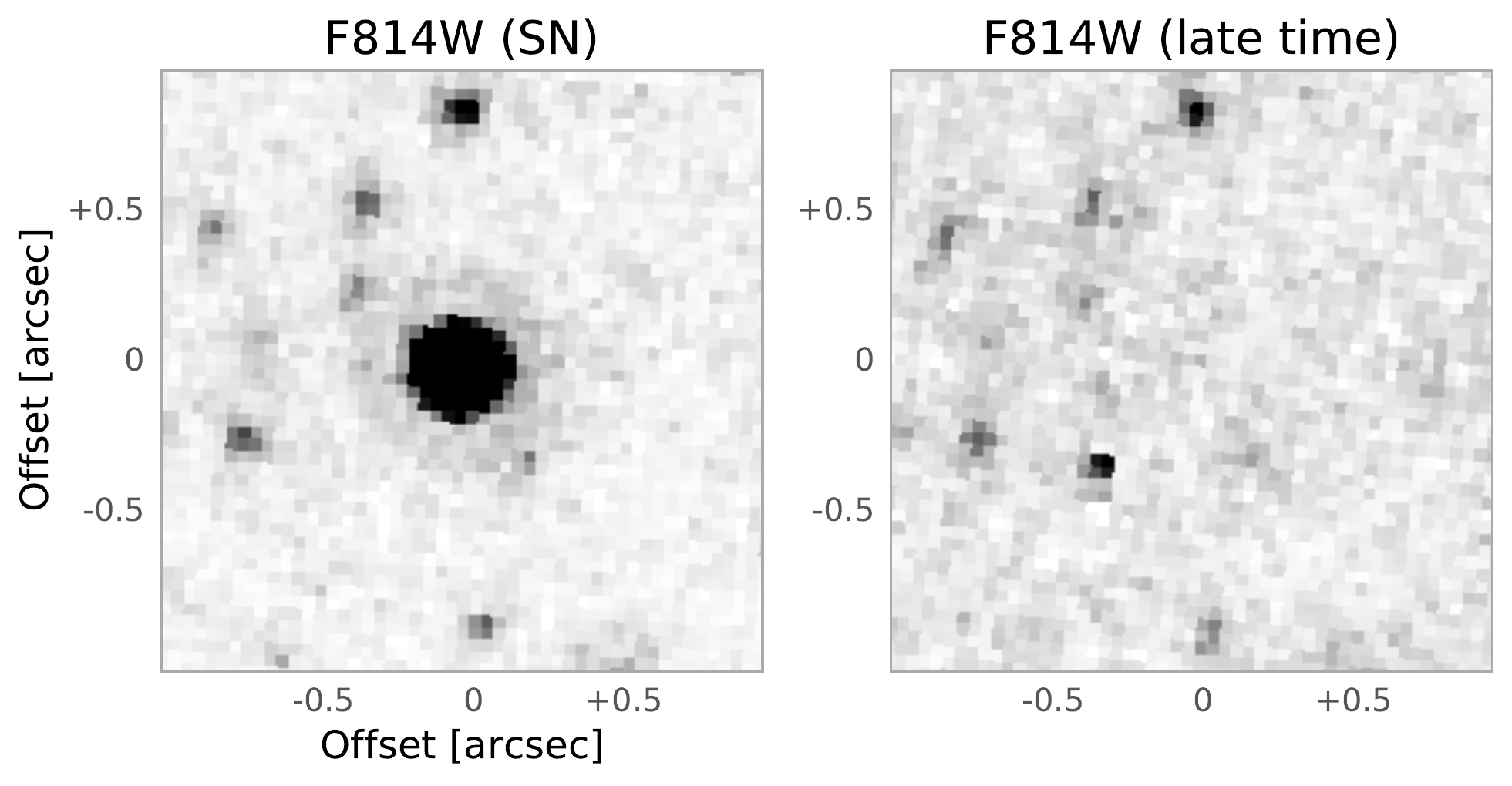}
\caption{Sections of the {\it HST} images of the host of SN~2018zd. Each panel is centred on the SN location, and oriented so that North is up and East is to the left. The upper three panels show pre-explosion data (the progenitor candidate is the faint source in the centre top panel); the bottom two panels show late time images with the SN present and after it has faded.}

\label{fig:progenitor}
\end{figure*}

The archival\footnote{\url{https://archive.stsci.edu/}} \textit{Hubble Space Telescope (HST)}  images which covered the site of SN~2018zd are listed in Table \ref{tab:pre-exp} and shown in Fig. \ref{fig:progenitor}. In addition, images of SN~2018zd itself were taken with {\it HST} on 2019 May 19, and these were used to derive precisely the location of the progenitor on the pre-explosion frames. Finally, a set of late time WFC3 {\it F814W} and {\it F606W} were taken on 2021 Feb 7, and we use these to test for the disappearance of a progenitor candidate.

\begin{table*}
\centering
\caption{Log of archival pre-explosion observations of the site of SN~2018zd. Dates refer to the start of the exposure. Upper limits correspond to magnitudes where either 68 per cent ($\sim 1\sigma$) or 95 per cent ($\sim 2\sigma$) of artificial sources are recovered within 0.1 mag of their input magnitude (in the Vega system).}
\label{tab:pre-exp}
\begin{tabular}{lccccrc} 
	\hline
	Telescope & Instrument  & Filter    & Exposure (s)  & Date         & Mag    & Proposal ID     \\
	\hline
	{\it HST} & ACS 	    & \textit{F658N}	    & 2x350 	& 2004-04-10.9 & $m_{1\sigma}>22.3$; $m_{2\sigma}>21.2$    & 9788 \\
    {\it HST} &	ACS  	    & \textit{F814W}	    & 1x120	    & 2004-04-10.9 & 24.65$\pm$0.14   & 9788 \\
	{\it HST} & WFC3 	    & \textit{F225W}     & 4x375	    & 2013-03-08.0 & $m_{1\sigma}>24.5$; $m_{2\sigma}>23.7$ &  13007\\
    \hline
\end{tabular}
\end{table*}

While there is a relatively deep (1500~s) pre-explosion UV image taken with the WFC3 {\it F225W} filter, this is unfortunately not constraining for a cool RSG progenitor, and no sources were detected close to the SN location in this image. There is also a 700~s ACS {\it F658N} narrow band pre-explosion image with no detectable source. We hence calculate a detection limit through artificial star tests for these images using the {\sc dolphot} package \citep{Dolp00}, which we report in Table \ref{tab:pre-exp}. Limits were calculated by injecting progressively fainter sources, to find the magnitude at which less than a certain fraction of injected sources were recovered by {\sc dolphot}. We report both 68 and 95 per cent completeness in Table \ref{tab:pre-exp}, which are approximately equivalent to a 1 and 2$\sigma$ limit respectively.

While the pre-explosion ACS {\it F814W} image is relatively shallow (120~s), it is potentially useful for identifying a red supergiant progenitor. To precisely locate the SN position on this image, we compared it to the 2019 WFC3 {\it F814W} image of SN~2018zd. Fifteen common sources were used to determine a coordinate transformation between the two images; the rms uncertainty on this transformation was found to be only 14~mas ($<$0.3 pixels in the pre-explosion image). A faint source is seen at the position of the SN, with a signal-to-noise ratio (SNR) of 7.5 (see Fig. \ref{fig:progenitor}). We measure the position of this source and find it to be only 4~mas from the transformed SN location, comfortably inside our astrometric uncertainty, making it a credible candidate for the progenitor.

\begin{figure}
\includegraphics[width=\columnwidth]{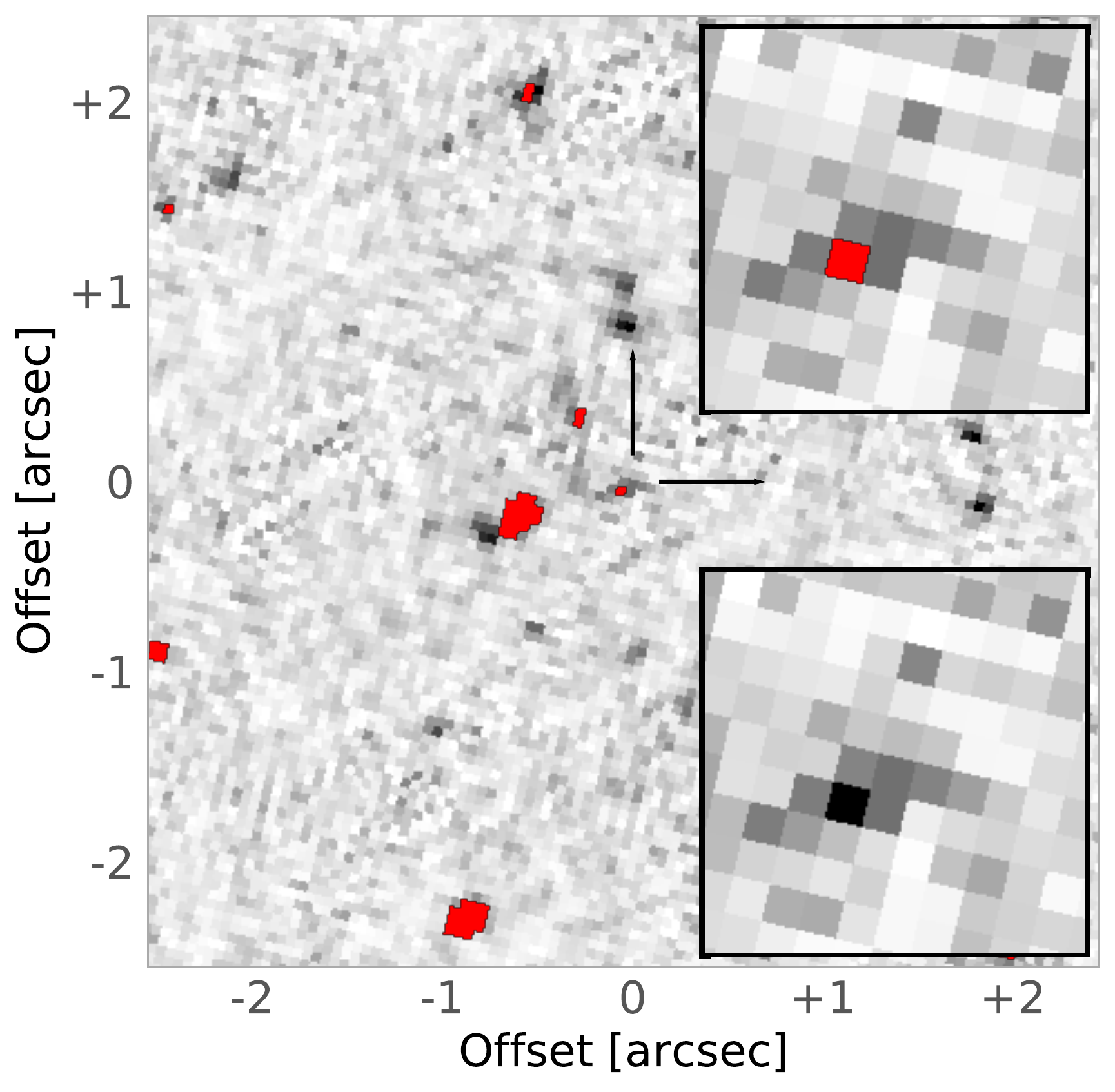}
\caption{Section of the pre-explosion \textit{F814W HST} image around the site of SN~2018zd. Pixels marked in red have been masked as cosmic ray hits by {\sc lacosmic}. In the lower inset we show 10$\times$10 pixels around the progenitor candidate, in the upper inset the same region with the cosmic ray mask applied. The progenitor is indicated om the main panel with tick marks.}
\label{fig:lacosmic}
\end{figure}

\begin{figure}
\includegraphics[width=\columnwidth]{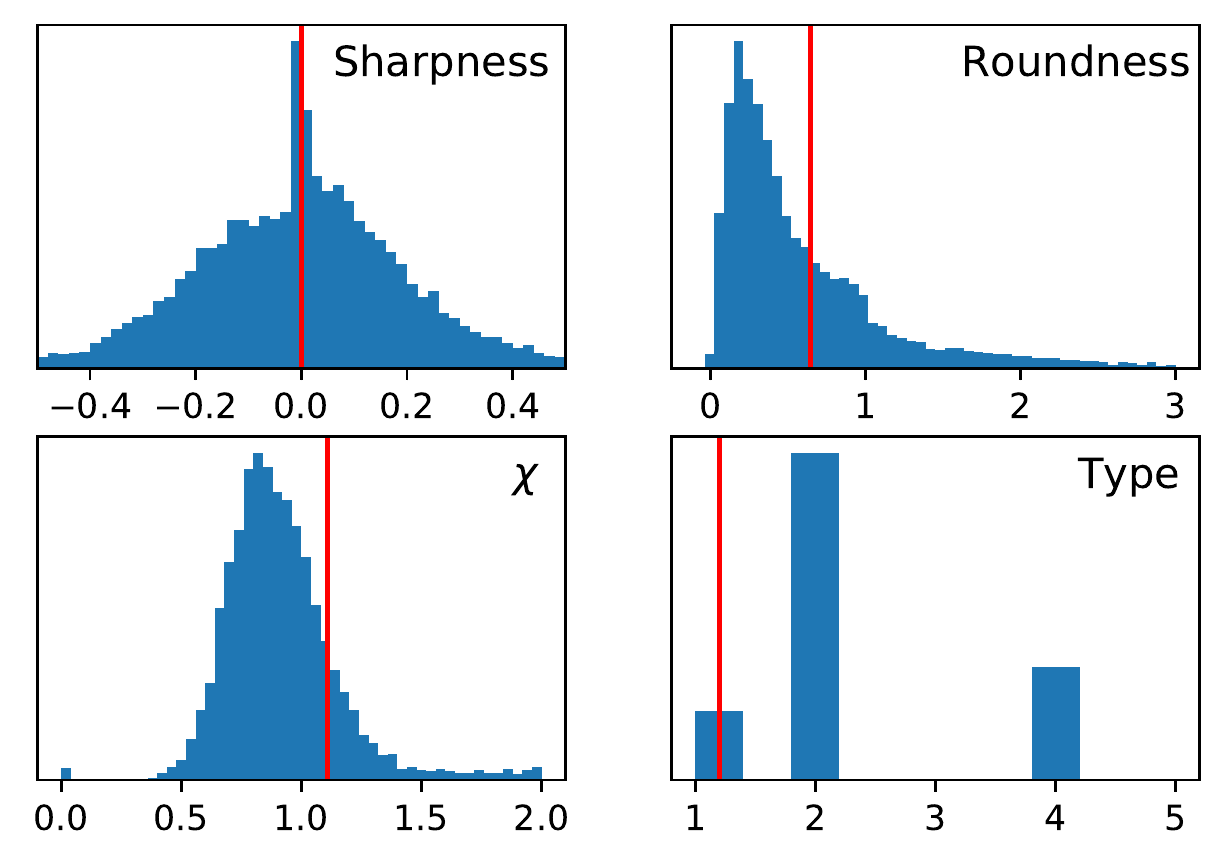}
\caption{Parameters returned by {\sc dolphot} for the progenitor candidate in \textit{F814W} (red vertical line), compared to the distribution of parameters for all other sources in the field. The panels show the $\chi^2$ of the fitted PSF, the sharpness parameter (which should be between $-0.3$ and $0.3$ for a PSF), and the roundness (which is higher for sources that are elongated). In addition, we show the {\sc dolphot} object type, where `1' is a point source, `2' is too faint to determine its type, and `4' is a likely cosmic ray.}
\label{fig:F814W_param}
\end{figure}

Using {\sc dolphot}, we measure a magnitude of {\it F814W}~=~24.65$\pm$0.14 mag for the progenitor candidate, somewhat brighter than the value measured by \citetalias{Hiramatsu2020_18zd} of 25.05$\pm$0.15 mag. 
While it is tempting to attribute this source to the progenitor of SN~2018zd, there is a possibility that the ``source'' may be due to a cosmic ray. This is of particular concern as there is only a single \textit{F814W} exposure. Close examination of the source shows that it is somewhat ``peaky'', with a single pixel that dominates the flux (Fig. \ref{fig:lacosmic}). 
Even more concerning, {\sc lacosmic} \citep{Dokkum_2001_lacosmic} flags this pixel as a cosmic ray (when using the parameters recommended for under-sampled {\it HST} data; viz. {\sc sigclip}=4.5; {\sc sigfrac}=0.3; {\sc objlim=4}; {\sc niter=4}).
In Fig. \ref{fig:F814W_param} we compare the fit parameters returned by {\sc dolphot} for the progenitor candidate to those of the rest of the sources in the field. The $\chi^2$ of the fit is somewhat high, but still reasonable for a source that is satisfactorily modelled by the point-spread function (PSF). Similarly, the sharpness of the source is consistent with a PSF, while the source itself is classed as ``Type 1'' (meaning a likely stellar source) by {\sc dolphot}.

In summary, {\sc lacosmic} suggests that the progenitor candidate is a cosmic ray, while the output from {\sc dolphot} indicates it is most likely a genuine astrophysical source. For the remainder of our analysis of the progenitor in Section~\ref{sec:Progenitor_analysis}, we assume that the progenitor candidate is real. However, we also consider briefly the implications if this is not a stellar source, in which case the measured magnitude can be regarded as an upper limit.

The late time images taken in 2021 Feb are more than 1000 days after explosion, and we can expect the SN to have faded by this epoch. We do not see a source coincident with SN~2018zd (Fig. \ref{fig:progenitor}), and we calculate a limiting magnitude $m_{F814W}>25.3$. Unfortunately this does not answer the question of whether the source seen prior to explosion in \textit{F814W} is real. Both a cosmic ray or a former SN progenitor are expected to be gone in this image.

\subsubsection{Infrared pre-explosion imaging}

We also searched ground-based telescope archives for any deep, pre-explosion images covering the site of SN~2018zd. The Canada-France-Hawaii Telescope observed NGC~2146 with the Wide-field InfraRed Camera (WIRCam; \citealt{Puget_2004_WIRCam}) on 2011 December 8 in the {\it Ks} filter. The pipeline-reduced (bias-subtracted, flat-fielded and sky-subtracted) images were downloaded from the CFHT archive\footnote{\url{https://www.cadc-ccda.hia-iha.nrc-cnrc.gc.ca/en/cfht/}}. WIRCam is a mosaic imager, and the location of SN~2018zd fell on different chips through these observations. We created a separate stacked image for each of the four WIRCam chips, giving a total exposure time of 4$\times$240~s. No source was visible at the location of SN~2018zd in these data, and as the images are not particularly deep (especially in comparison to the {\it HST} data), we do not consider them any further.

We similarly examined 7 WHT+INGRID and WHT+LIRIS $K_s$ images taken between October 2002 and March 2005. These images were taken as part of a long term survey searching for dust obscured SNe in nearby starburst galaxies \citep{Mattila04}, and each have a total exposure times of around $\sim 500$~s. As these images were taken  with different instruments, pointing and seeing, we did not stack them, but we note that we see no source at the position of SN~2018zd.

The {\it Spitzer Space Telescope} also observed the site of SN~2018zd in 2004, 2007 and 2011 (Program IDs 59, 14098, 40410, 80089) as the host galaxy NGC~2146 is a nearby starburst galaxy\footnote{We note that NGC~2146 is classed as a Luminous Infrared Galaxy (LIRG) by \citealp{Sanders03}. However, NGC~2146 is only slightly above the luminosity threshold, and as its distance is quite uncertain (Sect. \ref{sec:Distance}) we regard it as a marginal case.}. We examined the IRAC images taken in $Ch1$ and $Ch2$ (3.6 and 4.5 $\upmu$m respectively), but found no point source at the location of SN~2018zd. Due to the poor spatial resolution of {\it Spitzer} as well as the shallow nature of these observations, they are of limited use to constrain a progenitor.

\subsubsection{Wide-field survey imaging}
\label{sec:Wide-field}

Along with deep imaging from {\it HST} and large ground-based telescopes, we checked for any source at the position of SN~2018zd from ongoing transient surveys. While such surveys are too shallow to detect the quiescent progenitor of a SN at this distance, they offer the possibility of detecting (or ruling out) precursor outbursts that may be expected for SNe with circumstellar interaction \citep[e.g.][]{Fras13}. Furthermore, fortuitously timed observations may help constrain the explosion epoch.

The latest pre-discovery observations from the PanSTARRS Survey for Transients \citep[][]{Hube15} were taken around one month prior to the explosion of SN~2018zd, and hence do not provide a useful constraint on the explosion epoch. No precursor outbursts or eruptions were detected at this position in the course of the PanSTARRS survey (S. Smartt, private communication).

Unfortunately, the first observations of the field from the Zwicky Transient Facility \citep{Belm19} listed in the LASAIR alert broker \citep{Smit19} are from September 2018, six months after discovery. We also searched the ZTF image archive at the NASA/IPAC Infrared Science Archive (IRSA)\footnote{\url{https://irsa.ipac.caltech.edu/Missions/ztf.html}} to check if template images had been taken prior to this, but the earliest images were taken on the night of 2018 March 19, one month after discovery.
The Palomar Transient Factory \citep{Law09} survey data available through IRSA consists of three shallow observations of the field from 2011 July and 2013 Oct, with no detection of a source at the position of SN~2018zd.

The Asteroid Terrestrial-impact Last Alert System \citep[ATLAS;][]{Tonry18,Smith20} observed the field of SN~2018zd, and while these data do not provide particularly restrictive limits on the explosion epoch, they do help delineate the early rise of the SN (shown in Section \ref{sec:Explosion-epoch}). Similarly, ASAS-SN (All-Sky Automated Survey for Supernovae; \citealp{ASAS_SN_Shappee2014}, \citealp{ASAS_SN_Kochanek2017}) did not detect any historic outbursts for SN~2018zd, but are included in Section~\ref{sec:Explosion-epoch}.

\subsection{Followup observations}
\label{sec:Followup}

\subsubsection{Photometry}
\label{sec:Phot-followup}

SN~2018zd was discovered on 2018 March 2, by K. Itagaki at the coordinates RA = 06:18:03.18, DEC = +78:22:00.90 (J2000). Our intensive multi-band follow-up campaign began on 2018 March 6, approximately 4 days before the object reached maximum light. This campaign lasted just over a year, with extensive coverage in UV, optical and NIR bands using several telescopes. Optical imaging was obtained with the ALFOSC\footnote{Alhambra Faint Object Spectrograph and Camera} and StanCam instruments on the the 2.56~m Nordic Optical Telescope (NOT) on La Palma under the NOT Unbiased Transient Survey (NUTS)\footnote{\url{https://nuts.sn.ie/}},  the AFOSC\footnote{Asiago Faint Object Spectrograph and Camera} instrument on the 1.82~m Copernico Telescope and the 67/92~cm Schmidt Telescope, both at Mt. Ekar (Asiago, Italy), and through various facilities open to the ASAS-SN project, namely the 0.5~m DEdicated MONitor of EXotransits and Transients (DEMONEXT; \citealp{DEMONEXT_2018}) and the 0.5~m Iowa Robotic Telescope (both at the Winer Observatory, AZ, USA), the Las Cumbres Observatory Global Telescope Network (LCOGT; \citealp{Brown2013_LCO}), and the Apogee Alta U230 camera at Post Observatory SRO (CA, USA). 

The images were pre-reduced by various means depending on their origin. The Nordic Optical Telescope near-infrared Camera and spectrograph (NOTCam) NIR imaging was reduced using a modified version of the $notcam$ IRAF package following standard NIR reduction methods, including flat field correction, sky subtraction, and aligning and coadding the consecutive dithered exposures per filter. ALFOSC data were reduced using the {\sc alfoscgui}\footnote{\url{http://sngroup.oapd.inaf.it/foscgui.html}} pipeline. Images obtained through ASAS-SN using the automated ASAS-SN
pipeline. The StanCam, AFOSC, and Schmidt images were reduced using standard techniques within {\sc iraf}\footnote{\url{https://iraf.net/}} \citep{IRAF_1986}. In all cases, images were overscan and bias-subtracted, and flat-fielded. Cosmic rays were masked from long exposures using the {\sc lacosmic} package, and multiple consecutive exposures taken with the same filter were coadded.

\begin{figure*}
	\includegraphics[width=0.8\textwidth]{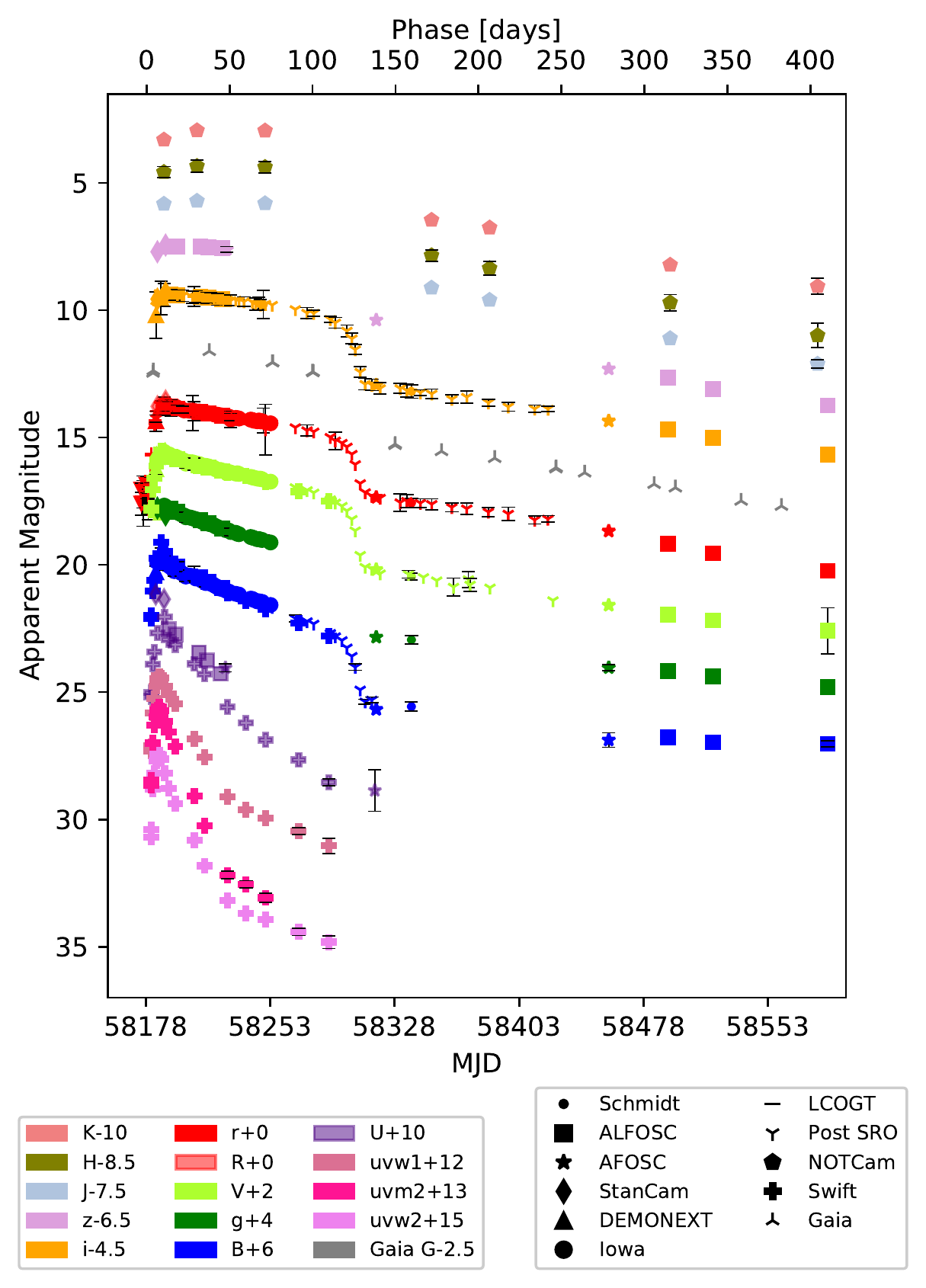}
\caption{Multi-band light curves of SN~2018zd. Phase 0 is the estimated explosion epoch, MJD=58178.5. The data has not been corrected for extinction. Uncertainties only plotted when greater than 0.1 mag.}%
    \label{fig:LC_18zd}
\end{figure*}

Aperture photometry was carried out with a pipeline using PyRAF and Astropy packages, and the results are reported in Tables \ref{tab:opt_phot_table} and \ref{tab:nir_phot_table}, and shown in Fig. \ref{fig:LC_18zd}. The pipeline automatically calculates the full width at half maximum (FWHM) for each image and chooses sources from a given catalogue in the field of view to minimize uncertainties and calibrate the average zero-point. For these data, {\it griz} bands were calibrated using Pan-STARRS1 \citep{Panstarrs_2016_Chambers} data, and NIR data was calibrated with the 2MASS catalogue \citep{2mass_Skrutskie2006}. To calculate Johnson-Cousins $BVRI$ band magnitudes from the Pan-STARRS catalogue, the colour transformations described in \cite{jordi06} were adopted.

SN~2018zd was observed by the {\it Neil Gehrels Swift Observatory} (hereafter {\it Swift}; \citealp{Gehrels_2004_Swift}) UVOT instrument \citep{Roming_UVOT_2005} between 2018 March 8 and 2018 June 25 in all six filters: {\it V} (5468 Å), {\it B} (4392 Å), {\it U} (3465 Å), {\it UVW1} (2600 Å), {\it UVM2} (2246 Å) and {\it UVW2} (1928 Å). While UV data were already published by \citetalias{Hiramatsu2020_18zd} and \citetalias{Zhan20}, we re-analyse them here for consistency. Aperture photometry reported in Table \ref{tab:swift_phot_table} was conducted on the images downloaded from the {\it Swift} data centre\footnote{\url{https://www.swift.ac.uk/swift\_portal/}}. A circular source extraction region of 5 arcsec radius was centered at the source, and the circular background extraction region was defined as a 28 arcsec radius region close to the source. The {\sc heasoft} task {\sc uvotproduct} was used to extract source and sky counts and return the apparent Vega/AB magnitudes in each band.

The field of SN~2018zd was observed also by the {\it Swift} UVOT instrument on 2013 February 17, enabling photometric calibration of the field with 8 reference stars in the {\it u}-band (Table \ref{tab:u_catalog}). 5 {\it u}-band images were downloaded from the {\it Swift} data centre and stacked using the {\sc heasoft} task {\sc uvotimsum}. The {\sc heasoft} task {\sc uvotsource} was used to perform aperture photometry on the reference stars using varying source region radii and a background region of radius 53 arcsec.

For consistency, we also re-analysed the discovery and pre-discovery images taken by K. Itagaki, as well as two images taken close in time by P. Wiggins. The images from Itagaki were taken with an KAF-1001E CCD detector, while those of Wiggins were taken with an SBIG ST-10 camera which has a KAF-3200ME CCD. In both cases, unfiltered images were calibrated to $r'$ band as the closest match to the peak quantum efficiency of the CCD. The Wiggins images were reduced using dark and flat fields, while these calibrations were not available for the Itagaki images. However, as SN~2018zd is centred in the field of view, and through judicious use of nearby sources for photometric calibration, we do not expect the lack of flat field correction to significantly affect the latter.

For the amateur data, {\sc snoopy}\footnote{
SNOoPy is a package for SN photometry using PSF fitting and/or template subtraction developed by E. Cappellaro. A package description can be found at \url{http://sngroup.oapd.inaf.it/ecsnoopy.html.}} was used to fit a PSF at the SN location. In cases where the SN could not be seen, progressively fainter artificial sources were injected at the SN location until they could no longer be recovered by eye (as these data re not particularly constraining on the explosion eproch, the exact method of calculating the limiting magnitude is not critical), and from this, a limiting magnitude was determined. This was performed to minimise the uncertainty on the explosion epoch, which is discussed in detail in Section \ref{sec:Explosion-epoch}.

\subsubsection{Spectroscopy}
\label{sec:Spec-followup}
\begin{figure*}
	\includegraphics[width=0.98\textwidth]{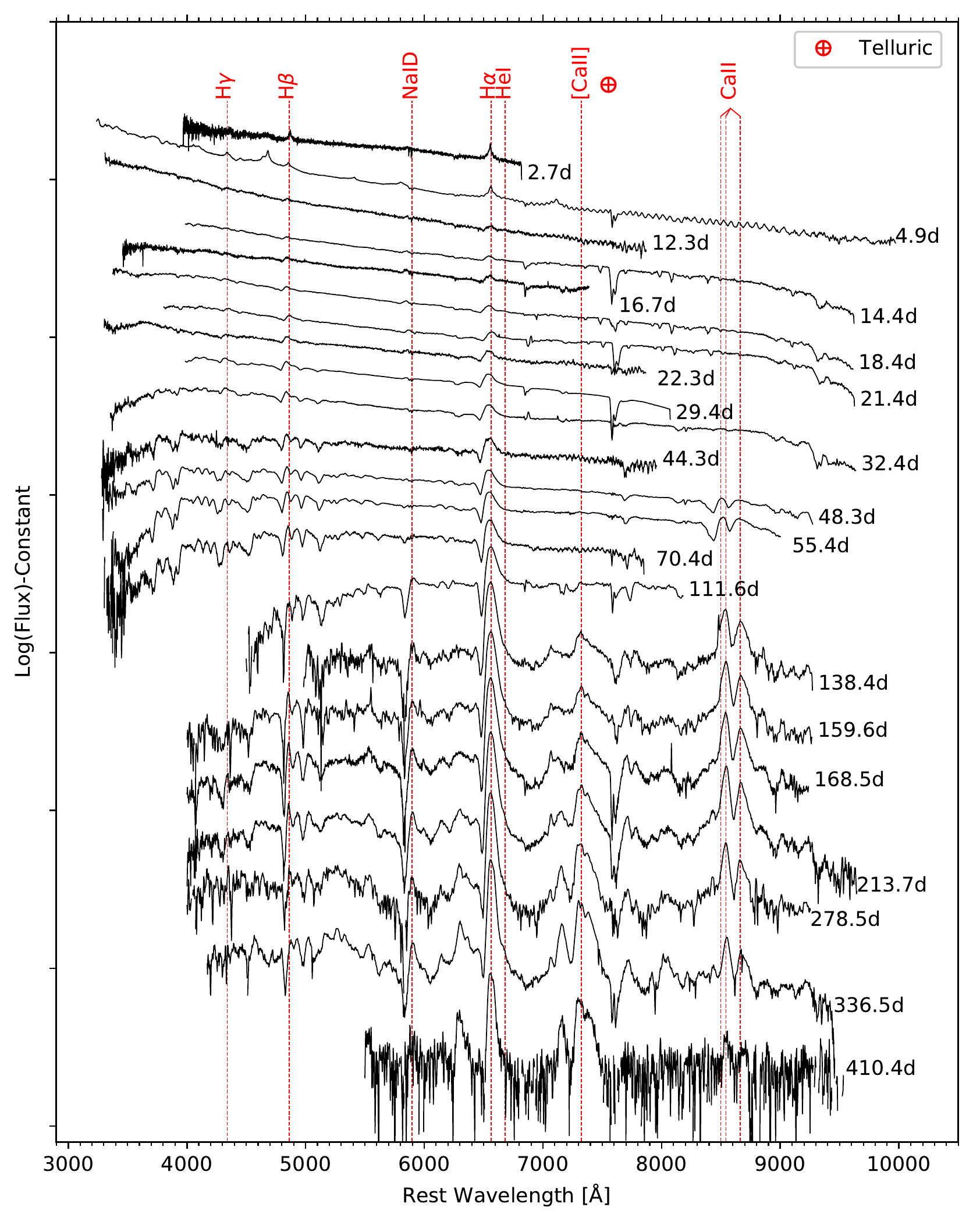}
    \caption{Sequence of optical spectra of SN~2018zd. The spectra have been corrected for the redshift of the host galaxy and the total estimated extinction. The phases are with respect to the estimated explosion epoch. The $\oplus$ symbol indicates the position of significant telluric absorption. A select number of lines are highlighted here, the high ionisation lines are highlighted in Fig. \ref{fig:early_spec}.}
    \label{fig:all_spectra}
\end{figure*}

We obtained a series of 22 optical spectra of SN~2018zd from a number of telescopes, spanning from +2.7~d to +410.4~d with respect to the estimated epoch of explosion.
The details of the telescopes and instrument configuration used are listed in Table~\ref{tab:spectra}. Spectra were obtained through NUTS with the ALFOSC and FIES instruments on the NOT, with the Boller \& Chivens instrument on the 1.22m Galileo telescope \citep{Rafanelli_2012_GalileoTelescope} and the AFOSC instrument on the 1.82m Reflector at the Asiago Observatory, the OSMOS \citep{Martini_2011_OSMOS} instrument on the Hiltner telescope at the MDM Observatory, with the FAST \citep{Fabricant_1998_FAST} instrument on the Tillinghast telescope at the Fred Lawrence Whipple Observatory (FLWO), with ACAM \citep{Benn_2008_ACAM} on the WHT on La Palma, and the Double Spectrograph (DBSP) on the Hale telescope at the Palomar observatory. We also include a spectrum taken with the FLOYDS instrument on the 2m telescope on Haleakala on the same day as the classification spectrum, downloaded from the Transient Name Server (TNS)\footnote{\url{https://www.wis-tns.org/}} and plotted in Fig. \ref{fig:all_spectra}. 

\begin{table*}
	\centering
	\caption{Table of spectroscopic observations. Phase is in days with respect to our adopted explosion epoch of MJD=58178.5.}
	\label{tab:spectra}
	\begin{tabular}{ccccc} 
		\hline
		Date       & MJD        & Phase & Telescope/Instrument      & Grating/Grism\\
		\hline
		2018-03-04 & 58181.2    & 2.7   & Hiltner(2.4m)/OSMOS       & Blue\\
		2018-03-12 & 58189.0    & 10.5  & NOT(2.5m)/FIES            & Fib1\\
        2018-03-13 & 58190.8    & 12.3  & GALILEO(1.22m)/Boller \& Chivens      & 300tr\\
        2018-03-15 & 58192.9    & 14.4  & NOT(2.5m)/ALFOSC         & Gr4\\
        2018-03-17 & 58195.2    & 16.7  & Tillinghast(1.5m)/FAST    & FAST\\
        2018-03-19 & 58196.9    & 18.4  & NOT(2.5m)/ALFOSC          & Gr4\\
        2018-03-22 & 58199.9    & 21.4  & NOT(2.5m)/ALFOSC          & Gr4\\
        2018-03-23 & 58200.8    & 22.3  & GALILEO(1.22m)/Boller \& Chivens      & 300tr\\
        2018-03-30 & 58207.9    & 29.4  & WHT(4.2m)/ACAM            & V400\\ 
        2018-04-02 & 58210.9    & 32.4  & NOT(2.5m)/ALFOSC          & Gr4\\
        2018-04-14 & 58222.8    & 44.3  & GALILEO(1.22m)/Boller \& Chivens      & 300tr\\
        2018-04-18 & 58226.8    & 48.3  & COPERNICO(1.82m)/AFOSC    & VPH6\\ 
        2018-04-25 & 58233.9    & 55.4  & NOT(2.5m)/ALFOSC          & Gr4\\
        2018-05-10 & 58248.9    & 70.4  & GALILEO(1.22m)/Boller \& Chivens      & 300tr\\ 
        2018-06-21 & 58290.1    & 111.6 & COPERNICO(1.82m)/AFOSC    & GR04\\
        2018-07-17 & 58316.9   & 138.4 & COPERNICO(1.82m)/AFOSC     & VPH7\\
        2018-08-08 & 58338.1    & 159.6 &COPERNICO(1.82m)/AFOSC     & VPH6\\
        2018-08-17 & 58347.0    & 168.5 & Hale(5m)/DBSP             & DBSP\\
        2018-09-30 & 58392.2    & 213.7 & NOT(2.5m)/ALFOSC          & Gr4\\
        2018-12-04 & 58456.9    & 278.5 & COPERNICO(1.82m)/AFOSC    & VPH6\\
        2019-01-31 & 58514.9    & 336.5 & WHT(4.2m)/ACAM            & V400\\
        2019-04-15 & 58588.9    & 410.4 & NOT(2.5m)/ALFOSC          & Gr4 \\ 
		\hline
	\end{tabular}
\end{table*}

The spectra were reduced in the standard manner. The frames were bias and overscan subtracted, and divided by a normalised flat field image. One dimensional spectra were optimally extracted, and wavelength calibration was performed using a contemporaneous arc lamp spectrum. In some cases a small shift (typically a few angstroms) was made to bring the night sky emission lines into agreement with their expected wavelengths. The spectra were flux calibrated against a sensitivity curve derived from spectrophotometric standards, and where possible a telluric correction was made using standard stars taken at comparable airmass. These steps were done using {\sc iraf} \citep{IRAF_1986}, by the {\sc foscgui} pipeline for the ALFOSC data, and by the FIEStool\footnote{\url{http://www.not.iac.es/instruments/fies/fiestool/FIEStool.html}} pipeline for the FIES spectrum.
In cases where we make quantitative comparison to spectral models, observed spectra were first scaled to match contemporaneous photometric measurements.
The final sequence is presented in Fig. \ref{fig:all_spectra}. Spectra will be released through the Weizmann Interactive Supernova data REPository \citep{Yaron_2012_WISEREP}.

\section{Distance and Extinction}
\label{sec:Dist_and Extinction}
\subsection{The distance towards NGC~2146}
\label{sec:Distance}
The distance to NGC~2146 is vital to the progenitor mass, ejected Ni mass and explosion energy of SN~2018zd, and hence to its physical interpretation.

The recessional velocity of NGC 2146 (after correction for infall on Virgo, Great Attractor and the Shapley supercluster following \citealp{Mould00}) as listed in the NASA Extragalactic Database (NED\footnote{\url{https://ned.ipac.caltech.edu/}}) is 1219~\kms\ \citep{RCBG}. For a Hubble Constant $H_0 = 74$ \kms\ Mpc$^{-1}$ \citep{Ries19}, this implies a distance of 16.5 Mpc. However, NGC~2146 is not sufficiently distant for us to safely neglect the effect of peculiar velocities (the typical peculiar velocity of a galaxy is 200-300 \kms; \citealp[e.g.][]{Tama11}). In particular, as the correction for infall on Virgo is large, we are cautious of this kinematic distance.

A number of distances determined using the Tully-Fisher relation are listed in NED, ranging from 10.3 to 39.7 Mpc. This large spread can be potentially attributed to the recent history of NGC 2146, which has experienced strong tidal interaction with another galaxy in the last Gyr \citep{Tara01}. It is known that the Tully-Fisher relation has a larger scatter for interacting galaxies. We note however that \cite{Tutu97} have argued that Tully-Fisher distances based on the CO line will be more reliable than H{\sc i}-derived measurements for interacting galaxies. The most recent CO Tully-Fisher relation distance for NGC 2146 is 21.3 Mpc \citep{Tutu97}. While these authors do not quote an uncertainty for this distance, the error on Tully-Fisher distances can be as high as 0.4 mag for individual galaxies \citep{Mast06}. 

Aside from kinematic estimates, \cite{Adam12} suggest that both photometry of the tip of the red giant branch (TRGB) and the size of globular clusters indicate a distance of $\sim18$~Mpc for NGC 2146. Since there are additional $HST$ images that were not analysed by \citeauthor{Adam12}, we revisit the question of the TRGB distance. A detailed description of our analysis can be found in Appendix \ref{sec:TRGB}, but unfortunately we can not place a strong constraint on the distance to NGC~2146 using the TRGB.
We also re-examined the Globular Cluster distances reported by \cite{Adam12}. By fitting a Moffat function to NGC 2146 globular clusters measured in HST+ACS HRC {\it F435W}, {\it F555W} and {\it F814W} images taken in 2005, using the {\sc ishape} package \citep{Lars99} and taking the typical globular cluster radius of 2.66~pc from \cite{Jord05}, we derive a distance of 14.6$\pm$2.3~Mpc.

\begin{figure}
\includegraphics[width=\columnwidth]{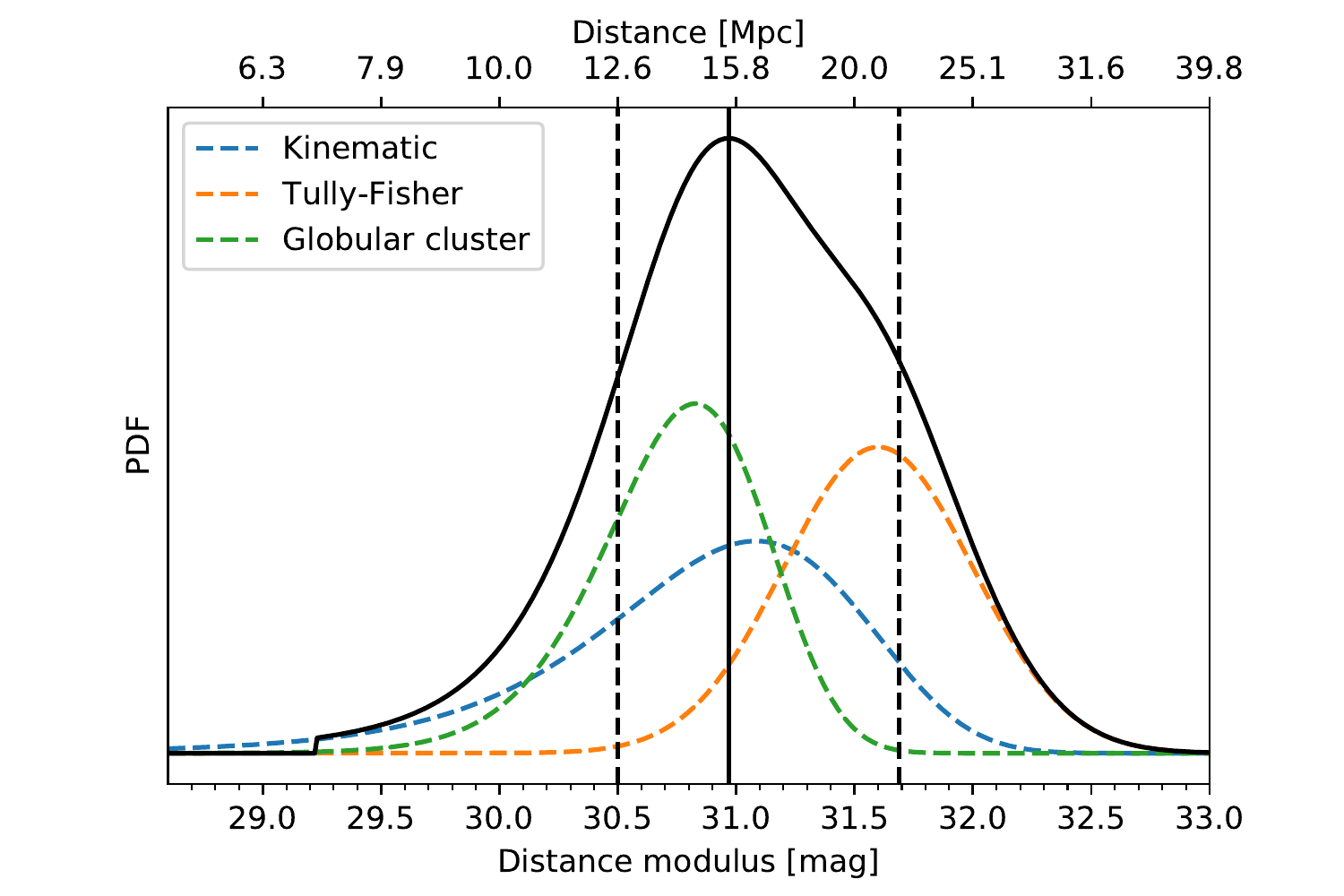}
\caption{PDF for each of the distance measurements considered, along with the joint PDF for NGC~2146 (shown as a solid black line). The solid vertical line marks the maximum of the joint PDF, while the dashed vertical lines mark the 16th and 84th percentiles of the distribution.}
\label{fig:dist}
\end{figure}

We hence have three independent distance estimates from three separate techniques (kinematic, Tully-Fisher and globular cluster radii). The question of how to combine these three independent estimates is somewhat fraught. If all three are unaffected by systematic errors, then we would simply multiply the probability distributions together to obtain the joint probability density function (PDF) for the distance to NGC~2146. However, it is quite possible that one or more of these measurements is systematically biased (for example, if the Tully-Fisher distance is affected by tidal interaction with another galaxy). To account for this possibility, we {\it add} the individual PDFs together, as this does not suppress the PDF for distances which one of the three estimators rules out entirely. We combine the PDF for each of these estimators in Fig. \ref{fig:dist}, and use the peak of the joint probability to infer a distance modulus of $m-M=30.97^{+0.72}_{-0.47}$ mag for NGC 2146, where the uncertainty is obtained from the 16th and 84th percentiles of the PDF. This distance modulus corresponds to a distance of $15.6^{+6.1}_{-3.0}$~Mpc, which is consistent with the distance of $18.4\pm4.5$~Mpc adopted by \citetalias{Zhan20}, but significantly further than the $9.6\pm0.1$~Mpc adopted by \citetalias{Hiramatsu2020_18zd}. In fact, we find there is only a 4 per cent probability that the distance to NGC~2146 is {\it less} than 9.8~Mpc. We will discuss the consequences of this further in Sect. \ref{sec:Conclusions}, however we emphasise here that any parameters calculated for SN~2018zd (e.g. luminosity, progenitor mass, ejected Ni mass) which depend on the distance should be viewed with caution.

\subsection{Extinction}
\label{sec:Extinction}
The \cite{Schl11} dust map gives a moderate colour excess $E(B - V)_{\rm{MW}}$ of $0.083~\pm~0.001$~mag for Galactic extinction in the direction of SN~2018zd, corresponding to $A_V^{\rm{MW}}=0.26$~mag for a standard $R_V=3.1$ dust law \citep{Cardelli89}. The extinction within NGC~2146 is harder to establish. NGC~2146 has an inclination angle of $i=37.4$ deg \citep{Makarov14} and prominent dust lanes are visible within the core of the galaxy. However, as SN~2018zd lies far from these dust lanes at the very tip of a spiral arm, one may expect the extinction to be relatively modest.

We attempt to estimate the host galaxy extinction through analysis of the Na~{\sc i}~D lines, details of which can be found in Appendix \ref{sec:Sodium_lines}, however, due to inconsistency and the known problems with these lines, we instead employ the value for total extinction from \citetalias{Boian_2020_18ZD}, $E(B - V)_{\rm{tot}}$=0.19 mag. This is calculated by selecting the best fit to the spectral energy distribution (SED) of SN~2018zd. This value is comparable to the values published in \citetalias{Hiramatsu2020_18zd} and \citetalias{Zhan20}, 0.17$\pm$0.03 and 0.17$\pm$0.05 respectively. 

\section{Photometric Analysis}
\label{sec:Phot-Analysis}

\subsection{Explosion epoch and early rise}
\label{sec:Explosion-epoch}
In light of the possible early interaction and evidence for confined CSM close to the progenitor, it is of great interest to constrain the early light curve. Not only does this establish how long the high ionization lines persist for, but it potentially allows us to search for signatures of a cooling tail from shock breakout. Indeed, \citetalias{Zhan20} suggested that the first detection by Itagaki caught the shock breakout from SN~2018zd.

\begin{figure}
	\includegraphics[width=\columnwidth]{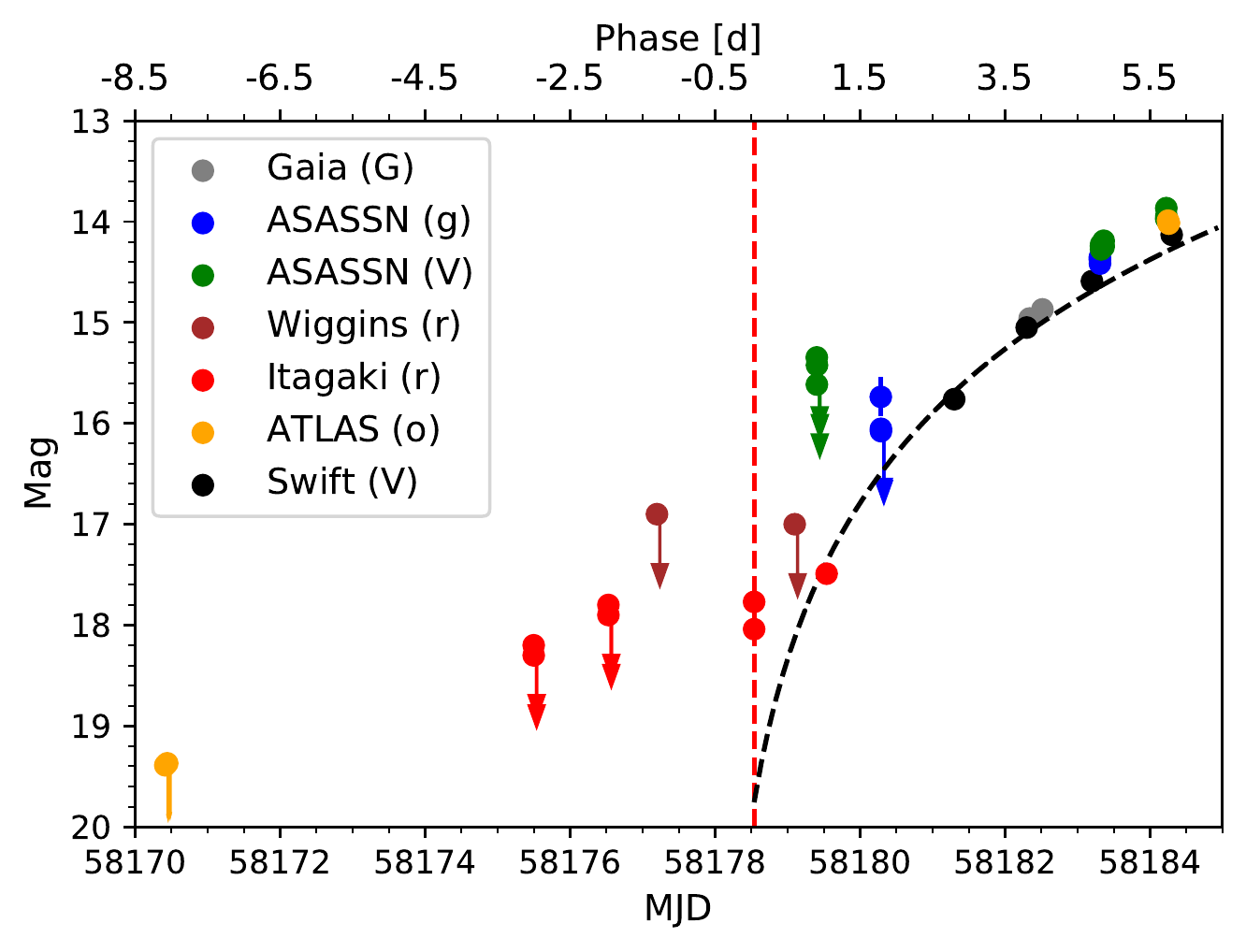}
    \caption{Detections and upper limits for SN~2018zd in a two week period around the explosion epoch. Upper limits to the magnitude are indicated with arrows. The dashed black curve shows the expected rise of an expanding fireball ($\propto t^2$), while the red dashed line marks our adopted explosion epoch (the first detection by Itagaki).}
\label{fig:rise}
\end{figure}

The first images on which we can confidently detect SN~2018zd are two frames taken back-to-back by Itagaki on MJD~58178.54. We detect SN~2018zd in both frames with a magnitude consistent with 0.3~mag. However,  we are unable to constrain the explosion epoch using the pre-discovery images from Itagaki and Wiggins as they are too shallow compared to the expected rise of the SN (Fig. \ref{fig:rise}). Fitting the early part of the light curve we find reasonable agreement with a $F \propto t^2$ rise, and an explosion epoch of MJD 58178.5. 

\subsection{Photometric Evolution}
\label{sec:Phot-evo}

Despite the initial classification of SN~2018zd as a Type IIn SN due to the strong, narrow H$\alpha$ lines and high ionisation features in its early spectra, the photometric evolution of SN~2018zd shows all of the typical features of a Type IIP SN light curve. The signatures of early interaction in the spectra disappear around the time of peak brightness, after which SN~2018zd evolves as a typical Type IIP. This early interaction provides an extra energy source, hence comparison to typical Type IIP events and those that also display signs of early interaction is required to fully interpret its evolution. The peak \textit{V}-band magnitude of 13.51$\pm$0.03~mag was reached approximately 8 days after explosion. The {\it g}, {\it V}, {\it r}, {\it i} bands then decline slowly throughout the plateau phase until +130~d.

\begin{figure}
	\includegraphics[width=\columnwidth]{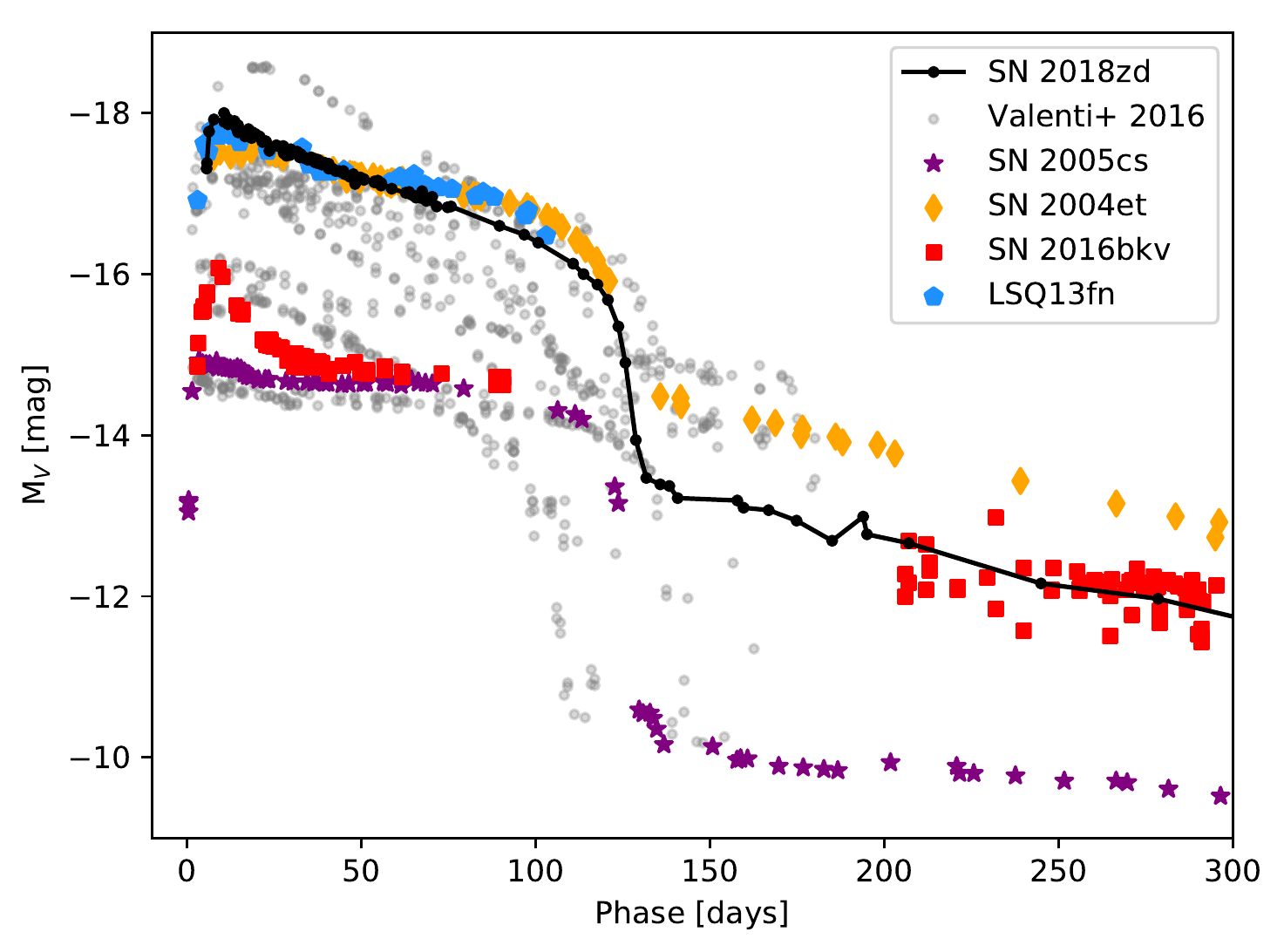}
    \caption{$V$-band light curve of SN~2018zd compared to a sample of Type II SNe. Data: Valenti sample (transparent grey dots) - \protect\citet{Valenti2016}, SN~2005cs - \protect\citet{Pastorello2005cs}, LSQ13fn - \protect\cite{PolshawLSQ13fn}, SN~2004et - \protect\cite{Maguire2004et}, SN~2016bkv - \protect\citet{Hosseinzadeh16bkv}. All light curves have been corrected for reddening, phase 0 is the estimated explosion epoch.}%
    \label{fig:LC_comparisons}
\end{figure}

To compare the{\it V}-band decline rate during the plateau to the other SNe, we followed the prescriptions of \citet{Valenti2016} and computed {\sc s$_{50V}$} (the decline of the $V$-band light curve in mag per 50~d) for SN~2018zd to be 0.89. We also estimated the parameters $a_0$, $w_0$, and $t_{PT}$ as in \cite{Valenti2016}, and found that SN~2018zd falls within the range of values typical for their sample of Type IIP SNe, as can be seen in Fig. \ref{fig:LC_comparisons}.

We also calculated the decline rate of SN~2018zd as prescribed in \citet{Anderson2014} for the variable $s2$ (defined as the decline rate of the $V$-band magnitude per
100 days of the second, shallower slope in the light curve), with $t_{tran}$ (the epoch of transition between the two slopes of the $V$-band plateau) and $t_{end}$ (the end of the plateau phase) were estimated by eye for SN~2018zd, to account for the extra luminosity and steeper decline in the first $\sim$10 days after peak luminosity due to interaction. However, as evidenced in Fig. \ref{fig:LC_comparisons}, SN~2018zd does not show a drastic change in the decline rate of the \textit{V}--band light curve during the plateau, therefore we were satisfied to use $s_{50V}$.
The {\it z}--band photometry is too sparse to calculate its decline rate. The NIR data is also sparse, but it is evident that SN~2018zd exhibits a similar NIR evolution to SN~2004et \citep{Maguire2004et} during its early evolution. To test for the presence of an IR excess in SN~2018zd, we fitted the SED at each epoch where NIR photometry was available with a single blackbody. We found that this reproduced the SED satisfactorily (although the fit does become poorer at late times as the SN becomes nebular). We conclude that there is no strong evidence of an IR excess up to 404~d from these data.

Fig. \ref{fig:LC_comparisons} shows the {\it V}-band lightcurve of SN~2018zd compared to a sample of Type IIP SNe. The \citet{Valenti2016} sample was chosen to provide a comparison to a wide range of Type II light curves, while the individual comparison objects were chosen because they had specific similarities with SN~2018zd, or because they provide a useful comparison. LSQ13fn \citep{PolshawLSQ13fn} and SN~2016bkv \citep{Hosseinzadeh16bkv, Nakaoka_2018_16bkv} both display similar high ionisation features in their early spectra. LSQ13fn has similar peak and plateau magnitudes to SN~2018zd, although unfortunately the decline to the radioactive tail was not observed. SN~2016bkv shows a fast rise and decline from a peak brightness $\sim$2~mag fainter than SN~2018zd. The drop from the plateau was not observed, but later time data (+200d) shows that SN~2016bkv settled onto the radioactive tail $\sim$1~mag fainter than SN~2018zd as shown in Fig. \ref{fig:BOL_LC_COMPS}.
 
SN~2018zd reaches a brighter peak magnitude and declines faster during the plateau phase than SN~2004et, but reaches the end of this phase at a comparable phase and magnitude. These differences may be attributed to the extra energy source provided by the interaction at early times, which would peak in the UV bands.
SN~2005cs \citep{Pastorello2005cs} was chosen as a representative of the underluminous, low-energy, $^{56}$Ni-poor Type IIP SN which form one of the most debated core-collapse supernova sub-groups. This comparison is made to demonstrate that while SN~2018zd has a low estimated $^{56}$Ni mass and low photospheric velocities, it is different from events like SN~2005cs. 

Taken individually, none of the photometric properties of SN~2018zd are particularly extreme. The plateau decline is slightly steeper, the drop onto the Ni tail slightly larger, and the Ni mass is slightly lower than other Type IIP SNe. However, as we will show in Sect. \ref{sec:SCM}, these slight differences are sufficient that SN~2018zd is an outlier to the standard-candle relation for Type IIP SNe. While the lightcurve of SN~2018zd is clearly not {\it dominated} by CSM interaction (the spectral signatures of this persist for only the first few days, and the $\sim$120~day plateau with a clear drop at the end suggests recombination is the dominant energy source), the difference in this SN might be accounted for by a small ongoing contribution from CSM interaction.

\subsection{Colour evolution}
\label{sec:colour_evo}
\begin{figure}
	\includegraphics[width=\columnwidth]{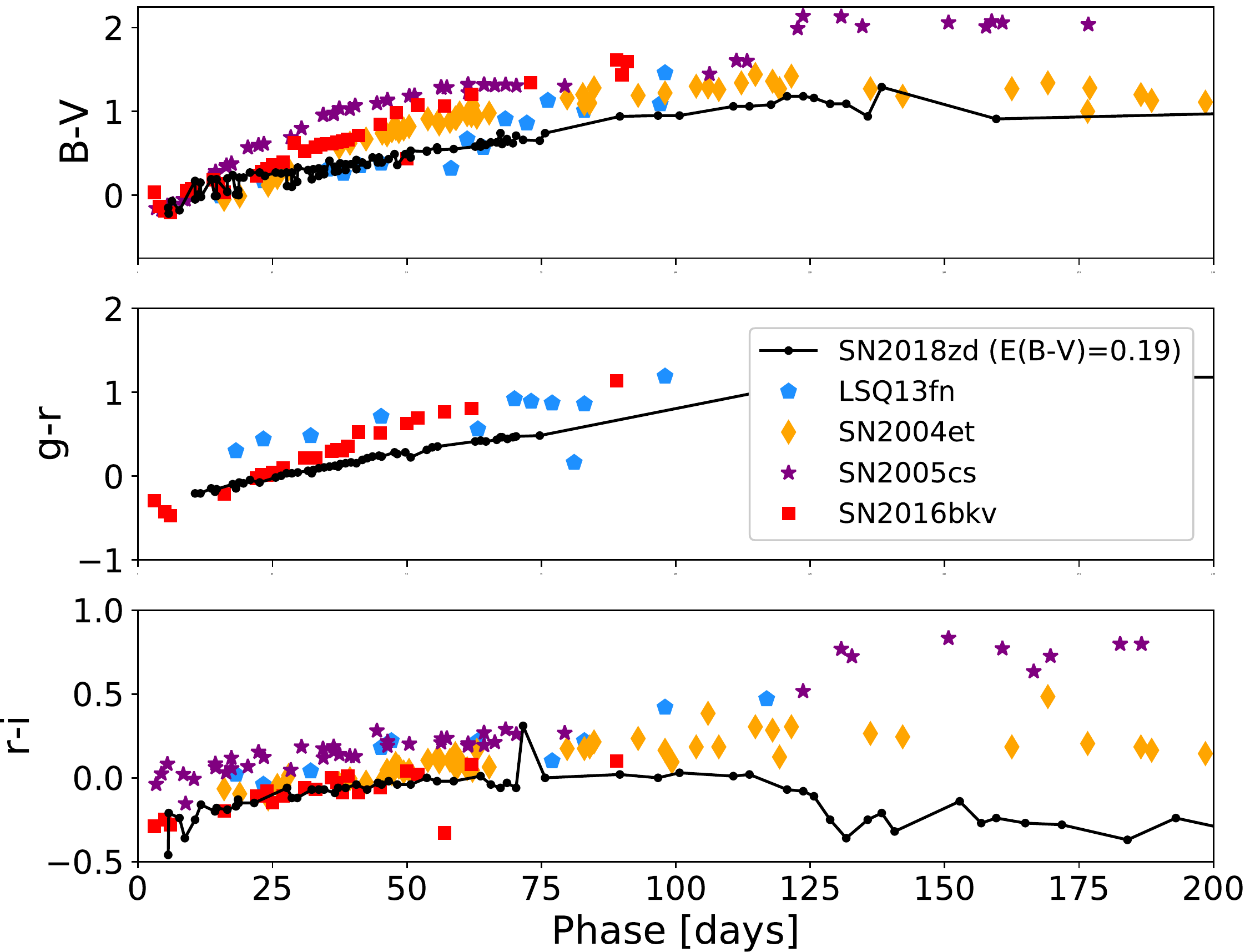}
    \caption{Colour evolution of SN~2018zd and comparison objects. Phase 0 is reference to the estimated explosion epoch, and the photometry has been corrected for reddening. In the $r-i$ (bottom) panel, SN~2004et and SN~2005cs is the $R-I$ colour evolution, with $R$ and $I$ photometry calibrated against the Vegamag system.}
\label{fig:Colour_plots}
\end{figure}

Fig. \ref{fig:Colour_plots} shows the dereddened $B-V$, $g-r$ and $r-i$ colour evolution of SN~2018zd and comparison objects up to +200~d post explosion. For the comparison objects, the $E(B - V)$ values are taken from the literature. SN~2018zd is bluer than the comparison objects throughout its evolution, with the possible exception of SN~2016bkv. For approximately the first week after the explosion, we see the colour getting steadily bluer in the $B-V$ colour plot, and this behaviour is also seen in the UV colour plot of $uvw2-uvw1$ (inset of Fig. \ref{fig:SUPERBOL_BOL_LC} (b)). At this same phase, the temperature is seen to be increasing, and the narrow emission lines are still present in the spectra. 
During the tail phase (>+130~d), the $r-i$ colour becomes bluer. This could be caused by SN~2018zd having a comparably weaker Ca II feature, which could be explained by SN~2018zd residing in a low metallicity environment, or by evolution in the line fluxes (e.g. H$\alpha$).

\cite{Rodriguez2020_LLEVs} showed that there are a family of objects, they term luminous, low expansion velocity Type~II SNe (LLEV's), that all have blue $B-V$ colours compared to normal Type~II SNe. They include LSQ13fn in their sample, and it is evident here that SN~2018zd has a similar colour evolution. The two main contributors to bluer colours could be a low metallicity environment, as the bluer bands are not subjected to blanketing by metal lines, and extra UV flux from the SN ejecta interacting with CSM. \cite{Rodriguez2020_LLEVs} show that in objects that can be classed as LLEV's, metallicity of the progenitor has less of an impact on the observed characteristics, and the majority of the observed differences can be primarily attributed to CSM interaction.

\subsection{Bolometric luminosity and nickel mass}
\label{sec:Bol_LC}

\begin{figure*}
	\includegraphics[width=0.7\textwidth]{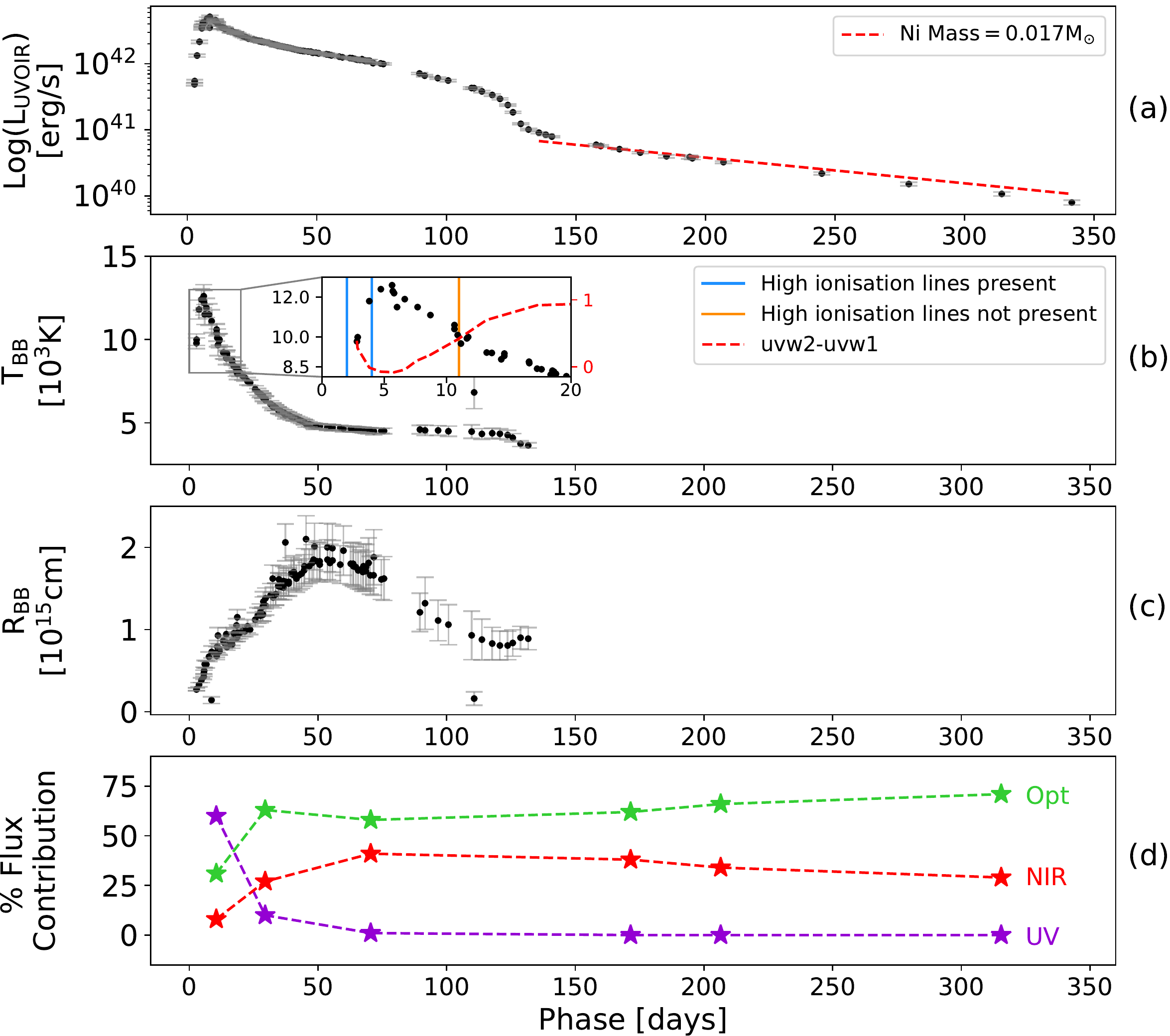}
    \caption{(a) Bolometric light curve of SN~2018zd computed using {\sc superbol}. Red dashed line is the expected tail luminosity from the calculated Ni mass. (b) The temperature evolution computed from {\sc superbol}. Inset focuses on time around peak luminosity, vertical lines denote the presence or absence of high ionization features in the spectra, the red dashed line is the vertically scaled \textit{uvw2-uvw1} colour evolution. (c) The radius evolution computed with {\sc superbol}. (d) Percentage flux contributions from UV, optical, and NIR.}
    \label{fig:SUPERBOL_BOL_LC}
\end{figure*}

The bolometric light curve in Fig. \ref{fig:SUPERBOL_BOL_LC} was computed with {\sc superbol}\footnote{\url{https://github.com/mnicholl/superbol}} \citep{Nich18} using all the data available in Tables \ref{tab:opt_phot_table}, \ref{tab:nir_phot_table} and \ref{tab:swift_phot_table}, and interpolating between epochs by invoking constant colour terms. We then computed multiple pseudo-bolometric light curves spanning just the optical bands, the UV + optical, and the optical + NIR. When comparing the extrapolated bolometric light curves produced by {\sc superbol} from the reduced wavelength ranges to the full bolometric light curve, we find that they are in good agreement. 
The bolometric light curve extrapolated from only the optical data was predictably the least accurate, overestimating the peak and tail luminosities. These tests give us confidence in the extrapolation of the light curve in the NIR and UV bands over epochs when data are sparse. 

{\sc superbol} was also employed to compare the separate contributions of the UV, optical, and NIR luminosities to the bolometric light curve (Fig. \ref{fig:SUPERBOL_BOL_LC} (d)). As expected, the UV contributes approximately 50 per cent at very early times and decreased exponentially over the first 75 days, while the NIR contribution increases from approximately 10 per cent to 50 per cent over the same time span. The optical component contributes approximately 30 per cent for the first 3 weeks, thereafter contributing approximately 60 per cent for the duration of the event.

The blackbody temperature at early times is relatively low, with {\sc superbol} returning a peak of only $\sim$12 kK. This is in part due to the extreme nature of the temporal evolution of temperature at early times, which could be dampened by using the {\it V}-band data (the band with the most data points) as the reference band and invoking constant colours when interpolating each light curve. We also note that as we do not directly observe the shock breakout as it flash ionizes the lines, it is not necessary for the temperature at the time of the spectrum to match the temperature needed to produce the lines. We also ran {\sc superbol} on just the {\it Swift} data\footnote{While the {\it UVW2} and {\it UVW1} filters have a ``red leak'' we do not expect this to be significant at early times when the SED of SN~2018zd is very blue.}, and used the {\it UVW1} {\it Swift} filter, centered on 2600 \AA, as the reference band, and show the resulting temperature evolution in Fig. \ref{fig:Swift_temp_evo}. Fitting the early temperature evolution still proves difficult, with large uncertainties at early times. By fitting just the bluer bands we see that SN~2018zd reaches the required temperatures to explain the high ionisation lines within the uncertainties. This can be understood by considering that interaction with CSM will provide a strong source of ionising X-ray flux. We can also note that by the mid-plateau phase (+50~d) the temperature evolution computed from just the bluest bands and the evolution computed from all the {\it Swift} data converge, implying that the evolution around the peak phase is dominated by the flux from the UV, and that this extra source of photons steadily decreases towards the end of the plateau. 

In Fig. \ref{fig:SUPERBOL_BOL_LC} (b) we include an inset showing the temperature evolution near peak, overplotted with vertical lines showing epochs of spectra and a red dashed line depicting the UV colour evolution with a shifted y-axis. The increasing temperature and blue trajectory of the UV colour for the first $\sim$5~d (also noted by \citetalias{Zhan20}) is not typically seen in Type II SNe with early enough observations (in fact, we are unaware of this in any other observed Type II SNe). This ``u-turn" can be attributed to CSM interaction at early times, when the photosphere is formed within the unshocked CSM, which provides an additional energy source as the shock front propagates through the CSM. This is supported by the presence of high ionisation lines in the spectra only during the epochs where the temperature is increasing. 

 We obtain an estimated $^{56}$Ni mass of 0.017$\pm$0.004~$\rm{M_{\odot}}$ for SN~2018zd using the prescriptions from \citet{Hamuy2003_NiMass}. Inputting only the values of the bolometric light curve from approximately +150~d to +200~d, as after this phase the light curve appears to decline faster than the expected radioactive decay rate, which could be due to incomplete $\gamma$-ray trapping. To estimate the length of time that the photosphere of SN~2018zd would remain optically thick, 
 we use eq. 58 in \cite{Jerkstrand_2017_Neb_Spec_HBSNe}. For a 6 \msun\ ejecta mass and a 1 foe explosion energy (reasonable for a low mass RSG exploding as a Fe core-collapse SN), we expect the ejecta to start to become optically thin at $\approx$~200~d, consistent with when we begin to see a faster decay.

We also attempt to estimate the ejected $^{56}$Ni using the steepness of decline from the plateau onto the radioactive tail as we have excellent coverage over this period. We measure $\frac{dM_V}{dt}=0.32$ mag~d$^{-1}$ from our photometry, and using eqn. 3 from \cite{Elmhamdi03} with this decline rate gives a $^{56}$Ni mass of $\sim$0.002~\msun.
This is considerably smaller than what we find from the tail phase luminosity of SN~2018zd, 
however we regard the latter as being a more reliable measurement as it is a direct measure of the luminosity from radioactive decay. It is unclear why the \citeauthor{Elmhamdi03} relation gives such a discrepant result (although we note that there is a scatter of $\sim$0.5~dex in the relation), but one may speculate that this may be due to a different level of mixing in the explosion of SN~2018zd.

The calculated $^{56}$Ni mass is comparable to the mass of 0.02~$\rm{M_{\odot}}$ found for LSQ13fn by \citet{PolshawLSQ13fn}. 
\citet{Maguire2004et} calculate a $^{56}$Ni mass of 0.04~$\rm{M_{\odot}}$ for SN~2004et, which settles onto the tail phase at $\sim$0.5 dex brighter than SN~2018zd. 
\begin{figure}
	\includegraphics[width=\columnwidth]{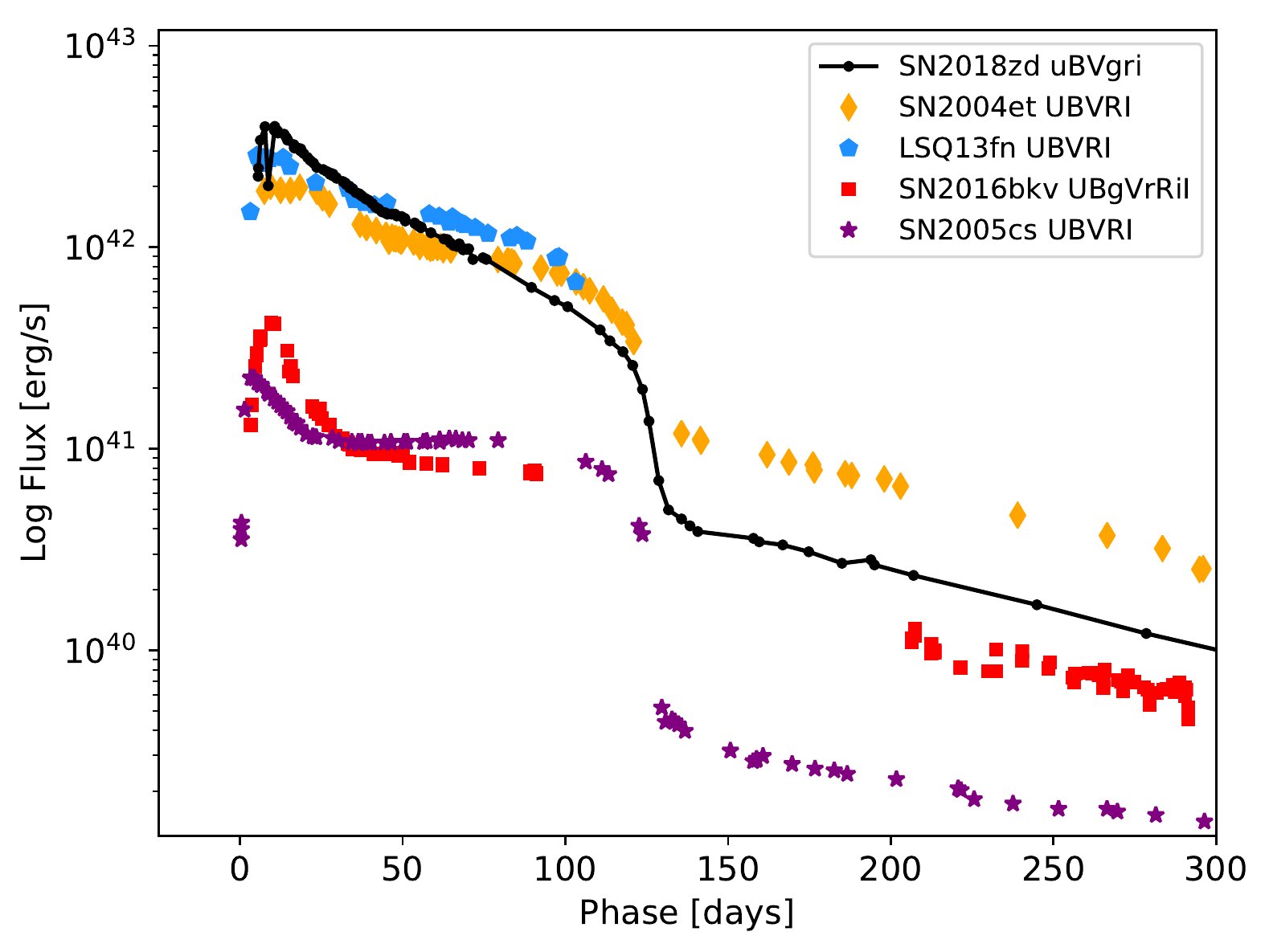}
    \caption{Comparison of the optical pseudo-bolometric light curves of SN~2018zd, SN~2004et, SN~2005cs, SN~2016bkv, and LSQ13fn. Phase 0 is the estimated explosion epoch.
    }
    \label{fig:BOL_LC_COMPS}
\end{figure}
Fig. \ref{fig:BOL_LC_COMPS} displays the pseudobolometric light curves of SN~2018zd and the comparison objects computed with {\sc superbol} from their available optical band photometry. The comparison objects were computed similarly to SN~2018zd, employing constant colour terms when extrapolating bands, and with parameters as listed in Table \ref{tab:SUPERBOL_params}.

It is clear that SN~2018zd is not in the regime of the low-luminosity Type~IIs. The mid-plateau magnitude and plateau duration are comparable to the archetypal SN~2004et, however there are some important discrepancies. The first is the increased peak luminosity and faster decline rate during the plateau. This is mainly the result of increased energy deposition from the early interaction, also evidenced by the blue colour at this epoch. We applied a similar analysis to the bolometric light curve as to the {\it V}-band light curve, following the prescriptions in \cite{Valenti2016}. While the \textit{V}-band plateau phase light curve of SN~2018zd could be reasonably described with a single slope value, for the bolometric light curve it is clear that there is a change in the rate of decline $\sim$45~d after explosion. We therefore calculate \textit{S1} and \textit{S2} as in \cite{Valenti2016}, and find the \textit{S2} value to be within the parameter space found therein. However, the \textit{S1} value for SN~2018zd is steeper than any in the \cite{Valenti2016} sample, which is to be expected from the observed strength of the interaction experienced at these epochs by SN~2018zd.
Secondly, SN~2018zd settles onto the radioactive tail phase $\sim$0.5~dex fainter than SN~2004et. \cite{Rodriguez2020_LLEVs} note that this enhanced plateau {\it V}-band magnitude seen in the objects in their sample can be attributed to an extra source of photons, still relevant at +50~d after explosion, which then becomes negligible during the radioactive tail phase, and conclude that the source of these photons is the early interaction observed in these objects. 

From the expansion of the blackbody radius shown in Fig. \ref{fig:SUPERBOL_BOL_LC} (c), we estimate the photospheric velocity to be approximately 4,000~km~s$^{-1}$ over the first 50 d after explosion. This is in agreement with the velocity derived from the Fe~{\sc ii} lines, discussed further in Section \ref{sec:Vel-evo}. After the first 50~d the photosphere is seen to recede inwards radially. 

\section{Spectroscopic Evolution}
\label{sec:Spec-evo}

We obtained spectra during all of the evolutionary phases. The first epoch of spectroscopy, taken $\sim$2~d after the explosion, displays a blue continuum with a number of strong emission features. There are several high ionisation lines present in the spectrum at $\sim$4~d, which disappear by the next spectrum taken a week later. The spectral evolution becomes typical of Type~IIP SNe thereafter. The Balmer series is easily identified, most clearly H$\alpha$ $\lambda$6563 and H$\beta$ $\lambda$4861. The blue continuum and Balmer series continue to dominate through the photospheric phase, displaying strong P Cygni line profiles similar to normal Type~IIP SNe. The full sequence of spectra are shown in Fig.~\ref{fig:all_spectra}. 

Towards the end of the plateau phase, we see the emergence of more features, including numerous Fe lines in the blue, a strong Na~{\sc i}~D P Cygni line, with [Ca {\sc II}] and the Ca {\sc II} triplet emerging slightly later, denoting the transition to the nebular phase. From approximately +140~d the nebular spectra are dominated by the calcium features and H$\alpha$. We also see the emergence of the [O {\sc I}] $\lambda\lambda 6300, 6364$ doublet. A more detailed description of the spectroscopic evolution is provided in the proceeding subsections. 
 
\subsection{Early Spectra with high ionisation lines}

\label{sec:Early-spec}
\begin{figure*}
	\includegraphics[width=1\textwidth]{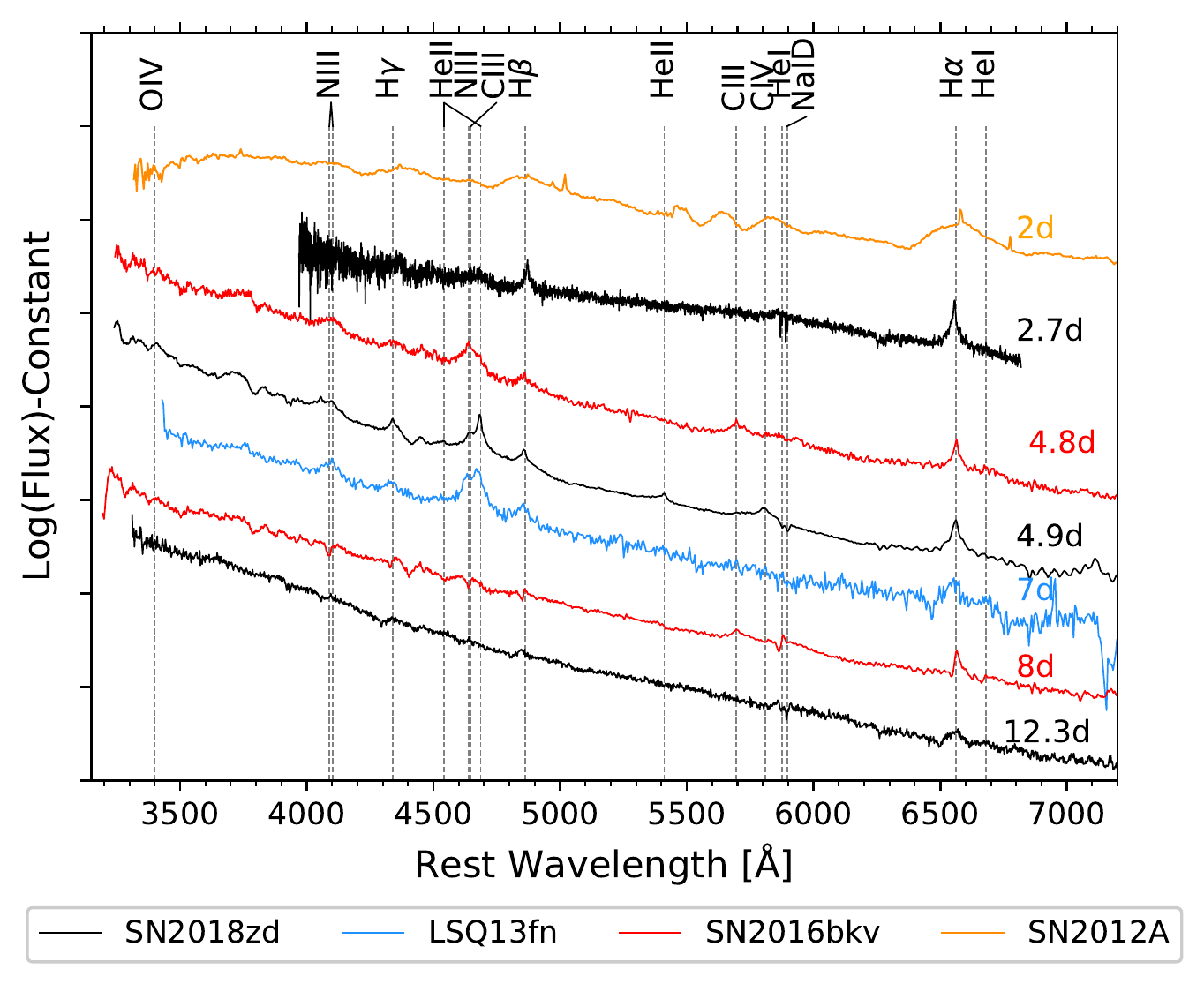}
    \caption{Earliest SN~2018zd spectra compared to SN~2016bkv and LSQ13fn to highlight similarities in the high ionisation features around 4650 \AA. We also compare to the Type IIP SN~2012A, a SN that does not exhibit these high ionisation lines at early times. 
    }
    \label{fig:early_spec}
\end{figure*}

Fig. \ref{fig:early_spec} shows the three earliest epochs of spectra we have for SN~2018zd to highlight the first 5 days of evolution. We also show similar epochs of spectra from SN~2016bkv and LSQ13fn, which also display this feature. Finally, SN~2012A is included as an example of a ``normal'' Type IIP SN which does not exhibit narrow high ionisation emission lines during its early evolution. 

We first estimate the mass loss rate from the strength of H$\alpha$, following \cite{Kochanek19} and assuming Case B recombination in a pure (i.e. metal-free) H wind. In our earliest spectrum we measure a flux in H$\alpha$ of $1.1\times 10^{51}$ photons~s$^{-1}$ which corresponds to a mass loss rate of $5^{+2}_{-1}\times 10^{-4}$~$\rm{M_\odot}$~yr$^{-1}$.

The evolution of the high ionisation lines during this period is of interest, with several features being identified in the spectrum at +4.9~d that were not visible in the spectrum at +2.7 d. The line at $\sim$5800~\AA\ which is at its strongest in the +4.9~d spectrum, was difficult to identify and it is partially blended with the Na~{\sc I}~D lines. We determined that it was not He~{\sc I} as there is no evidence of the He~{\sc I} line at 6678 \AA. We were also not convinced that it is \Ciii, although there is evidence of other \Ciii\ lines in the spectrum, as their peaks are not as drastically redshifted from their rest wavelength as we would need to invoke for this feature. Similar can be said for \Niii . We identify this line as \Civ, which agrees with the identification by \citetalias{Boian_2020_18ZD}, and also note the presence of a feature at $\sim$3400 \AA\  that we identify as possibly O {\sc iv}. 

These features are also seen in the winds of WC stars \citep{Crowther07_WR}, the subtype of which depends on the line ratios of \Ciii\ and \Civ\ lines and the emergence of the O~{\sc iv} feature. These highly ionised lines require temperatures of $\sim$40~kK, which is too high to be explained solely by shock breakout 4~d after explosion. Delayed shock breakout due to the presence of dense CSM could explain this evolution, and a detached shell of CSM was invoked to explain the very similar evolution of high ionisation lines seen in SN 2013fs (\citealp{Yaron2017_SN2013fs}; although see \citealt{Kochanek19} for an alternative interpretation that does not require a pre-SN outburst). 

What is remarkable about SN~2018zd (and indeed similar transients such as LSQ13fn) is that the narrow high ionization lines persist for days, compared to SN~2013fs where they lasted for only $\sim24$~hr. This implies that these lines form further out in the CSM of SN~2018zd compared to SN~2013fs. The outer edge of the CSM region where these lines form is set by the light travel time since shock breakout ($c \times t$). The inner edge is set by the SN ejecta velocity ($v \times t$), as inside this radius the fast supernova ejecta will have overrun the CSM. Taking a conservative value of 4,000~\kms\ for the fastest ejecta in the CSM, consistent with the velocity derived from the blackbody radius expansion (Section \ref{sec:Bol_LC}) and the Fe~{\sc ii} triplet (Section \ref{sec:Vel-evo}), we find that for our 4.9~d spectrum the CSM must lie $\sim$2500~\rsun\ from the progenitor.

The second puzzling aspect of the early emission is that the narrow metallic lines are not seen in our first spectrum at +2.7 days, where only H is visible. Since CNO lines are harder to ionize than H, one would expect them to be also present in the earliest spectrum (when there will be a hotter temperature and more ionizing photons). Moreover, for reasonable parameters for a red supergiant progenitor, one would expect the high ionisation lines (e.g. \Ciii) to be only present for the first $\sim$12 hours after explosion \citep{Kochanek19}.

One possibility that could explain these aspects is that the source of ionizing photons that excites the metallic lines in the CSM is not coming from shock breakout, but rather from a shock formed where the SN ejecta hits a denser region of CSM some days after explosion. There is no reason to expect such a CSM structure in a stellar wind (where the density should fall off $\propto r^{-2}$), however, it could be created by a pre-SN outburst. Assuming that the material lost in this putative outburst had a velocity of 500 \kms, this would have needed to occur around one month before the explosion of SN~2018zd to be consistent with the inferred location of the CSM. 

Alternatively, one could speculate that a dense CSM shell was formed in a colliding wind binary system \citep[e.g.][]{Prilutskii76}, as suggested for SN~2013fs by \cite{Kochanek19}. By necessity, such a binary companion would have to generate strong stellar winds, and avoid exploding before the (presumably lower mass) red supergiant that produced SN~2018zd. To explore this would require detailed modeling of binary evolution that is beyond the scope of this paper. A final possibility is that the dense CSM lies in a photo-ionization confined shell created by the radiation from nearby OB stars \citep{Mackey14}. However, the immediate environment of SN~2018zd (Fig. \ref{fig:progenitor}) does not show evidence for any young massive cluster that one would naturally associate with a group of OB stars.

By the time of our third spectrum at +12~d, the strong narrow emission features have disappeared, and the spectrum consists of a blue, featureless continuum, with a weak, relatively broad P Cygni line profile beginning to appear at H$\alpha$. The Balmer series continues to strengthen over the next $\sim$60~d as the continuum steadily becomes less blue. 

\subsection{Plateau-Phase Spectra}
\label{sec:Plateau-spec}
\begin{figure}
	\includegraphics[width=\columnwidth]{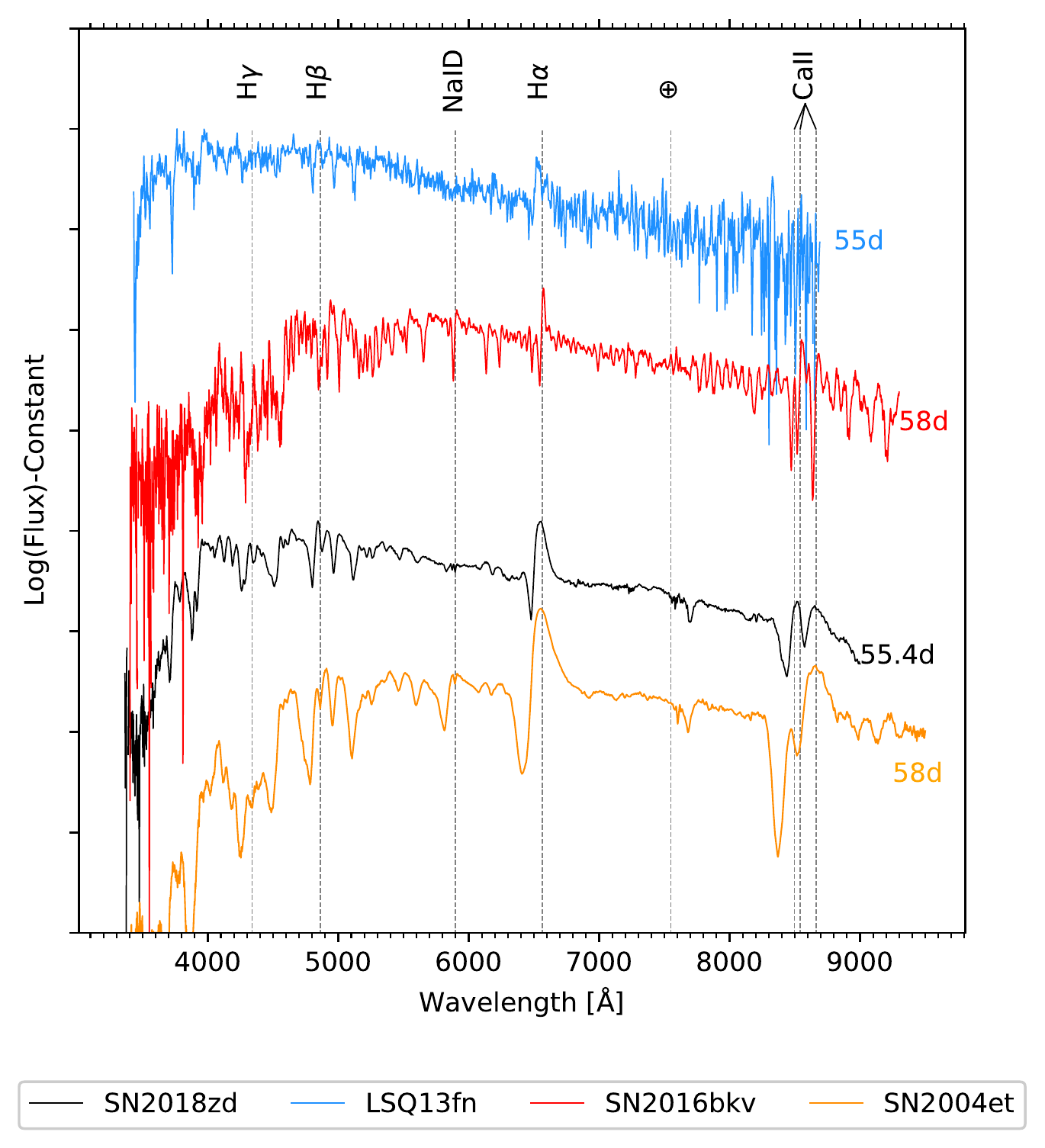}
    \caption{De-reddened spectra of SN~2018zd and comparison objects approximately at mid-plateau phase.} 
    \label{fig:plateau_spec}
\end{figure}
Fig. \ref{fig:plateau_spec} compares SN~2018zd during the plateau phase to LSQ13fn, SN~2016bkv and SN~2004et. During this phase, the spectra are similar to what is expected for a normal Type IIP SN such as SN~2004et. We have labeled the most prominent lines, including the Balmer series and the Ca~{\sc ii} triplet. We also see Fe~{\sc II} lines at <5200~\AA\, from which we determine a mid-plateau velocity of $\sim$3,100 km s$^{-1}$  discussed in detail in Section~\ref{sec:Vel-evo}. At this epoch, we would expect to see strong features of Sc~{\sc ii}, Ba~{\sc ii}, and Na~{\sc i}~D between 5200-6600~\AA\ in a normal Type IIP, however, these are absent. Weaker metallic lines suggest a low metallicity progenitor \citep{Anderson_2018_SN2015bs, Gutierrez_2018_Low_Lum_Host_TypeIIs}. Comparing the H$\alpha$ lines, SN~2018zd lies between SN~2016bkv and SN~2004et with respect to the depth and velocity of the absorption feature. We again note the absence of a strong Na~{\sc i}~D P Cygni feature in SN~2018zd, which is expected during this phase for normal Type IIP SNe and can be clearly seen in SN~2004et at 5890-5896~\AA.

The spectrum of SN~2018zd below 5000~\AA\ appears bluer than that of the comparison objects shown in Fig. \ref{fig:plateau_spec}. This region of the spectrum is strongly affected by the presence of many lines from Fe-group elements \citep{Turatto_1988Z}. We note that the bulge of the host galaxy is at roughly solar metallicity \citep{Aniano_2020_host_metallicity}, however, SN~2018zd is situated at the very edge of the spiral galaxy, and therefore would most likely have sub-solar metallicity. \cite{Anderson2016_II_metallicity} show that the equivalent width of Fe~{\sc ii} $\lambda$5018 \AA\ at 50~d post-explosion shows a statistically significant correlation with host HII-region oxygen abundance. We compared SN~2018zd to the correlation they find in their sample of Type II SNe with no cuts applied to the sample regarding light curve or spectral features. Fig. \ref{fig:anderson_metallicity_comp} shows that SN~2018zd would have an estimated 12+log[O/H] of approximately 8.32~dex, roughly equating to 0.2~$\rm{Z_{\odot}}$, or similar to the Large Magellanic Cloud (LMC). However, it must be noted that the scatter on this correlation is quite large, and therefore we put no constraint on SN~2018zd's metallicity other than it most likely sub-solar. This could partially explain why SN~2018zd is persistently bluer than the comparison objects, as it would be less affected by line blanketing in the blue part of the optical, where SNe are more sensitive to the composition of the photosphere.

The end of the plateau phase is denoted by the emergence and strengthening of the Ca~{\sc ii} features and the Na~{\sc i}~D line at approximately 111~d, before the spectra transition into the nebular phase.

\begin{figure}
	\includegraphics[width=\columnwidth]{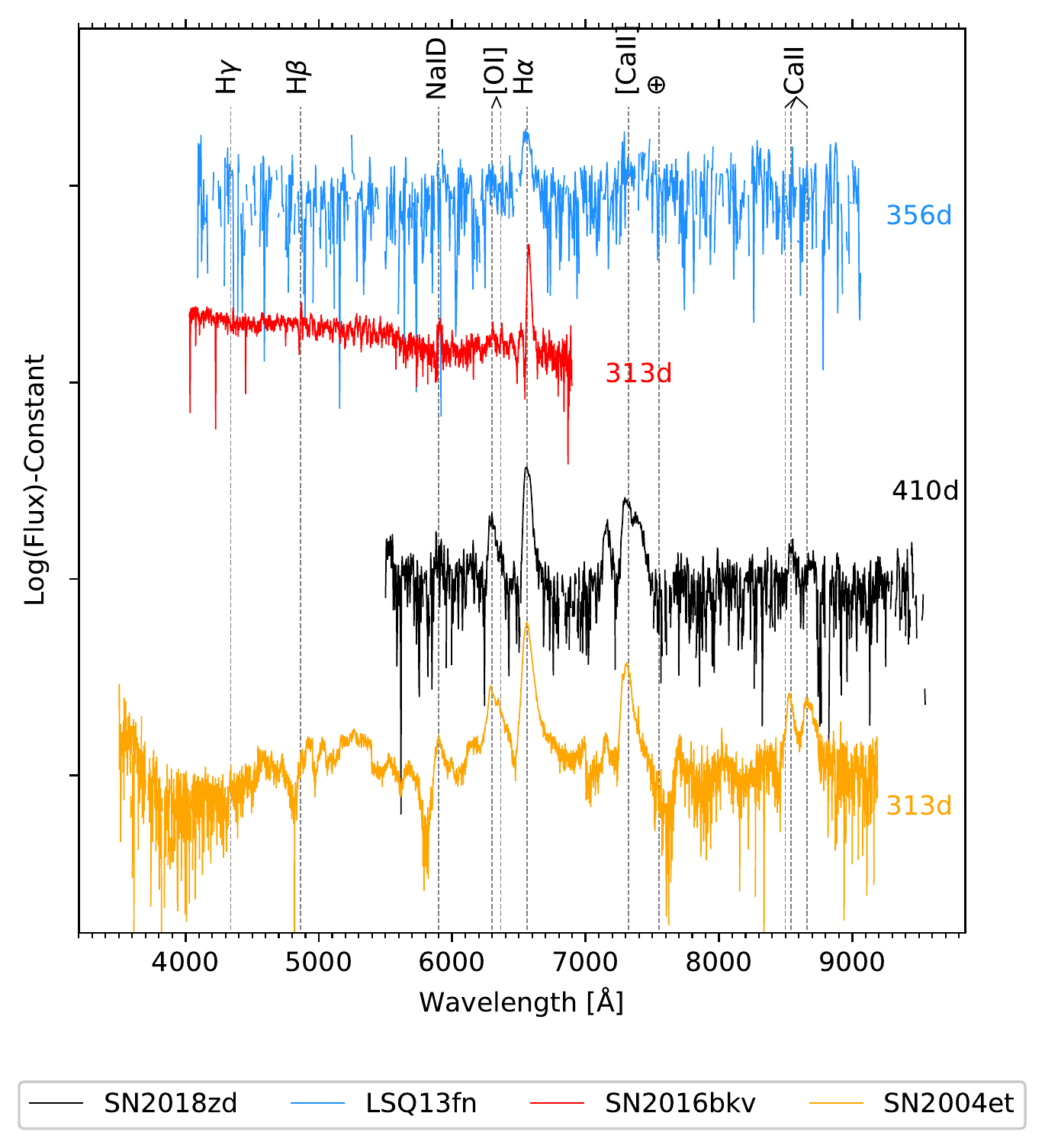}
    \caption{Spectrum of SN~2018zd and comparison objects during the nebular phase.}
    \label{fig:neb_spec}
\end{figure}

\subsection{Nebular-Phase Spectroscopy}
\label{sec:Nebular-spec}

\begin{figure}
	\includegraphics[width=\columnwidth]{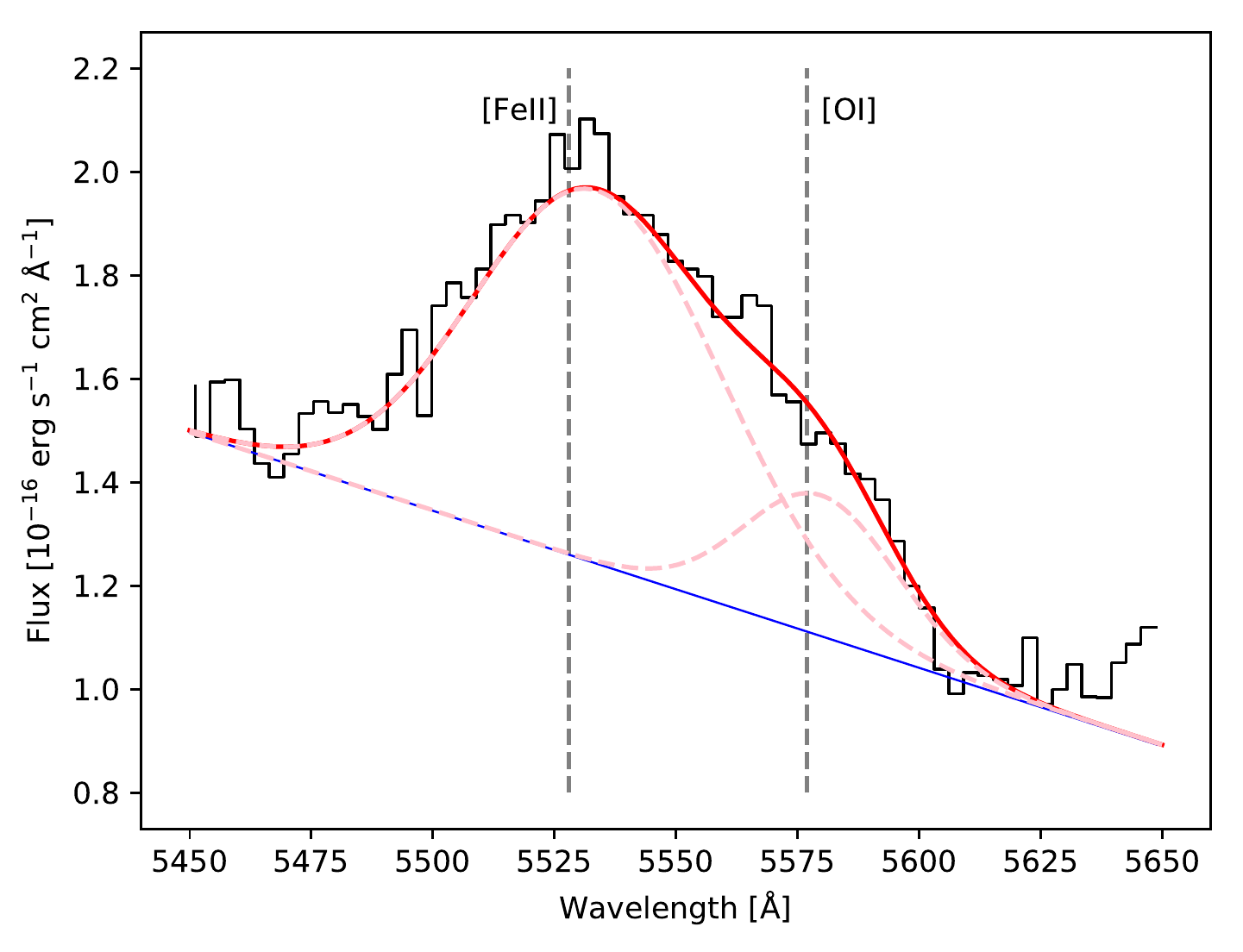}
    \caption{SN~2018zd (black, solid) between 5425-5675\AA\ at +214~d (dereddened and redshift corrected), along with the double Gaussian fit to [Fe~{\sc ii}] $\lambda$5528 and [O I] $\lambda$5577 (pink, dashed), and the chosen continuum fit (blue, solid). }
    \label{fig:O5577_splot_fit}
\end{figure}

The latest epoch of SN~2018zd spectroscopy is shown along with comparison objects in Fig. \ref{fig:neb_spec}. The features associated with the nebular phase are clearly present in the SN~2018zd spectrum, including [O~{\sc i}] $\lambda\lambda$6300,6364, [Fe~{\sc ii}] $\lambda$7155, and [Ca~{\sc ii}] $\lambda\lambda$7291, 7323. As the latest epoch of spectroscopy clearly shows that SN~2018zd is in its nebular phase, we can estimate the core oxygen mass from the [O~{\sc i}] $\lambda \lambda$6300,6364 doublet using the prescriptions in \cite{Jerkstrand_2014_OI_mass}. We obtain a value of 0.064$\pm$0.002 $\rm{M_{\odot}}$ by fitting the [O~{\sc i}] lines in the +214~d spectrum. This is an upper limit as it assumes that the luminosity is derived entirely from primordial oxygen, which is not the case for lower mass progenitors \citep{Maguire2012_IIP_NebSpec}. Deblending the [O~{\sc i}] $\lambda$5577 from the [Fe~{\sc ii}] $\lambda$5528 line at this epoch was difficult, and we show our fit in Fig. \ref{fig:O5577_splot_fit}. In the later epochs, local thermodynamic equilibrium (LTE) can not be assumed and only an upper limit to the oxygen mass can be derived. 

\begin{figure*}
	\includegraphics[width=2\columnwidth]{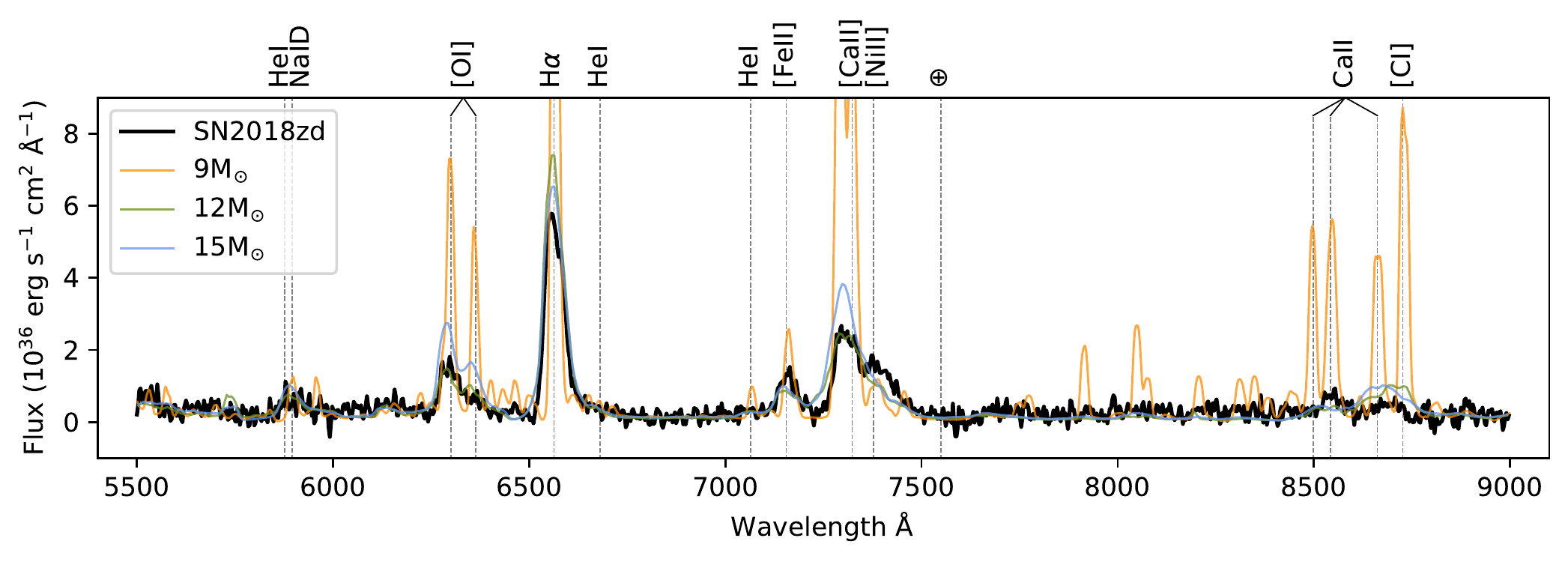}
    \caption{
    Spectral comparison of SN~2018zd at +410d to the models of \protect\cite{Jerkstrand_2015_Followup_NiFe_ratios} and \protect\cite{Jerkstrand_2018_9Msol}. The explosion energies and treatment of extinction differ between the 9 $\rm{M_{\odot}}$ and the 12-15 $\rm{M_{\odot}}$ models, accounting for the discrepancies seen between the models. The prescriptions for the Ni/Fe ratio presented in this paper are taken from the $\rm{15M_{\odot}}$ model.}
    \label{fig:jerkstrand_model_comps_ni58_inset}
\end{figure*}

Fig. \ref{fig:jerkstrand_model_comps_ni58_inset} shows the nebular phase spectra of SN~2018zd compared to the 12~$\rm{M_{\odot}}$ and 15~$\rm{M_{\odot}}$ models computed in \cite{Jerkstrand_2014_OI_mass} (we neglect the 19~$\rm{M_{\odot}}$ as a preliminary analysis showed that there were significant differences between it and SN~2018zd) and the 9~$\rm{M_{\odot}}$ model computed in \cite{Jerkstrand_2018_9Msol}. Fig. \ref{fig:Jerkstrand_OI_models} in the appendices focuses on the wavelength range of the nebular spectra containing the [O~{\sc i}] doublet. 

\begin{figure}
	\includegraphics[width=\columnwidth]{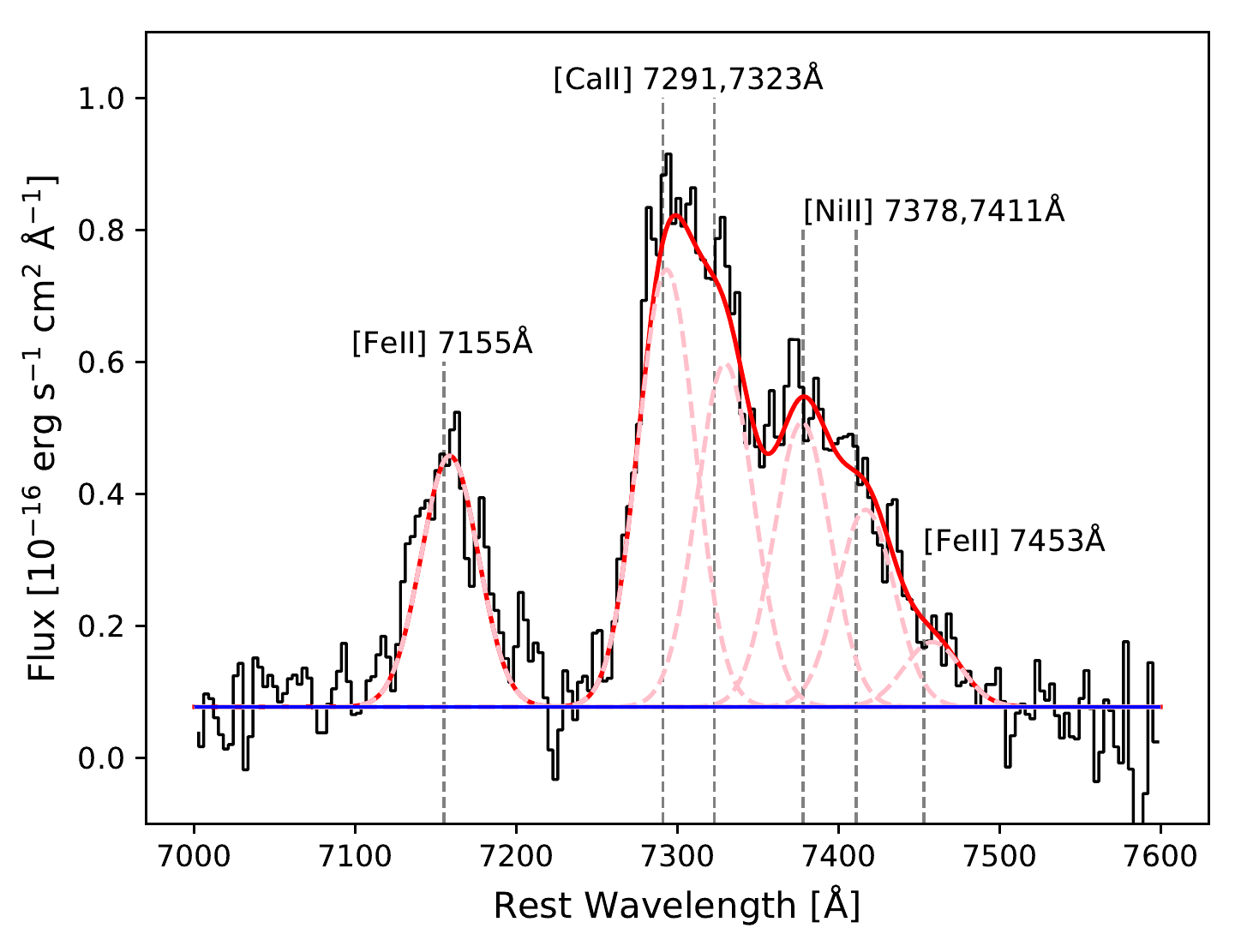}
    \caption{Dereddened and redshift corrected nebular spectrum of SN~2018zd at +410~d, along with the continuum (blue), Gaussian fits to the identified lines (pink dashed), and the sum total of the line fits (red). The Gaussian components have not been constrained to have the same FWHM values, but we find that similar values for all lines emerge from the fit ($\sim1700~\pm 110$~km~s$^{-1}$).}
    \label{fig:Ni58_line_ratios}
\end{figure}

We also identify a feature at $\sim$7378~\AA\ as [Ni~{\sc ii}], as shown in Fig. \ref{fig:Ni58_line_ratios}. This line is produced by $^{58}$Ni, whose abundance is a sensitive tracer of explosive burning conditions \citep{Jerkstrand_2015_Followup_NiFe_ratios}. The identification of distinct [Fe~{\sc ii}] 7155 and [Ni~{\sc ii}] 7378 lines, and their measured luminosities, can constrain the iron and nickel content with some analytic treatment. There are only a handful of CCSNe observed that display these features, which allow for the calculation of the Ni/Fe ratio. We use the same prescriptions for the line ratios in this region as in \cite{Jerkstrand2015_NiFe_2012ec} to estimate $n_{\mathrm{NiII}}/n_{\mathrm{FeII}}$, which is also expected to be an accurate estimator of the Ni/Fe ratio. We measure $L_{7378}/L_{7155}$~=~1.13, and following Eqn.~3 in \cite{Jerkstrand2015_NiFe_2012ec}, we estimate a temperature of $\sim3400$~K for SN~2018zd. This gives us an estimated ratio $n_{\mathrm{NiII}}/n_{\mathrm{FeII}}=0.09$, compared to the solar abundance ratio of 0.06, and compared to the range of 1-2 expected for ECSNe \citep{Wanajo_2009_ECSN_Nucleosynthesis}.

We repeated this analysis with the value of $L_{7378}/L_{7155}$=1.13-1.6 as reported by \citetalias{Hiramatsu2020_18zd}, also employing a lower nebular phase temperature of $\sim4000$K. From this the estimated Ni/Fe ratio is $\sim$0.1-0.14, higher than the estimate presented here, but still not in the range expected for ECSNe. 

We note however that this analysis is less accurate for SNe with very low $^{58}$Ni masses, as contamination by primordial Ni and Fe emission becomes more of a problem. There is also the caveat that this analysis only holds for non-local thermodynamic equilibrium (NLTE) when the lines are optically thin, which for a normal Type IIP SN would be hold from approximately +370~d. 

Therefore it is difficult to derive much information regarding the explosion properties of SN~2018zd from this feature, however, it can be safely said that the abundance is close to the solar value and much lower than the values expected from exotic explosion mechanisms (e.g. ECSNe).

\cite{Muller_Bravo_2020_2016aqf} performed a similar analysis for the low-luminosity Type II SN~2016aqf. They compare SN~2016afq to the models of \cite{Jerkstrand_2018_9Msol}, which we also compare SN~2018zd to in Fig. \ref{fig:jerkstrand_model_comps_ni58_inset}. These models predict the appearance of the He~{\sc i} $\lambda7065$ in SNe with low mass progenitors. We see this line weakly but clearly in the nebular spectra of SN~2018zd. We also note the tentative detection of [C~{\sc i}] $\lambda8727$ in SN~2018zd, which is an expected result of  He shell burning, although this is made difficult due to blending with the Ca~{\sc ii} triplet. The presence of these lines is evidence against SN~2018zd resulting from an ECSN, as ECSNe are expected to lack lines produced in the He layer \citep{Jerkstrand_2018_9Msol}. 

\subsection{Velocity Evolution}
\label{sec:Vel-evo}

\begin{figure}
	\includegraphics[width=\columnwidth]{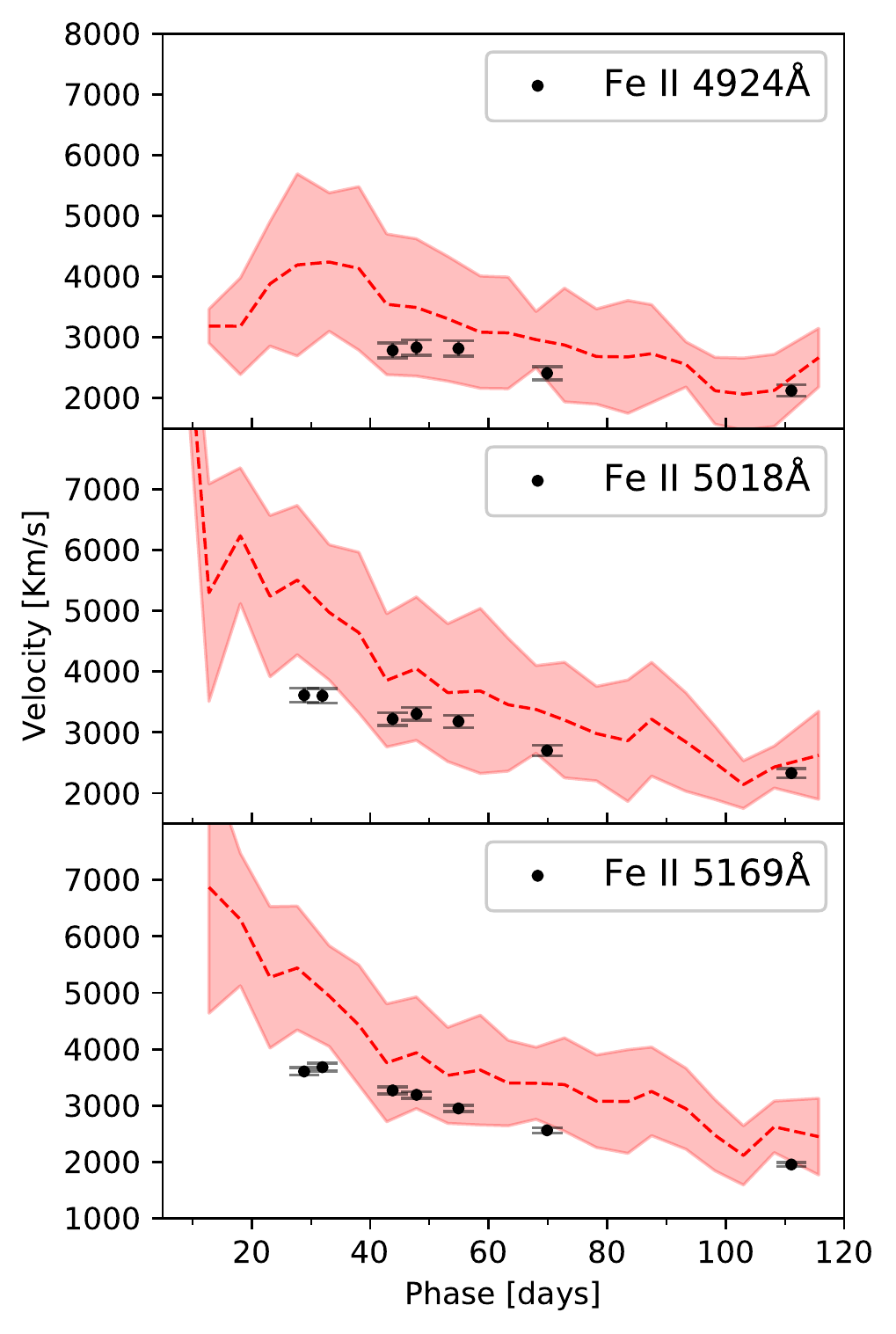}
    \caption{Velocity evolution of the Fe~{\sc ii} triplet. The black dots represent the velocity measurements of SN~2018zd, the red dashed line is the mean velocity of the sample from \citet{Gutierrez2017}, and the shaded region is the standard deviation of that sample.} 
    \label{fig:FeII_vel_evo}
\end{figure}

We mesasure the velocity evolution of SN~2018zd using the Fe~{\sc ii} triplet at 4924, 5018, 5169 \AA. We measured the velocity by fitting a Gaussian to the absorption feature and defining the velocities by the minimum. Following the example set by \citet{Gutierrez2017}, we did not use the average velocity of the triplet, as they show varying evolution, in part due to the Fe~{\sc ii} 4924~\AA\ line being blended with H$\beta$. Fig. \ref{fig:FeII_vel_evo} shows the evolution of each component separately, and compares SN~2018zd to the mean value and scatter found by \citet{Gutierrez2017}. SN~2018zd lies on the lower velocity end of this sample, in the regime of the LLEV's described by \cite{Rodriguez2020_LLEVs}.  We measured each line multiple times at several epochs and calculated the average standard deviation and applied this as a percentage error to the whole dataset, as the uncertainties will all be dominated by the choice for the continuum.

\citetalias{Zhan20} also analysed the velocity evolution of H$\alpha$, which the found to peak at $\rm{\sim8000\ km\ s^{-1}}$ $\sim$8~d after explosion. Following this they observe a rapid decrease in the photospheric velocity until it plateaus at $\rm{\sim3600\ km\ s^{-1}}$, which is comparable to what we observe in the Fe~{\sc ii} lines. \citetalias{Zhan20} switch to use the Fe~{\sc ii} $\rm{\lambda}$5169 line to measure the photospheric velocity after 20~d, and conclude that SN~2018zd displays mid-plateau photospheric velocities on the slow side for normal Type IIP SNe, but still faster than the low-velocity SNe IIP. \citetalias{Hiramatsu2020_18zd} obtained similar results,  and also suggested that a dense progenitor wind profile can help explain both the early-time luminosity excess and the suppressed early photospheric velocities in comparison to normal Type IIP SNe. 

\section{Discussion}
\label{sec:Discussion}

\subsection{Progenitor analysis}
\label{sec:Progenitor_analysis}

As discussed in Section \ref{sec:HST}, we detect a possible progenitor only in the $F814W$-band. As a consequence, we cannot determine the colour of the progenitor, and hence are unable to estimate a temperature or bolometric correction. Nonetheless, assuming SN~2018zd comes from the explosion of a RSG (a reasonable assumption given the H lines in spectra and plateau in the light curve), we can make some estimate of the progenitor luminosity and hence mass.

Given the measured {\it F814W} magnitude of $24.65\pm0.14$ mag, and our adopted distance and reddening (Section \ref{sec:Distance} and \ref{sec:Extinction}), we find an absolute magnitude of \textit{F814W=}$-6.66^{+0.73}_{-0.49}$~mag. Taking a range of possible bolometric corrections appropriate for a supergiant with temperature between 3400 and 4250~K, we find a luminosity log(L$/$L$_\odot$)=4.56$^{+0.16}_{-0.53}$~dex.
While the lower bound on this value should be viewed with caution (as it is strongly affected by the bolometric correction), the upper bound should be fairly robust. Unless the progenitor of SN~2018zd was enshrouded by dust before exploding (e.g. \citealp{Fraser2012}), it is quite unlikely to be more luminous than log(L$/$L$_\odot$)=4.7~dex. Comparing to stellar evolutionary models, this luminosity is consistent with a progenitor with a ZAMS mass of 8--10~$\rm{M_{\odot}}$ \citep{Eldridge2004}.

We cannot entirely rule out a super-Asymptotic Giant Branch (sAGB) progenitor as might be expected for an ECSN. While the luminosity of such a star would be closer to 5.0~dex due to the effects of second dredge-up, this flux would primarily emerge in the NIR bands (see for example fig. 3 in \citealp{Eldridge07}). In the absence of constraining NIR limits on the progenitor (Section \ref{sec:Pre-explosion}), we cannot exclude this scenario.

If we adopt the closer distance of 9.6~Mpc from \citetalias{Hiramatsu2020_18zd}, then the progenitor candidate would have a luminosity of around 4.0~dex (again, assuming a bolometric correction appropriate to a RSG, which may not be valid for a dust or CSM-enshrouded SAGB progenitor). This luminosity is fainter than one would expect for the SAGB progenitor of an ECSN (but could be explained by dust), and equally problematic for an Fe-CCSN progenitor.

Finally, we can also consider if the progenitor candidate is in fact a cosmic ray (Sect. \ref{sec:HST}). In this case, the true progenitor must be fainter than \textit{F814W}=24.65 mag, and hence the progenitor mass of 8--10 \msun\ inferred earlier in this section is an upper limit (i.e. the progenitor is $\lesssim$10 \msun).

\subsection{Mass-loss and CSM}
\label{sec:Mass-loss}

\begin{figure*}
	\includegraphics[width=\textwidth]{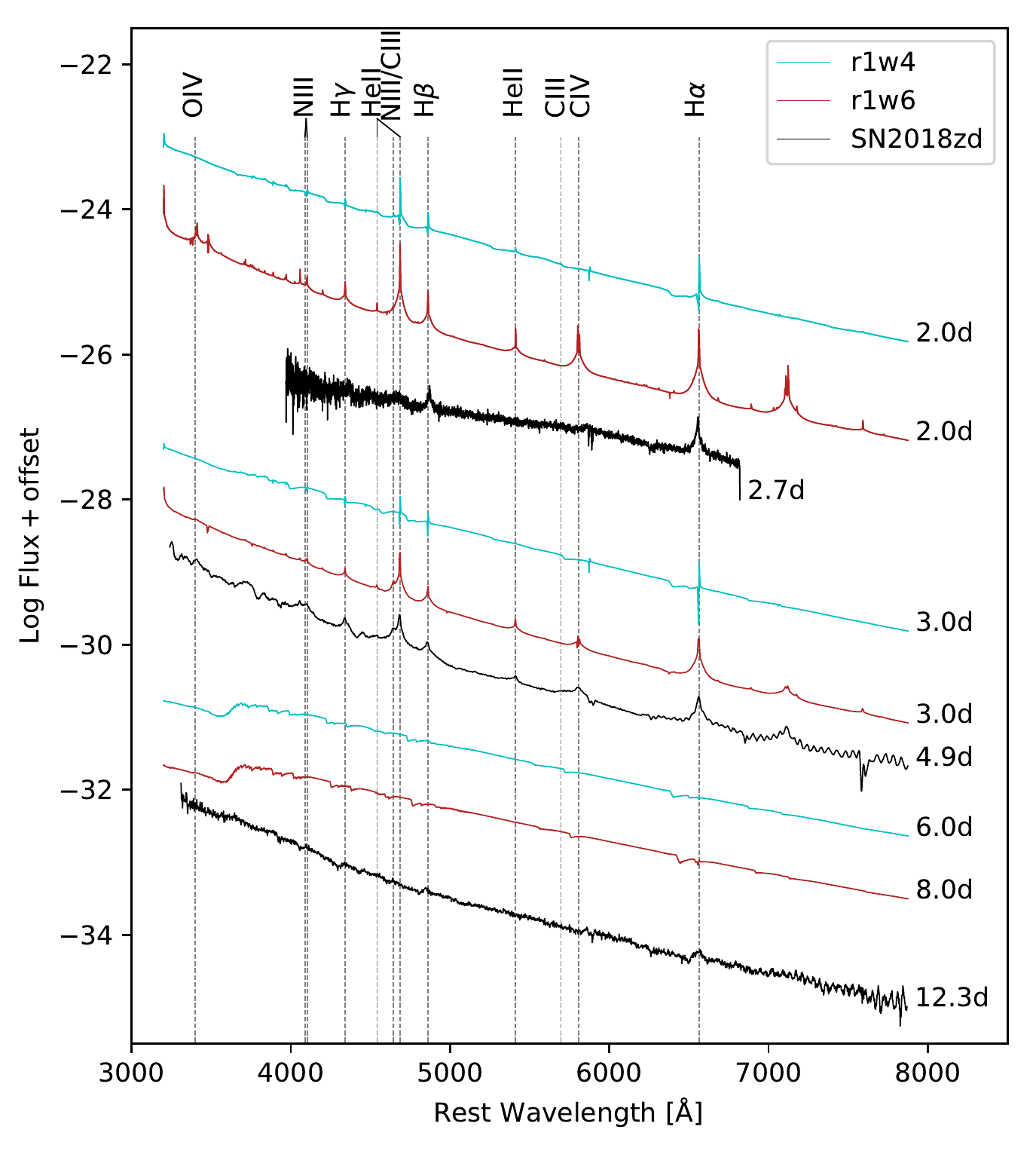}
    \caption{Comparison of the early spectral evolution of SN~2018zd to the models presented in \citet{Dessart2017}, where \textit{r1} corresponds to $R_\star$ = 501 $\rm{R_{\odot}}$, \textit{w4} means $\dot{M} = 1\times 10^{-3}$~$\rm{M_{\odot}}$ yr$^{-1}$ and \textit{w6} means $\dot{M} = 1\times 10^{-2}$~$\rm{M_{\odot}}$~yr$^{-1}$.}
    \label{fig:Dessart_models}
\end{figure*}

Fig. \ref{fig:Dessart_models} shows the earliest spectra compared to the models of \cite{Dessart2017}, which explored numerically the effects of a dense CSM shell around Type IIP SNe using radiation hydrodynamics and non-LTE radiation transfer. None of the models show the same early evolution as SN~2018zd, however two of the most similar are \textit{r1w4} and \textit{r1w6}, where \textit{r1} corresponds to $R_\star$ = 501 $\rm{R_{\odot}}$, \textit{w4} means $\dot{M} = 1\times 10^{-3}$~$\rm{M_{\odot}}$ yr$^{-1}$ and \textit{w6} means $\dot{M} = 1\times 10^{-2}$~$\rm{M_{\odot}}$~yr$^{-1}$. These mass-loss rates are much higher than the observed $10^{-6} - 10^{-4}$~$\rm{M_{\odot}}$~yr$^{-1}$ for most RSGs \citep[e.g][]{Willson2000}. Mass-loss rates this high could result in the progenitor being completely obscured by dust formation in the wind, helping to explain the difficulty of clearly detecting the progenitor in the pre-explosion images. \cite{Dessart2017} conclude that signatures of interaction in early-time spectra of RSG star explosions may actually be the norm, not the exception, and argue that inclusion of a super-wind phase prior to core-collapse may be superfluous. However, in objects such as SN~2018zd, the delay in the emergence of the high ionisation features, along with the early colour and temperature evolution, suggests that the emitting region is distinct from the bloated RSG envelope, indicating a period of enhanced mass loss shortly before explosion. 

\subsection{SN~2018zd and the standard candle method}
\label{sec:SCM}

\begin{figure}
	\includegraphics[width=\columnwidth]{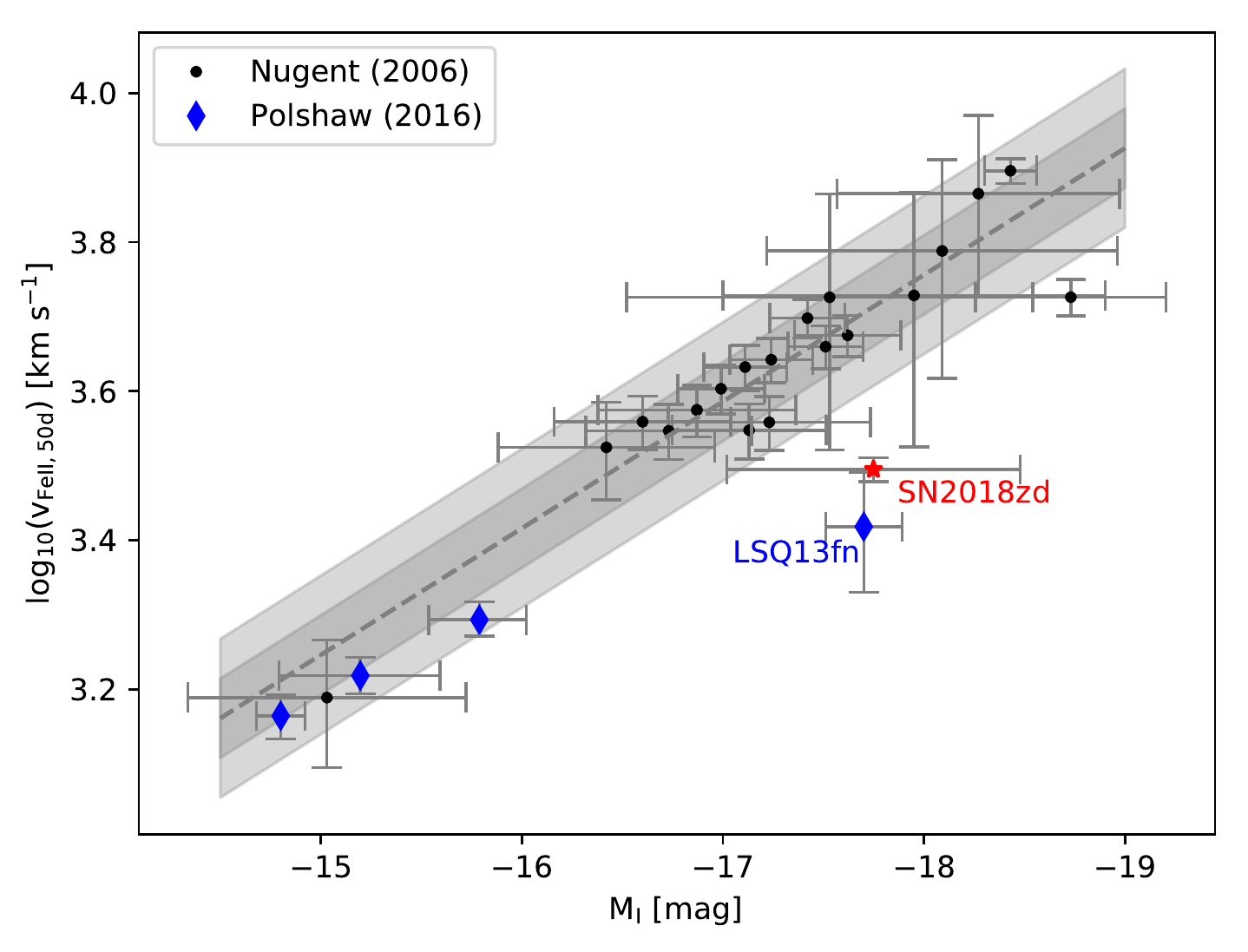}
    \caption{Velocity of the absorption minimum of the Fe~{\sc ii} $\lambda$5169 line of SN~2018zd and its absolute magnitude in the \textit{I}-band, both calculated at 50 days. The blue diamonds are the SNe from \citep{PolshawLSQ13fn}, including LSQ13fn. The black dots are a sample of Type IIP SNe taken from \citep{Nugent06_stnd_candle}. The grey shaded regions denote the 1$\sigma$ and 2$\sigma$ uncertainties calculated from the Nugent sample.}
    \label{fig:standardcandle}
\end{figure}

\cite{PolshawLSQ13fn} compared LSQ13fn to the sample in \cite{Nugent06_stnd_candle} and showed in their Fig. 8 that LSQ13fn did not conform to the standard candle relation for Type IIP SNe. We performed a similar analysis on SN~2018zd, computing $M_I$= $-17.7\pm$0.73~mag ($i$--band magnitude at +50~d) and $vFe_{50d}$=3125$\pm$115~km~s$^{-1}$ (velocity of Fe~{\sc ii} $\lambda$5169 measured at +50 d). The large uncertainty in the absolute magnitude, dominated by the uncertainty in the distance estimate, means SN~2018zd lies within the trend found by \cite{Nugent06_stnd_candle} within a 2-sigma uncertainty; however, as shown in Fig. \ref{fig:standardcandle}, SN~2018zd does trend towards the same outlier behaviour as LSQ13fn. It should be noted that if the extinction has been underestimated SN~2018zd would lie even further from the trend found by \cite{Nugent06_stnd_candle}. Using Equation 2 from \cite{Hamuy2002_stndcandleIIP}, a distance modulus of 27.9$\pm$0.7~mag was calculated for SN~2018zd, which differs considerably from the value of $\mu$=30.97~mag estimated from our distance estimate. 

We have noted throughout this paper the similarities of SN~2018zd to LSQ13fn and the LLEV's described by \cite{Rodriguez2020_LLEVs}. The defining characteristics of LLEV's are as follows:
\begin{itemize}
    \item signs of early interaction of the ejecta with CSM
    \item blue intrinsic $B-V$ colours 
    \item weakness of the metal lines
    \item low expansion velocities
    \item \textit{V}--band magnitudes 2--3 mag brighter than those expected from normal Type II SNe based on their expansion velocities
\end{itemize}

While SN~2018zd cannot conclusively be included in the LLEV sample, its characteristics do trend towards this behaviour, only not to the same extreme. \cite{Rodriguez2020_LLEVs} tested whether ejecta-CSM interaction is wholly responsible for converting a normal Type II SN into an LLEV by running hydrodynamical models. They found that by taking a 14 $\rm{M_{\odot}}$ progenitor model and attaching a dense CSM shell with a mass of 3.6 $\rm{M_{\odot}}$ they can recreate the observed luminosities and expansion velocities of SN 2009aj, one of the LLEV's in their sample, showing that CSM interaction moves SNe from the main locus of points towards the LLEV region around LSQ13fn. This supports our theory that SN~2018zd is the result of the iron core collapse of a RSG and CSM interaction, rather than the result of an exotic progenitor system or explosion. \citetalias{Hiramatsu2020_18zd} employ the SCM in estimating a distance to SN~2018zd, leading to a distance estimate of 9.6~Mpc. This reduced distance estimate results in a $^{56}$Ni mass of $\rm{(8.6\pm0.5)\times 10^{-3}\ M_{\odot}}$, which would bring SN~2018zd closer to the canonical $^{56}$Ni yield for ECSNe. 

While SN~2018zd hints that there may be a continuum between ``normal" Type IIs and LLEV-like objects, and possibly between LLEVs and IIn/IIs, more observations of similar objects are required to support this suggestion and fill the luminosity-expansion velocity plane.

Upcoming  large optical transient surveys (e.g. The Legacy Survey of Space and Time (LSST)) are predicted to observe thousands of Type II SNe annually, therefore there will be far too many Type IIP events with multiband light curves and no early spectroscopic coverage. This increases the risk of samples used to constrain the SCM being contaminated with objects experiencing unobserved early interaction. It would be judicious to attach a caveat to the SCM and other photometric distance estimate methods for Type IIP SNe, that in the absence of early spectra, strict cuts should be applied to the light curves in the sample. For example, cuts regarding early colour evolution, the drop in magnitude from the plateau to the tail phase, among others, could help curtail the number of 18zd-like events being included in such samples.

\subsection{Was SN~2018zd an electron capture SN?}
SN~2018zd has been suggested by \citetalias{Hiramatsu2020_18zd} to be the first confirmed example of an electron capture SN. While there are strong theoretical expectations that ECSNe should exist \citep{Miyaji80,Hillebrandt84,Nomoto_1984_ECSNe}, and the Crab SN is a strong contender for a SNR from an ECSN, there are thus far no SNe that are universally accepted to be ECSNe. One class of transients that has been linked to ECSNe are the intermediate-luminosity red transients \citep[ILRTs; e.g.][]{Prieto08,Botticella09,Cai_2021_ILRTS_ECSNe}. These transients have been observed to explode with low explosion energy, low ejected $^{56}$Ni masses, and typically in a dusty environment. \cite{Cai_2021_ILRTS_ECSNe} and others suggest that all ILRTs are regulated by the same mechanism, and that an ECSN explosion is a plausible scenario.

ECSNe have been modelled numerous times \citep[e.g.][]{Nomoto_1984_ECSNe,Woosley_2015_ECSNe}, and from these we expect their progenitors to be a low mass ($\sim 8$ ~\msun\ ) SAGB, that nonetheless may be quite luminous. They are also theorized to produce very low masses of $\rm{^{56}Ni}$, on the order of $\rm{10^{-3}M_{\odot}}$ \citep{Moriya_2014_ECSNe}. The $^{56}$Ni mass we calculate for SN~2018zd (0.017~\msun) is quite normal for an Fe CCSN, and indeed is much higher that that seen in SN~2005cs, the low-luminosity Type IIP SN we have compared to throughout this paper, for which an ECSN was clearly ruled out \citep{Eldridge07}.
We emphasise that the estimate of 0.0086~\msun\ of $^{56}$Ni from \citetalias{Hiramatsu2020_18zd} is dependent on the low distance estimate they calculate using the SCM.

ECSNe are also expected to display very strong signatures of stable Ni in their spectra \citep{Wanajo_2009_ECSN_Nucleosynthesis}. We discuss in Section \ref{sec:Nebular-spec} our analysis of the [Ni~{\sc ii}] $\lambda$7378 line. Our estimated Ni/Fe ratio is much lower than the theorized values for ECSNe \citep{Wanajo_2009_ECSN_Nucleosynthesis}. \cite{Jerkstrand_2018_9Msol} note that the effect of primordial contamination would be that the Ni/Fe ratio may be underestimated by $\sim$1/3, but this would still not push SN~2018zd into the realm of the ECSNe.

Finally, it has been theorised that ECSNe may be easier to identify by their lack of certain spectral lines than by the presence of specific lines. One such group of lines are those produced in the He burning zone in the progenitor. Numerous models of ECSNe progenitor evolution have shown that the He layer is reduced to almost nothing by 2nd dredge-up/dredge-out \citep{Jerkstrand_2018_9Msol}. We tentatively identify Mg~{\sc i}] $\lambda$4571, He~{\sc i} $\lambda$7065, and the Fe~{\sc i} lines in the 7900-8500\AA\ region, and confidently identify the [O~{\sc i}] $\lambda$6300,6364 doublet in the +336~d spectrum. We also tentatively identify the He~{\sc i} $\lambda$7065, Fe~{\sc i} $\lambda$5940, Fe~{\sc i} lines in the 7900-8500\AA\ region, and [C~{\sc i}] $\lambda$8727, as well as the [O~{\sc i}] doublet being clearly identified, in the +410~d spectrum. 

These lines, in particular the He~{\sc i} $\lambda$7065, Fe~{\sc i} lines, and the [C~{\sc i}] $\lambda$8727, are strongly linked to the He burning shell in the progenitor, therefore disfavoring the possibility of the progenitor being an SAGB, as expected for an ECSN. This, along with the solar-like production of stable Ni, places SN~2018zd in a similar scenario to SN~2005cs, for which the possibility of an ECSN was ruled out. \cite{Jerkstrand2015_NiFe_2012ec} present values of Ni/Fe ratios for a number of Fe CCSNe in their Table 4, and our value of 1.5 times the solar value fits well within this range. However, \cite{Muller_2017_LowMassFeCCSNe_and_ECSNe} caution that while progress has been made with regards to models of ECSNe and their nucleosynthesis constraints, there is still a lack of understanding surrounding the low-mass end of the CCSNe range, where neutrino transport and multi-dimensional effects may have a significant effect. While it may still be too early to declare {\it res judicata} for the case of SN~2018zd, we suggest that the evidence collected and presented thus far points clearly towards an Fe CCSN.

\section{Conclusions}
\label{sec:Conclusions}
The study of interacting supernovae has mainly been focused on classes of objects with long-lasting signatures of interaction, the hydrogen rich Type IIn's and the helium rich Type Ibn's. As more SNe are observed to experience brief, strong interaction with CSM only in their earliest epochs, we need to consider the effect this phenomenon has on our understanding of mass loss in massive stars during their final stages of evolution, and how it affects the entire evolution of the transient.

SN~2018zd is an interesting and well-observed object, and it has been argued by \citetalias{Hiramatsu2020_18zd} that it could be the result of an ECSN. We show that CSM interaction observed at early times in SN~2018zd can explain most of the discrepancies between it and a ``normal" Type IIP SN.  

The distance to SN~2018zd is salient to its nature. 
Difficulty in calculating the distance estimate is in part responsible for the differences between published results on SN~2018zd, which we have tried to highlight throughout this paper. A shorter distance estimate, computed from the SCM, is one of the main reasons for SN~2018zd being considered as an ECSN by \citetalias{Hiramatsu2020_18zd}. However, we have shown that the SCM is less reliable in objects effected by interaction with CSM, and cannot be used for this SN.

The tentative detection in the \textit{F814W} frame points to a probable cool, low mass RSG progenitor, which lends credence to our conclusion that SN~2018zd is the result of a normal Fe-CC supernova. A dense progenitor wind or CSM can delay and prolong the shock breakout, and if this material is detached from the progenitor, it can explain the evolution of the narrow high ionization lines in the early spectra and the early colour evolution seen in SN~2018zd.

If we were to overlook the signatures of interaction from the early spectra and consider only the photometry, which will be the scenario for the numerous objects observed by upcoming large optical surveys (e.g. LSST), it would be easy to consider SN~2018zd a ``normal" Type IIP SN and to include it in samples used to calibrate photometric relations derived from Type IIP SNe, leading to skewed results. One such method we have explored is the SCM, where we show that SN~2018zd and the comparable LSQ13fn both lie outside this relationship, because of the excess luminosity during the plateau phase contributed from early time interaction with CSM. Further work should be done on a larger sample of objects similar to SN~2018zd to develop improved selection criteria.

The analysis of SN~2018zd presented here reinforces the broad diversity observed in Type II SNe, and the importance of multiwavelength and early spectral observations, an especially important reminder as our photometric followup continues to outstrip our spectroscopic facilities.

\section*{Data availability statement}
The data underlying this article are available in the Weizmann Interactive Supernova Data Repository (for spectra) at \url{https://www.wiserep.org/}; and VizieR (for photometry) \url{https://vizier.u-strasbg.fr/viz-bin/VizieR}. {\it HST} data are available from the Mikulski Archive for Space Telescope (\url{https://archive.stsci.edu}). 

\section*{Acknowledgements}
MF is supported by a Royal Society - Science Foundation Ireland University Research Fellowship.\\
BJS is supported by NSF grants AST-1920392, AST-1911074, and AST-1908952.\\
MG is supported by the EU Horizon 2020 research and innovation programme under grant agreement No 101004719.\\
Support for JLP is provided in part by ANID through the Fondecyt regular grant 1191038 and through the Millennium Science Initiative grant ICN12009, awarded to the Millennium Institute of Astrophysics, MAS.\\
Support for TW-SH was provided by NASA through the NASA Hubble Fellowship grant HST-HF2-51458.001-A awarded by the Space Telescope Science Institute, which is operated by the Association of Universities for Research in Astronomy, Inc., for NASA, under contract NAS5-26555.\\
AR acknowledges support from ANID BECAS/DOCTORADO NACIONAL 21202412.\\
E.Congiu acknowledges support from ANID project Basal AFB-170002.\\
MS is supported by grants from the VILLUM FONDEN (grant number 28021) and  the Independent Research Fund Denmark (IRFD; 8021-00170B).\\
A.S.C acknowledge support from the G.R.E.A.T research environment, funded by {\em Vetenskapsr\aa det},  the Swedish Research Council, project number 2016-06012.\\
N.E.R. acknowledges support from MIUR, PRIN 2017 (grant 20179ZF5KS)\\
We acknowledge the use of public data from the Swift data archive. This work made use of data supplied by the UK Swift Science Data Centre at the University of Leicester.
Based on observations made with the NASA/ESA Hubble Space Telescope, obtained from the data archive at the Space Telescope Science Institute. STScI is operated by the Association of Universities for Research in Astronomy, Inc. under NASA contract NAS 5-26555.\\
This work is based in part on archival data obtained with the Spitzer Space Telescope, which was operated by the Jet Propulsion Laboratory, California Institute of Technology under a contract with NASA. Support for this work was provided by NASA.\\
Based on observations obtained with WIRCam, a joint project of CFHT, Taiwan, Korea, Canada, France, and the Canada-France-Hawaii Telescope (CFHT) which is operated by the National Research Council (NRC) of Canada, the Institut National des Sciences de l'Univers of the Centre National de la Recherche Scientifique of France, and the University of Hawaii.\\
This research uses data obtained through the Telescope Access Program (TAP).\\
NUTS is supported in part by IDA.\\
Based on observations made with the Nordic Optical Telescope, owned in collaboration by the University of Turku and Aarhus University, and operated jointly by Aarhus University, the University of Turku and the University of Oslo, representing Denmark, Finland and Norway, the University of Iceland and Stockholm University at the Observatorio del Roque de los Muchachos, La Palma, Spain, of the Instituto de Astrofisica de Canarias.\\
The William Herschel Telescope is operated on the island of La Palma by the Isaac Newton Group of Telescopes in the Spanish Observatorio del Roque de los Muchachos of the Instituto de Astrofísica de Canarias.\\ 
The data presented here are based in part on observations collected at Copernico and Schmidt telescopes (Asiago, Italy) of the INAF - Osservatorio Astronomico di Padova.\\
This paper is partially based on observations collected with the 1.22 m Galileo telescope of the Asiago Astrophysical Observatory, operated by the Department of Physics and Astronomy “G. Galilei” of the University of Padova.\\
The data presented here were obtained in part with ALFOSC, which is provided by the Instituto de Astrofisica de Andalucia (IAA) under a joint agreement with the University of Copenhagen and NOT.\\
We thank \'Osmar Rodr\'iguez, Anders Jerkstrand, Stephen J. Smartt, Ken Smith, and Angela Adamo for useful discussions. We are grateful to Christopher S. Kochanek for useful comments on the draft. We are grateful to Marco Fiaschi and Kristhell M. Lopez for their assistance with observations taken at the Asiago Astrophysical Observatory and the William Herschel Telescope respectively.\\
This research made use of Astropy,\footnote{\url{http://www.astropy.org}} a community-developed core Python package for Astronomy \citep{AstropyI, AstropyII}. 

\section*{Affiliations}
$^{1}$School of Physics, O'Brien Centre for Science North, University College Dublin, Belfield, Dublin 4, Ireland\\
$^{2}$INAF - Osservatorio Astronomico di Padova, Vicolo dell'Osservatorio 5, 35122 Padova, Italy\\
$^{3}$Kavli Institute for Astronomy and Astrophysics, Peking University, Yi He Huan Road 5, Hai Dian District, Beijing 100871, China\\
$^{4}$Department of Astronomy, The Ohio State University, 140 W. 18th Avenue, Columbus, OH 43210, USA\\
$^{5}$Center for Cosmology and AstroParticle Physics (CCAPP), The Ohio State University, 191 W.Woodruff Avenue, \\Columbus, OH 43210, USA\\
$^{6}$Department of Physics and Astronomy, University of Turku, FI- 20014 Turku, Finland\\
$^{7}$SRON, Netherlands Institute for Space Research, Sorbonnelaan, 2, NL-3584CA Utrecht, the Netherlands\\
$^{8}$Department of Astrophysics/IMAPP, Radboud University, P.O. Box 9010, 6500 GL Nijmegen, the Netherlands\\
$^{9}$Finnish Centre for Astronomy with ESO (FINCA), University of Turku, Vesilinnantie 5, FI-20014, Turku, Finland\\
$^{10}$Aalto University Mets{\"a}hovi Radio Observatory, Mets{\"a}hovintie 114, FI-02540 Kylm{\"a}l{\"a}, Finland\\
$^{11}$Departamento de Astronomía, Universidad de Chile, Camino del Observatorio 1515, Las Condes, Santiago, Chile\\
$^{12}$Department of Physics and Astronomy, Aarhus University, Ny Munkegade 120, DK-8000 Aarhus C, Denmark\\
$^{13}$Institute of Space Sciences (ICE, CSIC), Campus UAB, Carrer de Can Magrans s/n, 08193 Barcelona, Spain\\
$^{14}$Astronomical Observatory, University of Warsaw, Al. Ujazdowskie 4, 00-478 Warszawa, Poland\\
$^{15}$The Observatories of the Carnegie Institution for Science, 813 Santa Barbara St., Pasadena, CA 91101, USA\\
$^{16}$Yamagata Zao Observatory. 990-2492, 3-1-40 Teppomachi, Yamagata City, Japan\\
$^{17}$University of Iowa, Iowa City, IA 52242, USA\\
$^{18}$Dipartimento di Fisica e Astronomia 'G. Galilei' - Università di Padova, Vicolo dell'Osservatorio 3, 35122 Padova, Italy\\
$^{19}$Post Astronomy, Lexington, MA, USA\\
$^{20}$Núcleo de Astronomía de la Facultad de Ingeniería y Ciencias, Universidad Diego Portales, Av. Ejército 441 Santiago, Chile\\
$^{21}$Millennium Institute of Astrophysics, Nuncio Monsenor S\'{o}tero Sanz 100, Providencia, Santiago 8320000, Chile\\
$^{22}$Departamento de Ciencias F\'{i}sicas, Universidad Andres Bello, Avda. Rep\'{u}blica 252, Santiago 8320000, Chile\\
$^{23}$Space Telescope Science Institute, 3700 San Martin Drive, Baltimore, MD 21218, USA\\
$^{24}$The Oskar Klein Centre, Department of Physics, Stockholm University, AlbaNova, SE-106 91 Stockholm, Sweden \\
$^{25}$Institute for Astronomy, University of Hawaii, 2680 Woodlawn Drive, Honolulu, HI 96822, USA\\
$^{26}$Vanderbilt University, Department of Physics \& Astronomy, 6301 Stevenson Center Ln., Nashville, TN 37235, USA\\
$^{27}$Institute of Astronomy, University of Cambridge, Madingley Road, Cambridge CB3 0HA, UK\\
$^{28}$European Southern Observatory, Alonso de Córdova 3107, Casilla 19, Santiago, Chile\\
$^{29}$Wiggins Observatory, Tooele, Utah 84074, USA




\bibliographystyle{mnras}
\bibliography{references} 

\begin{thebibliography}{}
\makeatletter
\relax
\def\mn@urlcharsother{\let\do\@makeother \do\$\do\&\do\#\do\^\do\_\do\%\do\~}
\def\mn@doi{\begingroup\mn@urlcharsother \@ifnextchar [ {\mn@doi@}
  {\mn@doi@[]}}
\def\mn@doi@[#1]#2{\def\@tempa{#1}\ifx\@tempa\@empty \href
  {http://dx.doi.org/#2} {doi:#2}\else \href {http://dx.doi.org/#2} {#1}\fi
  \endgroup}
\def\mn@eprint#1#2{\mn@eprint@#1:#2::\@nil}
\def\mn@eprint@arXiv#1{\href {http://arxiv.org/abs/#1} {{\tt arXiv:#1}}}
\def\mn@eprint@dblp#1{\href {http://dblp.uni-trier.de/rec/bibtex/#1.xml}
  {dblp:#1}}
\def\mn@eprint@#1:#2:#3:#4\@nil{\def\@tempa {#1}\def\@tempb {#2}\def\@tempc
  {#3}\ifx \@tempc \@empty \let \@tempc \@tempb \let \@tempb \@tempa \fi \ifx
  \@tempb \@empty \def\@tempb {arXiv}\fi \@ifundefined
  {mn@eprint@\@tempb}{\@tempb:\@tempc}{\expandafter \expandafter \csname
  mn@eprint@\@tempb\endcsname \expandafter{\@tempc}}}

\bibitem[\protect\citeauthoryear{{Adamo} et~al.,}{{Adamo}
  et~al.}{2012}]{Adam12}
{Adamo} A.,  et~al., 2012, \mn@doi [\mnras] {10.1111/j.1365-2966.2012.21384.x},
  \href {https://ui.adsabs.harvard.edu/abs/2012MNRAS.426.1185A} {426, 1185}

\bibitem[\protect\citeauthoryear{{Anderson} et~al.,}{{Anderson}
  et~al.}{2014}]{Anderson2014}
{Anderson} J.~P.,  et~al., 2014, \mn@doi [\apj] {10.1088/0004-637X/786/1/67},
  \href {https://ui.adsabs.harvard.edu/abs/2014ApJ...786...67A} {786, 67}

\bibitem[\protect\citeauthoryear{{Anderson} et~al.,}{{Anderson}
  et~al.}{2016}]{Anderson2016_II_metallicity}
{Anderson} J.~P.,  et~al., 2016, \mn@doi [\aap] {10.1051/0004-6361/201527691},
  \href {https://ui.adsabs.harvard.edu/abs/2016A&A...589A.110A} {589, A110}

\bibitem[\protect\citeauthoryear{{Anderson} et~al.,}{{Anderson}
  et~al.}{2018}]{Anderson_2018_SN2015bs}
{Anderson} J.~P.,  et~al., 2018, \mn@doi [Nature Astronomy]
  {10.1038/s41550-018-0458-4}, \href
  {https://ui.adsabs.harvard.edu/abs/2018NatAs...2..574A} {2, 574}

\bibitem[\protect\citeauthoryear{{Aniano} et~al.,}{{Aniano}
  et~al.}{2020}]{Aniano_2020_host_metallicity}
{Aniano} G.,  et~al., 2020, \mn@doi [\apj] {10.3847/1538-4357/ab5fdb}, \href
  {https://ui.adsabs.harvard.edu/abs/2020ApJ...889..150A} {889, 150}

\bibitem[\protect\citeauthoryear{{Astropy Collaboration} et~al.,}{{Astropy
  Collaboration} et~al.}{2013}]{AstropyI}
{Astropy Collaboration} et~al., 2013, \mn@doi [\aap]
  {10.1051/0004-6361/201322068}, \href
  {https://ui.adsabs.harvard.edu/abs/2013A&A...558A..33A} {558, A33}

\bibitem[\protect\citeauthoryear{{Astropy Collaboration} et~al.,}{{Astropy
  Collaboration} et~al.}{2018}]{AstropyII}
{Astropy Collaboration} et~al., 2018, \mn@doi [\aj] {10.3847/1538-3881/aabc4f},
  \href {https://ui.adsabs.harvard.edu/abs/2018AJ....156..123A} {156, 123}

\bibitem[\protect\citeauthoryear{{Bellm} et~al.,}{{Bellm}
  et~al.}{2019}]{Belm19}
{Bellm} E.~C.,  et~al., 2019, \mn@doi [\pasp] {10.1088/1538-3873/aaecbe}, \href
  {https://ui.adsabs.harvard.edu/abs/2019PASP..131a8002B} {131, 018002}

\bibitem[\protect\citeauthoryear{{Benn}, {Dee}  \& {Ag{\'o}cs}}{{Benn}
  et~al.}{2008}]{Benn_2008_ACAM}
{Benn} C.,  {Dee} K.,   {Ag{\'o}cs} T.,  2008, in {McLean} I.~S.,  {Casali}
  M.~M.,  eds,  Society of Photo-Optical Instrumentation Engineers (SPIE)
  Conference Series Vol. 7014, Ground-based and Airborne Instrumentation for
  Astronomy II. p. 70146X, \mn@doi{10.1117/12.788694}

\bibitem[\protect\citeauthoryear{{Boian} \& {Groh}}{{Boian} \&
  {Groh}}{2020}]{Boian_2020_18ZD}
{Boian} I.,  {Groh} J.~H.,  2020, \mn@doi [\mnras] {10.1093/mnras/staa1540},
  \href {https://ui.adsabs.harvard.edu/abs/2020MNRAS.496.1325B} {496, 1325}

\bibitem[\protect\citeauthoryear{{Bose} et~al.,}{{Bose}
  et~al.}{2021}]{Bose_2021_2018gk}
{Bose} S.,  et~al., 2021, \mn@doi [\mnras] {10.1093/mnras/stab629}, \href
  {https://ui.adsabs.harvard.edu/abs/2021MNRAS.503.3472B} {503, 3472}

\bibitem[\protect\citeauthoryear{{Botticella} et~al.,}{{Botticella}
  et~al.}{2009}]{Botticella09}
{Botticella} M.~T.,  et~al., 2009, \mn@doi [\mnras]
  {10.1111/j.1365-2966.2009.15082.x}, \href
  {https://ui.adsabs.harvard.edu/abs/2009MNRAS.398.1041B} {398, 1041}

\bibitem[\protect\citeauthoryear{{Botticella} et~al.,}{{Botticella}
  et~al.}{2010}]{Botticella10}
{Botticella} M.~T.,  et~al., 2010, \mn@doi [\apjl]
  {10.1088/2041-8205/717/1/L52}, \href
  {https://ui.adsabs.harvard.edu/abs/2010ApJ...717L..52B} {717, L52}

\bibitem[\protect\citeauthoryear{{Brown} et~al.,}{{Brown}
  et~al.}{2013}]{Brown2013_LCO}
{Brown} T.~M.,  et~al., 2013, \mn@doi [\pasp] {10.1086/673168}, \href
  {https://ui.adsabs.harvard.edu/abs/2013PASP..125.1031B} {125, 1031}

\bibitem[\protect\citeauthoryear{{Cai} et~al.,}{{Cai}
  et~al.}{2021}]{Cai_2021_ILRTS_ECSNe}
{Cai} Y.~Z.,  et~al., 2021, arXiv e-prints, \href
  {https://ui.adsabs.harvard.edu/abs/2021arXiv210805087C} {p. arXiv:2108.05087}

\bibitem[\protect\citeauthoryear{{Cardelli}, {Clayton}  \& {Mathis}}{{Cardelli}
  et~al.}{1989}]{Cardelli89}
{Cardelli} J.~A.,  {Clayton} G.~C.,   {Mathis} J.~S.,  1989, \mn@doi [\apj]
  {10.1086/167900}, \href
  {https://ui.adsabs.harvard.edu/abs/1989ApJ...345..245C} {345, 245}

\bibitem[\protect\citeauthoryear{{Chambers} et~al.,}{{Chambers}
  et~al.}{2016}]{Panstarrs_2016_Chambers}
{Chambers} K.~C.,  et~al., 2016, arXiv e-prints, \href
  {https://ui.adsabs.harvard.edu/abs/2016arXiv161205560C} {p. arXiv:1612.05560}

\bibitem[\protect\citeauthoryear{{Crowther}}{{Crowther}}{2007}]{Crowther07_WR}
{Crowther} P.~A.,  2007, \mn@doi [\araa]
  {10.1146/annurev.astro.45.051806.110615}, \href
  {https://ui.adsabs.harvard.edu/abs/2007ARA&A..45..177C} {45, 177}

\bibitem[\protect\citeauthoryear{{Dalcanton} et~al.,}{{Dalcanton}
  et~al.}{2012}]{Dalc12}
{Dalcanton} J.~J.,  et~al., 2012, \mn@doi [\apjs] {10.1088/0067-0049/198/1/6},
  \href {https://ui.adsabs.harvard.edu/abs/2012ApJS..198....6D} {198, 6}

\bibitem[\protect\citeauthoryear{{Davis} et~al.,}{{Davis}
  et~al.}{2011}]{Tama11}
{Davis} T.~M.,  et~al., 2011, \mn@doi [\apj] {10.1088/0004-637X/741/1/67},
  \href {https://ui.adsabs.harvard.edu/abs/2011ApJ...741...67D} {741, 67}

\bibitem[\protect\citeauthoryear{{Dessart}, {John Hillier}  \&
  {Audit}}{{Dessart} et~al.}{2017}]{Dessart2017}
{Dessart} L.,  {John Hillier} D.,   {Audit} E.,  2017, \mn@doi [\aap]
  {10.1051/0004-6361/201730942}, \href
  {https://ui.adsabs.harvard.edu/abs/2017A&A...605A..83D} {605, A83}

\bibitem[\protect\citeauthoryear{{Dolphin}}{{Dolphin}}{2000}]{Dolp00}
{Dolphin} A.~E.,  2000, \mn@doi [\pasp] {10.1086/316630}, \href
  {https://ui.adsabs.harvard.edu/abs/2000PASP..112.1383D} {112, 1383}

\bibitem[\protect\citeauthoryear{{Eldridge} \& {Tout}}{{Eldridge} \&
  {Tout}}{2004}]{Eldridge2004}
{Eldridge} J.~J.,  {Tout} C.~A.,  2004, \mn@doi [\mnras]
  {10.1111/j.1365-2966.2004.07344.x}, \href
  {https://ui.adsabs.harvard.edu/abs/2004MNRAS.348..201E} {348, 201}

\bibitem[\protect\citeauthoryear{{Eldridge}, {Mattila}  \& {Smartt}}{{Eldridge}
  et~al.}{2007}]{Eldridge07}
{Eldridge} J.~J.,  {Mattila} S.,   {Smartt} S.~J.,  2007, \mn@doi [\mnras]
  {10.1111/j.1745-3933.2007.00285.x}, \href
  {https://ui.adsabs.harvard.edu/abs/2007MNRAS.376L..52E} {376, L52}

\bibitem[\protect\citeauthoryear{{Elmhamdi}, {Chugai}  \&
  {Danziger}}{{Elmhamdi} et~al.}{2003}]{Elmhamdi03}
{Elmhamdi} A.,  {Chugai} N.~N.,   {Danziger} I.~J.,  2003, \mn@doi [\aap]
  {10.1051/0004-6361:20030522}, \href
  {https://ui.adsabs.harvard.edu/abs/2003A&A...404.1077E} {404, 1077}

\bibitem[\protect\citeauthoryear{{Fabricant}, {Cheimets}, {Caldwell}  \&
  {Geary}}{{Fabricant} et~al.}{1998}]{Fabricant_1998_FAST}
{Fabricant} D.,  {Cheimets} P.,  {Caldwell} N.,   {Geary} J.,  1998, \mn@doi
  [\pasp] {10.1086/316111}, \href
  {https://ui.adsabs.harvard.edu/abs/1998PASP..110...79F} {110, 79}

\bibitem[\protect\citeauthoryear{{Filippenko}}{{Filippenko}}{1997}]{Filippenko_1997}
{Filippenko} A.~V.,  1997, \mn@doi [\araa] {10.1146/annurev.astro.35.1.309},
  \href {https://ui.adsabs.harvard.edu/abs/1997ARA&A..35..309F} {35, 309}

\bibitem[\protect\citeauthoryear{{Fraser}}{{Fraser}}{2020}]{Fras20}
{Fraser} M.,  2020, \mn@doi [Royal Society Open Science] {10.1098/rsos.200467},
  \href {https://ui.adsabs.harvard.edu/abs/2020RSOS....700467F} {7, 200467}

\bibitem[\protect\citeauthoryear{{Fraser} et~al.,}{{Fraser}
  et~al.}{2012}]{Fraser2012}
{Fraser} M.,  et~al., 2012, \mn@doi [\apjl] {10.1088/2041-8205/759/1/L13},
  \href {https://ui.adsabs.harvard.edu/abs/2012ApJ...759L..13F} {759, L13}

\bibitem[\protect\citeauthoryear{{Fraser} et~al.,}{{Fraser}
  et~al.}{2013}]{Fras13}
{Fraser} M.,  et~al., 2013, \mn@doi [\apjl] {10.1088/2041-8205/779/1/L8}, \href
  {https://ui.adsabs.harvard.edu/abs/2013ApJ...779L...8F} {779, L8}

\bibitem[\protect\citeauthoryear{{Gal-Yam}}{{Gal-Yam}}{2017}]{2017hsn..book..195G}
{Gal-Yam} A.,  2017, {Observational and Physical Classification of Supernovae}.
Springer, p.~195, \mn@doi{10.1007/978-3-319-21846-5\_35}

\bibitem[\protect\citeauthoryear{{Gal-Yam} et~al.,}{{Gal-Yam}
  et~al.}{2014}]{Gal-Yam_2014_13cu_FlashSpec}
{Gal-Yam} A.,  et~al., 2014, \mn@doi [\nat] {10.1038/nature13304}, \href
  {https://ui.adsabs.harvard.edu/abs/2014Natur.509..471G} {509, 471}

\bibitem[\protect\citeauthoryear{{Gall} et~al.,}{{Gall}
  et~al.}{2018}]{Gall_2018_IIP_HubbleDiagram}
{Gall} E.~E.~E.,  et~al., 2018, \mn@doi [\aap] {10.1051/0004-6361/201731271},
  \href {https://ui.adsabs.harvard.edu/abs/2018A&A...611A..25G} {611, A25}

\bibitem[\protect\citeauthoryear{{Gehrels} et~al.,}{{Gehrels}
  et~al.}{2004}]{Gehrels_2004_Swift}
{Gehrels} N.,  et~al., 2004, \mn@doi [\apj] {10.1086/422091}, \href
  {https://ui.adsabs.harvard.edu/abs/2004ApJ...611.1005G} {611, 1005}

\bibitem[\protect\citeauthoryear{{Groh}}{{Groh}}{2014}]{Groh_2014_13cu}
{Groh} J.~H.,  2014, \mn@doi [\aap] {10.1051/0004-6361/201424852}, \href
  {https://ui.adsabs.harvard.edu/abs/2014A&A...572L..11G} {572, L11}

\bibitem[\protect\citeauthoryear{{Guti{\'e}rrez} et~al.,}{{Guti{\'e}rrez}
  et~al.}{2017}]{Gutierrez2017}
{Guti{\'e}rrez} C.~P.,  et~al., 2017, \mn@doi [\apj]
  {10.3847/1538-4357/aa8f52}, \href
  {https://ui.adsabs.harvard.edu/abs/2017ApJ...850...89G} {850, 89}

\bibitem[\protect\citeauthoryear{{Guti{\'e}rrez} et~al.,}{{Guti{\'e}rrez}
  et~al.}{2018}]{Gutierrez_2018_Low_Lum_Host_TypeIIs}
{Guti{\'e}rrez} C.~P.,  et~al., 2018, \mn@doi [\mnras] {10.1093/mnras/sty1581},
  \href {https://ui.adsabs.harvard.edu/abs/2018MNRAS.479.3232G} {479, 3232}

\bibitem[\protect\citeauthoryear{{Hamuy}}{{Hamuy}}{2003}]{Hamuy2003_NiMass}
{Hamuy} M.,  2003, \mn@doi [\apj] {10.1086/344689}, \href
  {https://ui.adsabs.harvard.edu/abs/2003ApJ...582..905H} {582, 905}

\bibitem[\protect\citeauthoryear{{Hamuy} \& {Pinto}}{{Hamuy} \&
  {Pinto}}{2002}]{Hamuy2002_stndcandleIIP}
{Hamuy} M.,  {Pinto} P.~A.,  2002, \mn@doi [\apjl] {10.1086/339676}, \href
  {https://ui.adsabs.harvard.edu/abs/2002ApJ...566L..63H} {566, L63}

\bibitem[\protect\citeauthoryear{{Hillebrandt}, {Nomoto}  \&
  {Wolff}}{{Hillebrandt} et~al.}{1984}]{Hillebrandt84}
{Hillebrandt} W.,  {Nomoto} K.,   {Wolff} R.~G.,  1984, \aap, \href
  {https://ui.adsabs.harvard.edu/abs/1984A&A...133..175H} {133, 175}

\bibitem[\protect\citeauthoryear{{Hillier} \& {Miller}}{{Hillier} \&
  {Miller}}{1998}]{Hillier_1998_CMFGEN}
{Hillier} D.~J.,  {Miller} D.~L.,  1998, \mn@doi [\apj] {10.1086/305350}, \href
  {https://ui.adsabs.harvard.edu/abs/1998ApJ...496..407H} {496, 407}

\bibitem[\protect\citeauthoryear{{Hiramatsu} et~al.,}{{Hiramatsu}
  et~al.}{2020}]{Hiramatsu2020_18zd}
{Hiramatsu} D.,  et~al., 2020, arXiv e-prints, \href
  {https://ui.adsabs.harvard.edu/abs/2020arXiv201102176H} {p. arXiv:2011.02176}

\bibitem[\protect\citeauthoryear{{Hosseinzadeh} et~al.,}{{Hosseinzadeh}
  et~al.}{2018}]{Hosseinzadeh16bkv}
{Hosseinzadeh} G.,  et~al., 2018, \mn@doi [\apj] {10.3847/1538-4357/aac5f6},
  \href {https://ui.adsabs.harvard.edu/abs/2018ApJ...861...63H} {861, 63}

\bibitem[\protect\citeauthoryear{{Huber} et~al.,}{{Huber}
  et~al.}{2015}]{Hube15}
{Huber} M.,  et~al., 2015, The Astronomer's Telegram, \href
  {https://ui.adsabs.harvard.edu/abs/2015ATel.7153....1H} {7153, 1}

\bibitem[\protect\citeauthoryear{{Inserra} et~al.,}{{Inserra}
  et~al.}{2012}]{Inserra12}
{Inserra} C.,  et~al., 2012, \mn@doi [\mnras]
  {10.1111/j.1365-2966.2012.20685.x}, \href
  {https://ui.adsabs.harvard.edu/abs/2012MNRAS.422.1122I} {422, 1122}

\bibitem[\protect\citeauthoryear{{Jerkstrand}}{{Jerkstrand}}{2017}]{Jerkstrand_2017_Neb_Spec_HBSNe}
{Jerkstrand} A.,  2017, {Spectra of Supernovae in the Nebular Phase}.
Springer, p.~795, \mn@doi{10.1007/978-3-319-21846-5\_29}

\bibitem[\protect\citeauthoryear{{Jerkstrand}, {Smartt}, {Fraser}, {Fransson},
  {Sollerman}, {Taddia}  \& {Kotak}}{{Jerkstrand}
  et~al.}{2014}]{Jerkstrand_2014_OI_mass}
{Jerkstrand} A.,  {Smartt} S.~J.,  {Fraser} M.,  {Fransson} C.,  {Sollerman}
  J.,  {Taddia} F.,   {Kotak} R.,  2014, \mn@doi [\mnras]
  {10.1093/mnras/stu221}, \href
  {https://ui.adsabs.harvard.edu/abs/2014MNRAS.439.3694J} {439, 3694}

\bibitem[\protect\citeauthoryear{{Jerkstrand} et~al.,}{{Jerkstrand}
  et~al.}{2015a}]{Jerkstrand2015_NiFe_2012ec}
{Jerkstrand} A.,  et~al., 2015a, \mn@doi [\mnras] {10.1093/mnras/stv087}, \href
  {https://ui.adsabs.harvard.edu/abs/2015MNRAS.448.2482J} {448, 2482}

\bibitem[\protect\citeauthoryear{{Jerkstrand} et~al.,}{{Jerkstrand}
  et~al.}{2015b}]{Jerkstrand_2015_Followup_NiFe_ratios}
{Jerkstrand} A.,  et~al., 2015b, \mn@doi [\apj] {10.1088/0004-637X/807/1/110},
  \href {https://ui.adsabs.harvard.edu/abs/2015ApJ...807..110J} {807, 110}

\bibitem[\protect\citeauthoryear{{Jerkstrand}, {Ertl}, {Janka}, {M{\"u}ller},
  {Sukhbold}  \& {Woosley}}{{Jerkstrand} et~al.}{2018}]{Jerkstrand_2018_9Msol}
{Jerkstrand} A.,  {Ertl} T.,  {Janka} H.~T.,  {M{\"u}ller} E.,  {Sukhbold} T.,
   {Woosley} S.~E.,  2018, \mn@doi [\mnras] {10.1093/mnras/stx2877}, \href
  {https://ui.adsabs.harvard.edu/abs/2018MNRAS.475..277J} {475, 277}

\bibitem[\protect\citeauthoryear{{Jord{\'a}n} et~al.,}{{Jord{\'a}n}
  et~al.}{2005}]{Jord05}
{Jord{\'a}n} A.,  et~al., 2005, \mn@doi [\apj] {10.1086/497092}, \href
  {https://ui.adsabs.harvard.edu/abs/2005ApJ...634.1002J} {634, 1002}

\bibitem[\protect\citeauthoryear{{Jordi}, {Grebel}  \& {Ammon}}{{Jordi}
  et~al.}{2006}]{jordi06}
{Jordi} K.,  {Grebel} E.~K.,   {Ammon} K.,  2006, \mn@doi [\aap]
  {10.1051/0004-6361:20066082}, \href
  {https://ui.adsabs.harvard.edu/abs/2006A&A...460..339J} {460, 339}

\bibitem[\protect\citeauthoryear{{Khazov} et~al.,}{{Khazov}
  et~al.}{2016}]{Khazov_2016_flashspec_SNe_II}
{Khazov} D.,  et~al., 2016, \mn@doi [\apj] {10.3847/0004-637X/818/1/3}, \href
  {https://ui.adsabs.harvard.edu/abs/2016ApJ...818....3K} {818, 3}

\bibitem[\protect\citeauthoryear{{Kochanek}}{{Kochanek}}{2019}]{Kochanek19}
{Kochanek} C.~S.,  2019, \mn@doi [\mnras] {10.1093/mnras/sty3363}, \href
  {https://ui.adsabs.harvard.edu/abs/2019MNRAS.483.3762K} {483, 3762}

\bibitem[\protect\citeauthoryear{{Kochanek} et~al.,}{{Kochanek}
  et~al.}{2017}]{ASAS_SN_Kochanek2017}
{Kochanek} C.~S.,  et~al., 2017, \mn@doi [\pasp] {10.1088/1538-3873/aa80d9},
  \href {https://ui.adsabs.harvard.edu/abs/2017PASP..129j4502K} {129, 104502}

\bibitem[\protect\citeauthoryear{{Larsen}}{{Larsen}}{1999}]{Lars99}
{Larsen} S.~S.,  1999, \mn@doi [\aaps] {10.1051/aas:1999509}, \href
  {https://ui.adsabs.harvard.edu/abs/1999A&AS..139..393L} {139, 393}

\bibitem[\protect\citeauthoryear{{Law} et~al.,}{{Law} et~al.}{2009}]{Law09}
{Law} N.~M.,  et~al., 2009, \mn@doi [\pasp] {10.1086/648598}, \href
  {https://ui.adsabs.harvard.edu/abs/2009PASP..121.1395L} {121, 1395}

\bibitem[\protect\citeauthoryear{{Mackey}, {Mohamed}, {Gvaramadze}, {Kotak},
  {Langer}, {Meyer}, {Moriya}  \& {Neilson}}{{Mackey} et~al.}{2014}]{Mackey14}
{Mackey} J.,  {Mohamed} S.,  {Gvaramadze} V.~V.,  {Kotak} R.,  {Langer} N.,
  {Meyer} D. M.~A.,  {Moriya} T.~J.,   {Neilson} H.~R.,  2014, \mn@doi [\nat]
  {10.1038/nature13522}, \href
  {https://ui.adsabs.harvard.edu/abs/2014Natur.512..282M} {512, 282}

\bibitem[\protect\citeauthoryear{{Maguire} et~al.,}{{Maguire}
  et~al.}{2010}]{Maguire2004et}
{Maguire} K.,  et~al., 2010, \mn@doi [\mnras]
  {10.1111/j.1365-2966.2010.16332.x}, \href
  {https://ui.adsabs.harvard.edu/abs/2010MNRAS.404..981M} {404, 981}

\bibitem[\protect\citeauthoryear{{Maguire} et~al.,}{{Maguire}
  et~al.}{2012}]{Maguire2012_IIP_NebSpec}
{Maguire} K.,  et~al., 2012, \mn@doi [\mnras]
  {10.1111/j.1365-2966.2011.20276.x}, \href
  {https://ui.adsabs.harvard.edu/abs/2012MNRAS.420.3451M} {420, 3451}

\bibitem[\protect\citeauthoryear{{Makarov}, {Makarova}, {Rizzi}, {Tully},
  {Dolphin}, {Sakai}  \& {Shaya}}{{Makarov} et~al.}{2006}]{Maka06}
{Makarov} D.,  {Makarova} L.,  {Rizzi} L.,  {Tully} R.~B.,  {Dolphin} A.~E.,
  {Sakai} S.,   {Shaya} E.~J.,  2006, \mn@doi [\aj] {10.1086/508925}, \href
  {https://ui.adsabs.harvard.edu/abs/2006AJ....132.2729M} {132, 2729}

\bibitem[\protect\citeauthoryear{{Makarov}, {Prugniel}, {Terekhova}, {Courtois}
   \& {Vauglin}}{{Makarov} et~al.}{2014}]{Makarov14}
{Makarov} D.,  {Prugniel} P.,  {Terekhova} N.,  {Courtois} H.,   {Vauglin} I.,
  2014, \mn@doi [\aap] {10.1051/0004-6361/201423496}, \href
  {http://adsabs.harvard.edu/abs/2014A%26A...570A..13M} {570, A13}

\bibitem[\protect\citeauthoryear{{Martini} et~al.,}{{Martini}
  et~al.}{2011}]{Martini_2011_OSMOS}
{Martini} P.,  et~al., 2011, \mn@doi [\pasp] {10.1086/658357}, \href
  {https://ui.adsabs.harvard.edu/abs/2011PASP..123..187M} {123, 187}

\bibitem[\protect\citeauthoryear{{Masters}, {Springob}, {Haynes}  \&
  {Giovanelli}}{{Masters} et~al.}{2006}]{Mast06}
{Masters} K.~L.,  {Springob} C.~M.,  {Haynes} M.~P.,   {Giovanelli} R.,  2006,
  \mn@doi [\apj] {10.1086/508924}, \href
  {https://ui.adsabs.harvard.edu/abs/2006ApJ...653..861M} {653, 861}

\bibitem[\protect\citeauthoryear{{Mattila}, {Meikle}  \& {Greimel}}{{Mattila}
  et~al.}{2004}]{Mattila04}
{Mattila} S.,  {Meikle} W.~P.~S.,   {Greimel} R.,  2004, \mn@doi [\nar]
  {10.1016/j.newar.2003.12.033}, \href
  {https://ui.adsabs.harvard.edu/abs/2004NewAR..48..595M} {48, 595}

\bibitem[\protect\citeauthoryear{{Miyaji}, {Nomoto}, {Yokoi}  \&
  {Sugimoto}}{{Miyaji} et~al.}{1980}]{Miyaji80}
{Miyaji} S.,  {Nomoto} K.,  {Yokoi} K.,   {Sugimoto} D.,  1980, \pasj, \href
  {https://ui.adsabs.harvard.edu/abs/1980PASJ...32..303M} {32, 303}

\bibitem[\protect\citeauthoryear{{Moriya}, {Tominaga}, {Langer}, {Nomoto},
  {Blinnikov}  \& {Sorokina}}{{Moriya} et~al.}{2014}]{Moriya_2014_ECSNe}
{Moriya} T.~J.,  {Tominaga} N.,  {Langer} N.,  {Nomoto} K.,  {Blinnikov} S.~I.,
    {Sorokina} E.~I.,  2014, \mn@doi [\aap] {10.1051/0004-6361/201424264},
  \href {https://ui.adsabs.harvard.edu/abs/2014A&A...569A..57M} {569, A57}

\bibitem[\protect\citeauthoryear{{Mould} et~al.,}{{Mould}
  et~al.}{2000}]{Mould00}
{Mould} J.~R.,  et~al., 2000, \mn@doi [\apj] {10.1086/308304}, \href
  {https://ui.adsabs.harvard.edu/abs/2000ApJ...529..786M} {529, 786}

\bibitem[\protect\citeauthoryear{{M{\"u}ller-Bravo} et~al.,}{{M{\"u}ller-Bravo}
  et~al.}{2020}]{Muller_Bravo_2020_2016aqf}
{M{\"u}ller-Bravo} T.~E.,  et~al., 2020, \mn@doi [\mnras]
  {10.1093/mnras/staa1932}, \href
  {https://ui.adsabs.harvard.edu/abs/2020MNRAS.497..361M} {497, 361}

\bibitem[\protect\citeauthoryear{{M{\"u}ller}, {Wanajo}, {Janka}, {Heger},
  {Gay}  \& {Sim}}{{M{\"u}ller}
  et~al.}{2017}]{Muller_2017_LowMassFeCCSNe_and_ECSNe}
{M{\"u}ller} B.,  {Wanajo} S.,  {Janka} H.~T.,  {Heger} A.,  {Gay} D.,   {Sim}
  S.~A.,  2017, \memsai, \href
  {https://ui.adsabs.harvard.edu/abs/2017MmSAI..88..288M} {88, 288}

\bibitem[\protect\citeauthoryear{{Nakaoka} et~al.,}{{Nakaoka}
  et~al.}{2018}]{Nakaoka_2018_16bkv}
{Nakaoka} T.,  et~al., 2018, \mn@doi [\apj] {10.3847/1538-4357/aabee7}, \href
  {https://ui.adsabs.harvard.edu/abs/2018ApJ...859...78N} {859, 78}

\bibitem[\protect\citeauthoryear{{Nicholl}}{{Nicholl}}{2018}]{Nich18}
{Nicholl} M.,  2018, \mn@doi [Research Notes of the American Astronomical
  Society] {10.3847/2515-5172/aaf799}, \href
  {https://ui.adsabs.harvard.edu/abs/2018RNAAS...2..230N} {2, 230}

\bibitem[\protect\citeauthoryear{{Nomoto}}{{Nomoto}}{1984}]{Nomoto_1984_ECSNe}
{Nomoto} K.,  1984, \mn@doi [\apj] {10.1086/161749}, \href
  {https://ui.adsabs.harvard.edu/abs/1984ApJ...277..791N} {277, 791}

\bibitem[\protect\citeauthoryear{{Nugent} \& {Hamuy}}{{Nugent} \&
  {Hamuy}}{2017}]{Nugent17}
{Nugent} P.,  {Hamuy} M.,  2017, in {Alsabti} A.~W.,  {Murdin} P.,  eds, ,
  Handbook of Supernovae.
Springer, p.~2671, \mn@doi{10.1007/978-3-319-21846-5_108}

\bibitem[\protect\citeauthoryear{{Nugent} et~al.,}{{Nugent}
  et~al.}{2006}]{Nugent06_stnd_candle}
{Nugent} P.,  et~al., 2006, \mn@doi [\apj] {10.1086/504413}, \href
  {https://ui.adsabs.harvard.edu/abs/2006ApJ...645..841N} {645, 841}

\bibitem[\protect\citeauthoryear{{Ouchi} \& {Maeda}}{{Ouchi} \&
  {Maeda}}{2021}]{Ouchi21_SN2009kf}
{Ouchi} R.,  {Maeda} K.,  2021, \mn@doi [\mnras] {10.1093/mnras/staa2527},
  \href {https://ui.adsabs.harvard.edu/abs/2021MNRAS.500.1889O} {500, 1889}

\bibitem[\protect\citeauthoryear{{Paraskeva}, {Bonanos}, {Liakos}, {Spetsieri}
  \& {Maund}}{{Paraskeva} et~al.}{2020}]{Paras20}
{Paraskeva} E.,  {Bonanos} A.~Z.,  {Liakos} A.,  {Spetsieri} Z.~T.,   {Maund}
  J.~R.,  2020, arXiv e-prints, \href
  {https://ui.adsabs.harvard.edu/abs/2020arXiv200708540P} {p. arXiv:2007.08540}

\bibitem[\protect\citeauthoryear{{Pastorello} et~al.,}{{Pastorello}
  et~al.}{2009}]{Pastorello2005cs}
{Pastorello} A.,  et~al., 2009, \mn@doi [\mnras]
  {10.1111/j.1365-2966.2009.14505.x}, \href
  {https://ui.adsabs.harvard.edu/abs/2009MNRAS.394.2266P} {394, 2266}

\bibitem[\protect\citeauthoryear{{Phillips} et~al.,}{{Phillips}
  et~al.}{2013}]{Phil13}
{Phillips} M.~M.,  et~al., 2013, \mn@doi [\apj] {10.1088/0004-637X/779/1/38},
  \href {https://ui.adsabs.harvard.edu/abs/2013ApJ...779...38P} {779, 38}

\bibitem[\protect\citeauthoryear{{Polshaw} et~al.,}{{Polshaw}
  et~al.}{2016}]{PolshawLSQ13fn}
{Polshaw} J.,  et~al., 2016, \mn@doi [\aap] {10.1051/0004-6361/201527682},
  \href {https://ui.adsabs.harvard.edu/abs/2016A&A...588A...1P} {588, A1}

\bibitem[\protect\citeauthoryear{{Poznanski}, {Prochaska}  \&
  {Bloom}}{{Poznanski} et~al.}{2012}]{Posz12}
{Poznanski} D.,  {Prochaska} J.~X.,   {Bloom} J.~S.,  2012, \mn@doi [\mnras]
  {10.1111/j.1365-2966.2012.21796.x}, \href
  {https://ui.adsabs.harvard.edu/abs/2012MNRAS.426.1465P} {426, 1465}

\bibitem[\protect\citeauthoryear{{Prieto} et~al.,}{{Prieto}
  et~al.}{2008}]{Prieto08}
{Prieto} J.~L.,  et~al., 2008, \mn@doi [\apjl] {10.1086/589922}, \href
  {https://ui.adsabs.harvard.edu/abs/2008ApJ...681L...9P} {681, L9}

\bibitem[\protect\citeauthoryear{{Prilutskii} \& {Usov}}{{Prilutskii} \&
  {Usov}}{1976}]{Prilutskii76}
{Prilutskii} O.~F.,  {Usov} V.~V.,  1976, \sovast, \href
  {https://ui.adsabs.harvard.edu/abs/1976SvA....20....2P} {20, 2}

\bibitem[\protect\citeauthoryear{{Puget} et~al.,}{{Puget}
  et~al.}{2004}]{Puget_2004_WIRCam}
{Puget} P.,  et~al., 2004, in {Moorwood} A. F.~M.,  {Iye} M.,  eds,  Society of
  Photo-Optical Instrumentation Engineers (SPIE) Conference Series Vol. 5492,
  Ground-based Instrumentation for Astronomy. pp 978--987,
  \mn@doi{10.1117/12.551097}

\bibitem[\protect\citeauthoryear{{Rafanelli} \& {Siviero}}{{Rafanelli} \&
  {Siviero}}{2012}]{Rafanelli_2012_GalileoTelescope}
{Rafanelli} P.,  {Siviero} A.,  2012, \mn@doi [Baltic Astronomy]
  {10.1515/astro-2017-0351}, \href
  {https://ui.adsabs.harvard.edu/abs/2012BaltA..21....1R} {21, 1}

\bibitem[\protect\citeauthoryear{{Riess}, {Casertano}, {Yuan}, {Macri}  \&
  {Scolnic}}{{Riess} et~al.}{2019}]{Ries19}
{Riess} A.~G.,  {Casertano} S.,  {Yuan} W.,  {Macri} L.~M.,   {Scolnic} D.,
  2019, \mn@doi [\apj] {10.3847/1538-4357/ab1422}, \href
  {https://ui.adsabs.harvard.edu/abs/2019ApJ...876...85R} {876, 85}

\bibitem[\protect\citeauthoryear{{Rodr{\'{i}}guez} et~al.,}{{Rodr{\'{i}}guez}
  et~al.}{2020}]{Rodriguez2020_LLEVs}
{Rodr{\'{i}}guez} {\'O}.,  et~al., 2020, \mn@doi [\mnras]
  {10.1093/mnras/staa1133}, \href
  {https://ui.adsabs.harvard.edu/abs/2020MNRAS.494.5882R} {494, 5882}

\bibitem[\protect\citeauthoryear{{Roming} et~al.,}{{Roming}
  et~al.}{2005}]{Roming_UVOT_2005}
{Roming} P. W.~A.,  et~al., 2005, \mn@doi [\ssr] {10.1007/s11214-005-5095-4},
  \href {https://ui.adsabs.harvard.edu/abs/2005SSRv..120...95R} {120, 95}

\bibitem[\protect\citeauthoryear{{Sanders}, {Mazzarella}, {Kim}, {Surace}  \&
  {Soifer}}{{Sanders} et~al.}{2003}]{Sanders03}
{Sanders} D.~B.,  {Mazzarella} J.~M.,  {Kim} D.~C.,  {Surace} J.~A.,   {Soifer}
  B.~T.,  2003, \mn@doi [\aj] {10.1086/376841}, \href
  {https://ui.adsabs.harvard.edu/abs/2003AJ....126.1607S} {126, 1607}

\bibitem[\protect\citeauthoryear{{Schlafly} \& {Finkbeiner}}{{Schlafly} \&
  {Finkbeiner}}{2011}]{Schl11}
{Schlafly} E.~F.,  {Finkbeiner} D.~P.,  2011, \mn@doi [\apj]
  {10.1088/0004-637X/737/2/103}, \href
  {https://ui.adsabs.harvard.edu/abs/2011ApJ...737..103S} {737, 103}

\bibitem[\protect\citeauthoryear{{Shappee} et~al.,}{{Shappee}
  et~al.}{2014}]{ASAS_SN_Shappee2014}
{Shappee} B.~J.,  et~al., 2014, \mn@doi [\apj] {10.1088/0004-637X/788/1/48},
  \href {https://ui.adsabs.harvard.edu/abs/2014ApJ...788...48S} {788, 48}

\bibitem[\protect\citeauthoryear{Skrutskie et~al.,}{Skrutskie
  et~al.}{2006}]{2mass_Skrutskie2006}
Skrutskie M.~F.,  et~al., 2006, \mn@doi [The Astronomical Journal]
  {10.1086/498708}, 131, 1163

\bibitem[\protect\citeauthoryear{{Smartt}}{{Smartt}}{2009}]{Smartt2009_progenitors}
{Smartt} S.~J.,  2009, \mn@doi [\araa] {10.1146/annurev-astro-082708-101737},
  \href {https://ui.adsabs.harvard.edu/abs/2009ARA&A..47...63S} {47, 63}

\bibitem[\protect\citeauthoryear{{Smith} et~al.,}{{Smith}
  et~al.}{2019}]{Smit19}
{Smith} K.~W.,  et~al., 2019, \mn@doi [Research Notes of the American
  Astronomical Society] {10.3847/2515-5172/ab020f}, \href
  {https://ui.adsabs.harvard.edu/abs/2019RNAAS...3...26S} {3, 26}

\bibitem[\protect\citeauthoryear{{Smith} et~al.,}{{Smith}
  et~al.}{2020}]{Smith20}
{Smith} K.~W.,  et~al., 2020, \mn@doi [\pasp] {10.1088/1538-3873/ab936e}, \href
  {https://ui.adsabs.harvard.edu/abs/2020PASP..132h5002S} {132, 085002}

\bibitem[\protect\citeauthoryear{{Sollerman}, {Cox}, {Mattila}, {Ehrenfreund},
  {Kaper}, {Leibundgut}  \& {Lundqvist}}{{Sollerman}
  et~al.}{2005}]{Sollerman_2005_DIBS}
{Sollerman} J.,  {Cox} N.,  {Mattila} S.,  {Ehrenfreund} P.,  {Kaper} L.,
  {Leibundgut} B.,   {Lundqvist} P.,  2005, \mn@doi [\aap]
  {10.1051/0004-6361:20041465}, \href
  {https://ui.adsabs.harvard.edu/abs/2005A&A...429..559S} {429, 559}

\bibitem[\protect\citeauthoryear{{Tak{\'a}ts} et~al.,}{{Tak{\'a}ts}
  et~al.}{2015}]{Takats15_2009ib}
{Tak{\'a}ts} K.,  et~al., 2015, \mn@doi [\mnras] {10.1093/mnras/stv857}, \href
  {https://ui.adsabs.harvard.edu/abs/2015MNRAS.450.3137T} {450, 3137}

\bibitem[\protect\citeauthoryear{{Taramopoulos}, {Payne}  \&
  {Briggs}}{{Taramopoulos} et~al.}{2001}]{Tara01}
{Taramopoulos} A.,  {Payne} H.,   {Briggs} F.~H.,  2001, \mn@doi [\aap]
  {10.1051/0004-6361:20000143}, \href
  {https://ui.adsabs.harvard.edu/abs/2001A&A...365..360T} {365, 360}

\bibitem[\protect\citeauthoryear{{Tody}}{{Tody}}{1986}]{IRAF_1986}
{Tody} D.,  1986, in {Crawford} D.~L.,  ed.,  Society of Photo-Optical
  Instrumentation Engineers (SPIE) Conference Series Vol. 627, Instrumentation
  in astronomy VI. p.~733, \mn@doi{10.1117/12.968154}

\bibitem[\protect\citeauthoryear{{Tonry} et~al.,}{{Tonry}
  et~al.}{2018}]{Tonry18}
{Tonry} J.~L.,  et~al., 2018, \mn@doi [\pasp] {10.1088/1538-3873/aabadf}, \href
  {https://ui.adsabs.harvard.edu/abs/2018PASP..130f4505T} {130, 064505}

\bibitem[\protect\citeauthoryear{{Turatto}, {Cappellaro}, {Danziger},
  {Benetti}, {Gouiffes}  \& {della Valle}}{{Turatto}
  et~al.}{1993}]{Turatto_1988Z}
{Turatto} M.,  {Cappellaro} E.,  {Danziger} I.~J.,  {Benetti} S.,  {Gouiffes}
  C.,   {della Valle} M.,  1993, \mn@doi [\mnras] {10.1093/mnras/262.1.128},
  \href {https://ui.adsabs.harvard.edu/abs/1993MNRAS.262..128T} {262, 128}

\bibitem[\protect\citeauthoryear{{Tutui} \& {Sofue}}{{Tutui} \&
  {Sofue}}{1997}]{Tutu97}
{Tutui} Y.,  {Sofue} Y.,  1997, \aap, \href
  {https://ui.adsabs.harvard.edu/abs/1997A&A...326..915T} {326, 915}

\bibitem[\protect\citeauthoryear{{Valenti} et~al.,}{{Valenti}
  et~al.}{2016}]{Valenti2016}
{Valenti} S.,  et~al., 2016, \mn@doi [\mnras] {10.1093/mnras/stw870}, \href
  {https://ui.adsabs.harvard.edu/abs/2016MNRAS.459.3939V} {459, 3939}

\bibitem[\protect\citeauthoryear{{Villanueva}, {Gaudi}, {Pogge}, {Eastman},
  {Stassun}, {Trueblood}  \& {Trueblood}}{{Villanueva}
  et~al.}{2018}]{DEMONEXT_2018}
{Villanueva} Steven J.,  {Gaudi} B.~S.,  {Pogge} R.~W.,  {Eastman} J.~D.,
  {Stassun} K.~G.,  {Trueblood} M.,   {Trueblood} P.,  2018, \mn@doi [\pasp]
  {10.1088/1538-3873/aa9603}, \href
  {https://ui.adsabs.harvard.edu/abs/2018PASP..130a5001V} {130, 015001}

\bibitem[\protect\citeauthoryear{{Wanajo}, {Nomoto}, {Janka}, {Kitaura}  \&
  {M{\"u}ller}}{{Wanajo} et~al.}{2009}]{Wanajo_2009_ECSN_Nucleosynthesis}
{Wanajo} S.,  {Nomoto} K.,  {Janka} H.~T.,  {Kitaura} F.~S.,   {M{\"u}ller} B.,
   2009, \mn@doi [\apj] {10.1088/0004-637X/695/1/208}, \href
  {https://ui.adsabs.harvard.edu/abs/2009ApJ...695..208W} {695, 208}

\bibitem[\protect\citeauthoryear{{Willson}}{{Willson}}{2000}]{Willson2000}
{Willson} L.~A.,  2000, \mn@doi [\araa] {10.1146/annurev.astro.38.1.573}, \href
  {https://ui.adsabs.harvard.edu/abs/2000ARA&A..38..573W} {38, 573}

\bibitem[\protect\citeauthoryear{{Woosley} \& {Heger}}{{Woosley} \&
  {Heger}}{2015}]{Woosley_2015_ECSNe}
{Woosley} S.~E.,  {Heger} A.,  2015, \mn@doi [\apj]
  {10.1088/0004-637X/810/1/34}, \href
  {https://ui.adsabs.harvard.edu/abs/2015ApJ...810...34W} {810, 34}

\bibitem[\protect\citeauthoryear{{Yaron} \& {Gal-Yam}}{{Yaron} \&
  {Gal-Yam}}{2012}]{Yaron_2012_WISEREP}
{Yaron} O.,  {Gal-Yam} A.,  2012, \mn@doi [\pasp] {10.1086/666656}, \href
  {https://ui.adsabs.harvard.edu/abs/2012PASP..124..668Y} {124, 668}

\bibitem[\protect\citeauthoryear{{Yaron} et~al.,}{{Yaron}
  et~al.}{2017}]{Yaron2017_SN2013fs}
{Yaron} O.,  et~al., 2017, \mn@doi [Nature Physics] {10.1038/nphys4025}, \href
  {https://ui.adsabs.harvard.edu/abs/2017NatPh..13..510Y} {13, 510}

\bibitem[\protect\citeauthoryear{{Zhang}, {Xu}  \& {Wang}}{{Zhang}
  et~al.}{2018a}]{ATel_18zd_11379}
{Zhang} J.,  {Xu} L.,   {Wang} X.,  2018a, The Astronomer's Telegram, \href
  {https://ui.adsabs.harvard.edu/abs/2018ATel11379....1Z} {11379, 1}

\bibitem[\protect\citeauthoryear{{Zhang}, {Xu}  \& {Wang}}{{Zhang}
  et~al.}{2018b}]{ATel_Zhang_18zd_IIn}
{Zhang} J.,  {Xu} L.,   {Wang} X.,  2018b, The Astronomer's Telegram, \href
  {https://ui.adsabs.harvard.edu/abs/2018ATel11379....1Z} {11379, 1}

\bibitem[\protect\citeauthoryear{{Zhang} et~al.,}{{Zhang}
  et~al.}{2020}]{Zhan20}
{Zhang} J.,  et~al., 2020, arXiv e-prints, \href
  {https://ui.adsabs.harvard.edu/abs/2020arXiv200714348Z} {p. arXiv:2007.14348}

\bibitem[\protect\citeauthoryear{{de Jaeger}, {Stahl}, {Zheng}, {Filippenko},
  {Riess}  \& {Galbany}}{{de Jaeger} et~al.}{2020}]{deJaeger_2020_Hubble}
{de Jaeger} T.,  {Stahl} B.~E.,  {Zheng} W.,  {Filippenko} A.~V.,  {Riess}
  A.~G.,   {Galbany} L.,  2020, \mn@doi [\mnras] {10.1093/mnras/staa1801},
  \href {https://ui.adsabs.harvard.edu/abs/2020MNRAS.496.3402D} {496, 3402}

\bibitem[\protect\citeauthoryear{{de Vaucouleurs}, {de Vaucouleurs}, {Corwin},
  {Buta}, {Paturel}  \& {Fouque}}{{de Vaucouleurs} et~al.}{1991}]{RCBG}
{de Vaucouleurs} G.,  {de Vaucouleurs} A.,  {Corwin} Herold~G. J.,  {Buta}
  R.~J.,  {Paturel} G.,   {Fouque} P.,  1991, {Third Reference Catalogue of
  Bright Galaxies}.
Springer

\bibitem[\protect\citeauthoryear{{van Dokkum}}{{van
  Dokkum}}{2001}]{Dokkum_2001_lacosmic}
{van Dokkum} P.~G.,  2001, \mn@doi [\pasp] {10.1086/323894}, \href
  {https://ui.adsabs.harvard.edu/abs/2001PASP..113.1420V} {113, 1420}

\makeatother
\end{thebibliography}



\appendix

\section{TRGB Distance Estimate towards NGC~2146}
\label{sec:TRGB}
ACS/HRC images were taken of the nucleus of NGC 2146 in 2005, and while relatively long exposure times were used, the small field of view and dusty nuclear environment makes these less suitable for a TRGB distance. Images taken of SN~2018zd itself in May 2019 (\textit{F555W} and \textit{F814W}), as well as WFC3-IR images of NGC~2146 taken in March 2011 (\textit{F110W} and \textit{F160W}), were investigated to see if they are sufficiently deep to detect the TRGB. We performed photometry on the WFC3-IR \textit{F110W} and \textit{F160W} filters using the {\sc dolphot} package. We apply the cuts on sharpness (<0.12), crowding (<0.48) and SNR used by \cite{Dalc12} to create a catalogue of high quality detections of sources in both filters. We find that these catalogues reach more than one magnitude deeper than the Hubble Source catalogue products, but even so, they are only able to probe the TRGB in the NIR out to $<10$ Mpc \citep{Dalc12}. However, since the closest Tully-Fisher distance and the kinematic distance to NGC 2146 are comparable to this distance, we used a Sobel filter to search for the TRGB in the histogram of source counts as a function of magnitude. 

\begin{figure}
\includegraphics[width=\columnwidth]{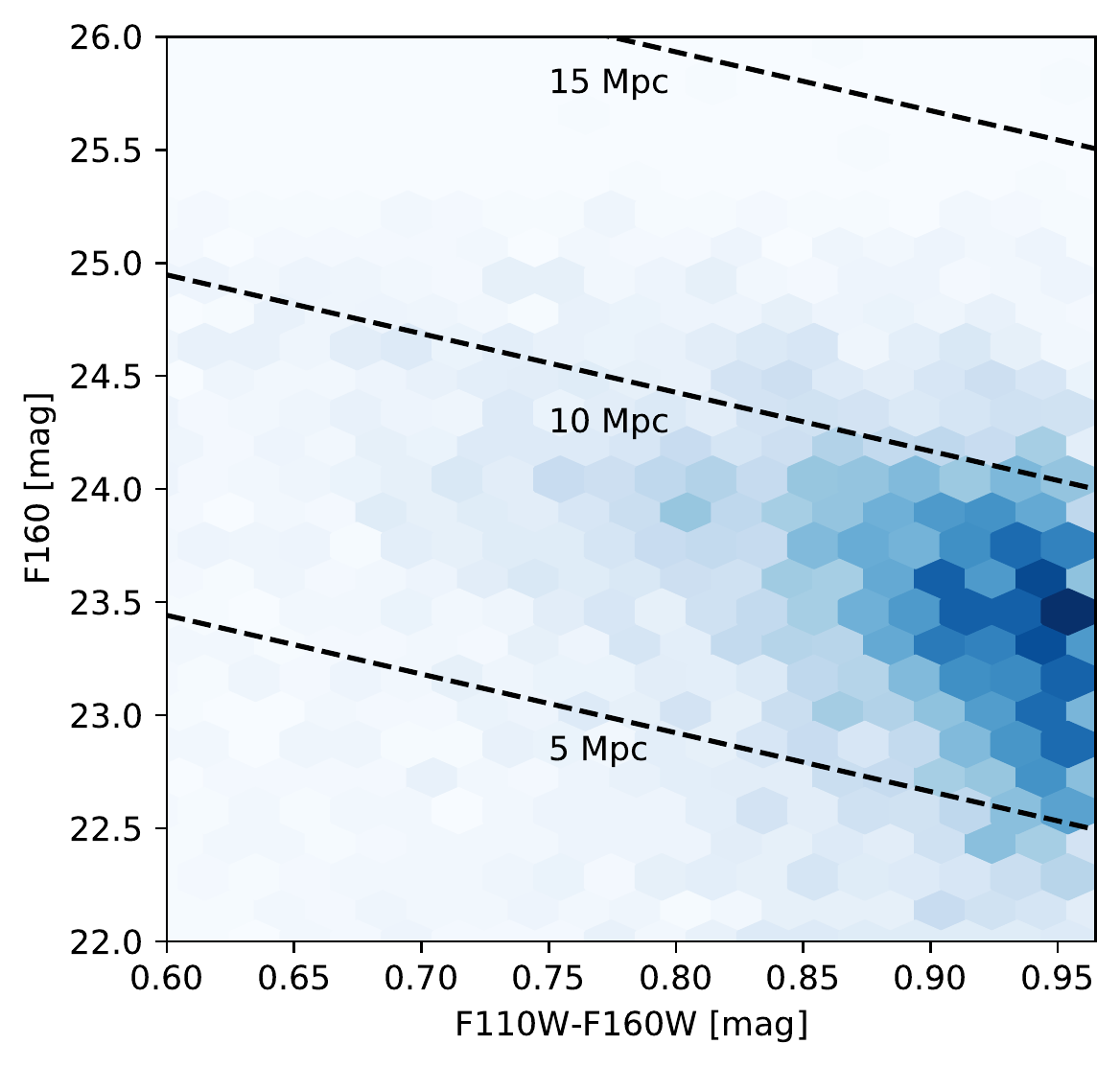}
\caption{Extinction corrected NIR colour-magnitude diagram for NGC~2146 using only sources that pass the quality cuts for point sources. Darker regions indicate more sources within each bin. The dashed lines show the expected location of the TRGB from \protect\cite{Dalc12} at 5, 10 and 15~Mpc.}
\label{fig:TRGB}
\end{figure}

To account for the strong dependence of the TRGB on colour in the NIR, we split the sources into 0.05 mag colour bins and run the Sobel filter separately on each. We find no peaks that are consistent with the detection of the TRGB, which unfortunately does not provide a strong constraint on the distance to NGC 2146. However, as can be seen in Fig. \ref{fig:TRGB}, these data would be sufficiently deep to confidently identify the TRGB at distances $\lesssim$7~Mpc, and so we can infer that the distance is greater than this value.%

Turning to the optical ({\it F555W} and {\it F814W}) images from 2019, these do not go sufficiently deep to detect the TRGB. TRGB distances from a single orbit of ACS imaging become unreliable at $m_I>$26 mag \citep{Maka06}. Assuming that the TRGB is $\sim-4$~mag, this implies that we cannot go beyond 10~Mpc with these data.

\section{Analysis of Na~{\sc i}~D}
\label{sec:Sodium_lines}

\begin{figure}
	\includegraphics[width=\columnwidth]{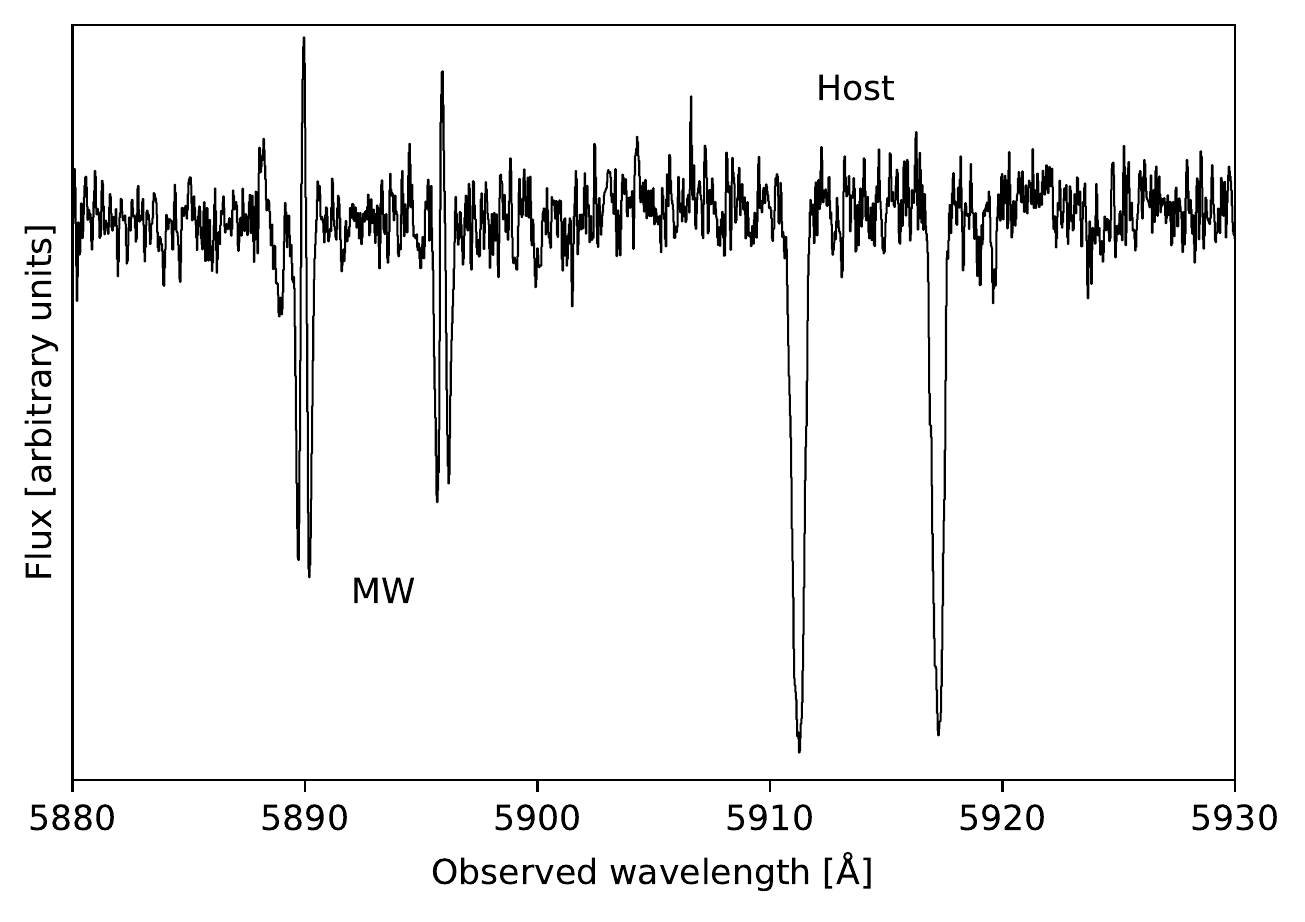}
    \caption{FIES spectrum of SN~2018zd in the region of NaD.}
    \label{fig:FIES}
\end{figure}

To estimate the host galaxy extinction, we attempted to use the strength of the NaD lines measured in our high-resolution FIES (R=25,000) spectrum. 
We measured the equivalent widths for the components of the doublet to be $EW_{D1}=0.519\pm0.007$~\AA\ and $EW_{D2}=0.594\pm0.014$~\AA. Applying the relations between NaD equivalent width and $E(B - V)$ from \cite{Posz12}, we find $E(B - V)=0.33_{-0.10}^{+0.14}$~mag and $E(B - V)=0.24_{-0.07}^{+0.10}$~mag from the D$_1$ and D$_2$ lines respectively. 
Unfortunately, the other interstellar lines such as K~{\sc i} $\lambda\lambda$7665, 7699 or the 5780~\AA\ diffuse interstellar band which are useful for determining extinction \citep[e.g.][]{Phil13, Sollerman_2005_DIBS} are not seen in the FIES spectrum (Fig. \ref{fig:FIES}). Similarly, the Balmer lines are not visible in the high-resolution spectra, and so we cannot use the Balmer decrement to measure extinction.

Taking the total of $E(B - V)_{tot}$=0.37 mag (Milky Way + host galaxy estimate from the Na~{\sc i}~D lines) corresponds to 
$A_V^{tot}=1.15$ mag. Applying this to the earliest spectra and fitting a blackbody function results in an unphysically high temperature.

\onecolumn
\section{Tables}

{\small
\begin{longtable}{cccccccccc} 
\hline
MJD&Phase (d)&$u$&$B$&$V$&$g$&$r$&$i$&$z$&Source \\
\hline
58179.5&1.0&-&-&-&-&17.74$\pm$0.16&-&-&Itagaki \\
58180.6&2.1&-&-&-&-&16.52$\pm$0.07&-&-&Itagaki \\
58181.4&2.9&-&-&-&-&15.69$\pm$0.02&-&-&Itagaki \\
58183.5&5.0&-&-&-&-&14.33$\pm$0.03&-&-&Itagaki \\
58184.1&5.6&-&14.30$\pm$0.24&14.25$\pm$0.22&-&14.36$\pm$0.40&14.70$\pm$0.91&-&DEMONEXT \\
58184.2&5.7&-&14.16$\pm$0.06&14.18$\pm$0.05&-&14.25$\pm$0.14&14.34$\pm$0.07&-&Post SRO \\
58184.5&6.0&-&-&-&-&13.97$\pm$0.06&-&-&Itagaki \\
58184.8&6.3&-&13.92$\pm$0.07&13.79$\pm$0.06&13.77$\pm$0.04&13.78$\pm$0.02$^{*}$&14.06$\pm$0.02&14.20$\pm$0.01&StanCam \\
58186.2&7.7&-&13.66$\pm$0.06&13.64$\pm$0.03&-&13.72$\pm$0.05&13.84$\pm$0.08&-&Post SRO \\
58187.2&8.7&-&13.78$\pm$0.44&15.95$\pm$0.05&-&13.78$\pm$0.21&14.02$\pm$0.65&-&DEMONEXT \\
\hline
\caption{Abbreviated optical photometry log of SN 2018zd (AB mag), full table is available as supplementary material. Phase is with respect to the estimated explosion epoch (MJD 58178.5). Values have not been dereddened. StanCam $R$--band magnitudes are reported in the $r$--band column and denoted with $^{*}$.} 
\label{tab:opt_phot_table}
\end{longtable}
}

{\small
\begin{longtable}{ccccc} 
\hline
MJD&Phase (d) &$J$&$H$&$K$ \\
\hline
58188.9&10.4&13.32$\pm$0.06&13.08$\pm$0.22&13.30$\pm$0.01 \\
58208.9&30.4&13.21$\pm$0.05&12.84$\pm$0.24&12.94$\pm$0.03 \\
58249.9&71.4&13.30$\pm$0.10&12.88$\pm$0.22&12.95$\pm$0.01 \\
58350.2&171.7&16.62$\pm$0.07&16.35$\pm$0.22&16.46$\pm$0.03 \\
58385.2&206.7&17.10$\pm$0.08&16.85$\pm$0.27&16.76$\pm$0.06 \\
58494.0&315.5&18.61$\pm$0.08&18.20$\pm$0.32&18.22$\pm$0.08 \\
58582.9&404.4&19.61$\pm$0.17&19.49$\pm$0.49&19.06$\pm$0.31 \\
\hline
\caption{Full NIR photometry log of SN~2018zd taken with the NOT. Phase is with respect to the estimated explosion epoch. Values have not been dereddened.} 
\label{tab:nir_phot_table}
\end{longtable}
}

{\small
\begin{longtable}{cccccccc} 
\hline
MJD&Phase (d)&$U$&$B$&$V$&$uvw1$&$uvw2$&$uvm2$ \\
\hline
58181.3&2.8&15.02$\pm$0.04&15.99$\pm$0.04&15.76$\pm$0.06&15.12$\pm$0.04&15.40$\pm$0.04&15.46$\pm$0.04 \\
58182.3&3.8&13.91$\pm$0.03&15.04$\pm$0.03&15.05$\pm$0.04&13.82$\pm$0.02&13.81$\pm$0.02&14.00$\pm$0.02 \\
58183.2&4.7&13.42$\pm$0.03&14.61$\pm$0.03&14.59$\pm$0.04&13.26$\pm$0.03&13.19$\pm$0.02&13.30$\pm$0.06 \\
58184.3&5.8&-&13.75$\pm$0.03&14.13$\pm$0.03&12.78$\pm$0.06&12.70$\pm$0.06&12.98$\pm$0.03 \\
58185.1&6.6&12.68$\pm$0.03&13.90$\pm$0.03&13.87$\pm$0.04&12.49$\pm$0.03&12.45$\pm$0.02&12.69$\pm$0.03 \\
58186.1&7.6&-&-&-&12.37$\pm$0.02&12.46$\pm$0.02&12.59$\pm$0.03 \\
58187.2&8.7&-&13.12$\pm$0.02&13.51$\pm$0.03&12.45$\pm$0.02&12.63$\pm$0.02&12.72$\pm$0.02 \\
58189.4&10.9&12.05$\pm$0.03&13.49$\pm$0.02&13.61$\pm$0.03&12.78$\pm$0.02&13.19$\pm$0.02&13.16$\pm$0.02 \\
58191.7&13.2&12.94$\pm$0.03&13.97$\pm$0.02&13.73$\pm$0.03&13.09$\pm$0.02&13.79$\pm$0.02&13.57$\pm$0.02 \\
58195.8&17.3&13.15$\pm$0.03&14.09$\pm$0.02&13.85$\pm$0.03&13.47$\pm$0.02&14.39$\pm$0.02&14.13$\pm$0.02 \\
\hline
\caption{Abbreviated \textit{Swift} photometry log of SN 2018zd (Vega mag), full table is available as supplementary material. Phase is with respect to the estimated explosion epoch. Values have not been dereddened.} 
\label{tab:swift_phot_table}
\end{longtable}
}

\begin{table}
\centering
\begin{tabular}{lcccc}
\hline
Object      & z         & $\mu$ & E(B-V) & Ref.\\
\hline
SN2018zd    & 0.003     & 30.97 & 0.19  & -     \\ 
SN2004et    & -         & 28.85 & 0.41  & 1\\
SN2005cs    & -         & 29.29 & 0.05  & 2\\
SN2016bkv   & 0.0019    & -     & 0.0   & 3\\
LSQ13fn     & 0.063     & -     & 0.054 & 4\\
\hline
\end{tabular}
\caption{SUPERBOL input parameters. 1-\protect\citep{Maguire2004et}, 2-\protect\citep{Pastorello2005cs}, 3-\protect\citep{Hosseinzadeh16bkv}, 4-\protect\citep{PolshawLSQ13fn}.}
\label{tab:SUPERBOL_params}
\end{table}

\begin{table}
\centering
\begin{tabular}{cccc}
\hline
R.A. (deg) & Dec. (deg) & $u$ (Vega mag) & $u$ (AB mag) \\
\hline
94.46850 & 78.33003 & 14.53$\pm$0.02 & 15.55$\pm$0.02 \\
94.43350 & 78.35502 & 15.49$\pm$0.02 & 16.51$\pm$0.02 \\
94.64375 & 78.38668 & 16.47$\pm$0.02 & 17.49$\pm$0.02 \\
94.86687 & 78.34745 & 14.90$\pm$0.02 & 15.92$\pm$0.02 \\
94.90050 & 78.35081 & 12.55$\pm$0.02 & 13.57$\pm$0.02 \\
94.31092 & 78.38581 & 15.41$\pm$0.02 & 16.43$\pm$0.02 \\
94.24829 & 78.40583 & 14.87$\pm$0.02 & 15.89$\pm$0.02 \\
94.38379 & 78.42454 & 14.70$\pm$0.02 & 15.72$\pm$0.02 \\
\hline
\end{tabular}
\caption{Sequence stars used to calibrate $u$--band photometry. 
}
\label{tab:u_catalog}
\end{table}

\section{Additional figures}

\begin{figure*}
	\includegraphics[width=0.5\columnwidth]{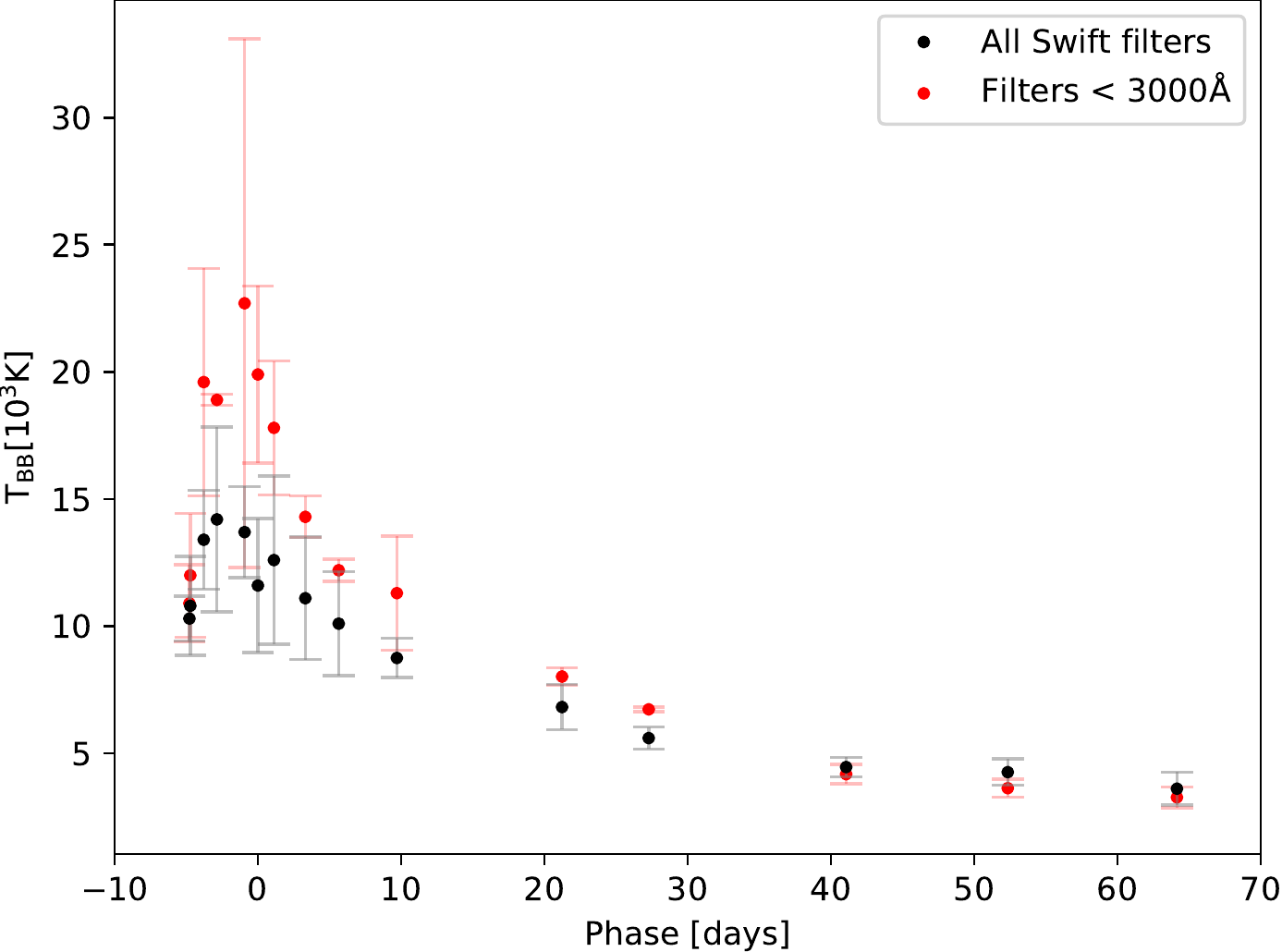}
    \caption{SUPERBOL output of the temperature evolution calculated using only the UV \textit{Swift} data.}
    \label{fig:Swift_temp_evo}
\end{figure*}

\begin{figure*}
	\includegraphics[width=0.5\columnwidth]{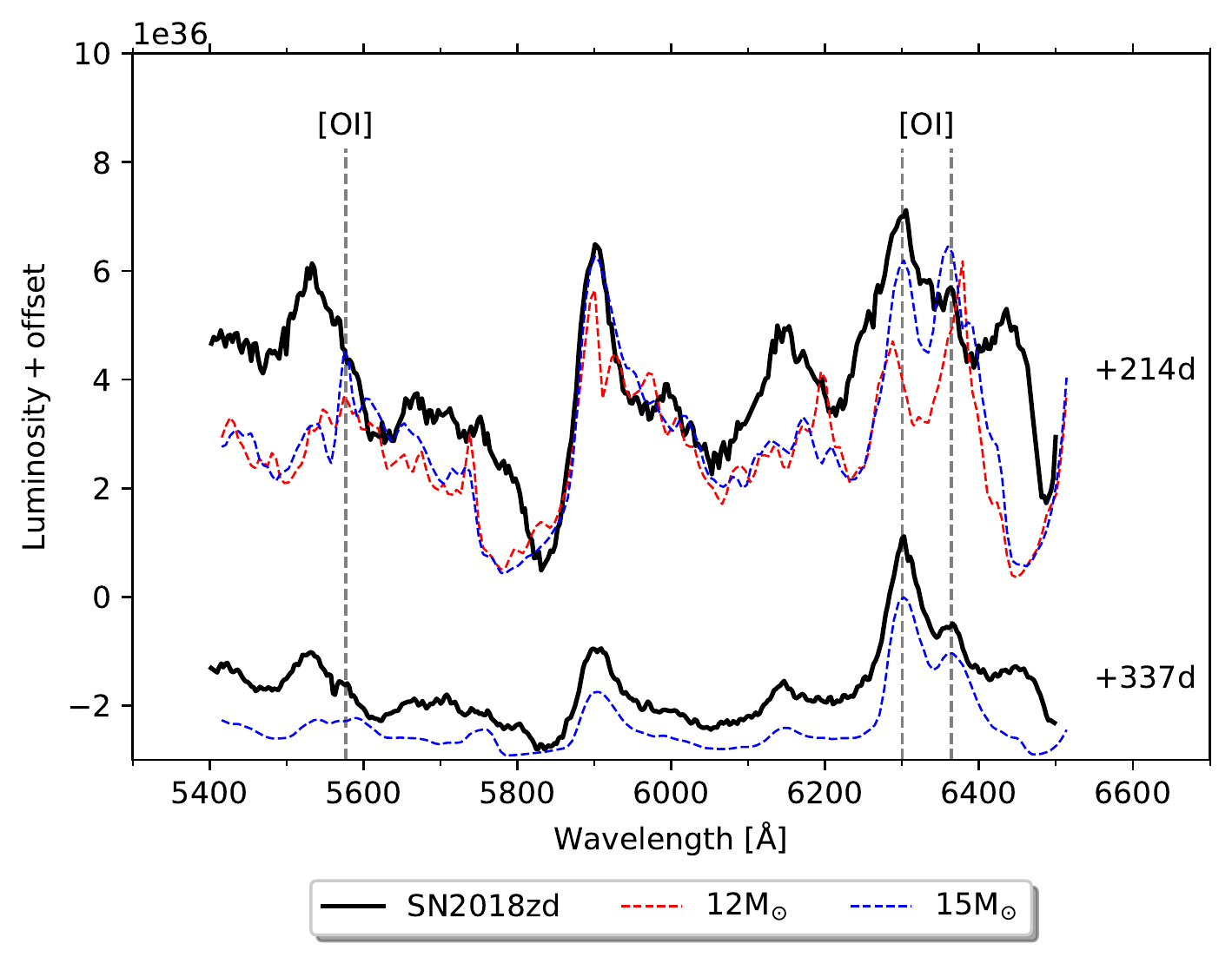}
    \caption{Comparison of the nebular-phase spectra of SN~2018zd at 214~d and 337~d (in black) to the \protect\cite{Jerkstrand_2014_OI_mass} model spectra corresponding to different progenitor masses (12~$\rm{M_{\odot}}$ at 212~d; 15 $\rm{M_{\odot}}$ at 212~d and 369~d). The model spectra are scaled to the distance of SN~2018zd and its $^{56}$Ni mass. The SN~2018zd flux calibration was checked against photometry.}
    \label{fig:Jerkstrand_OI_models}
\end{figure*}

\begin{figure*}
	\includegraphics[width=0.5\columnwidth]{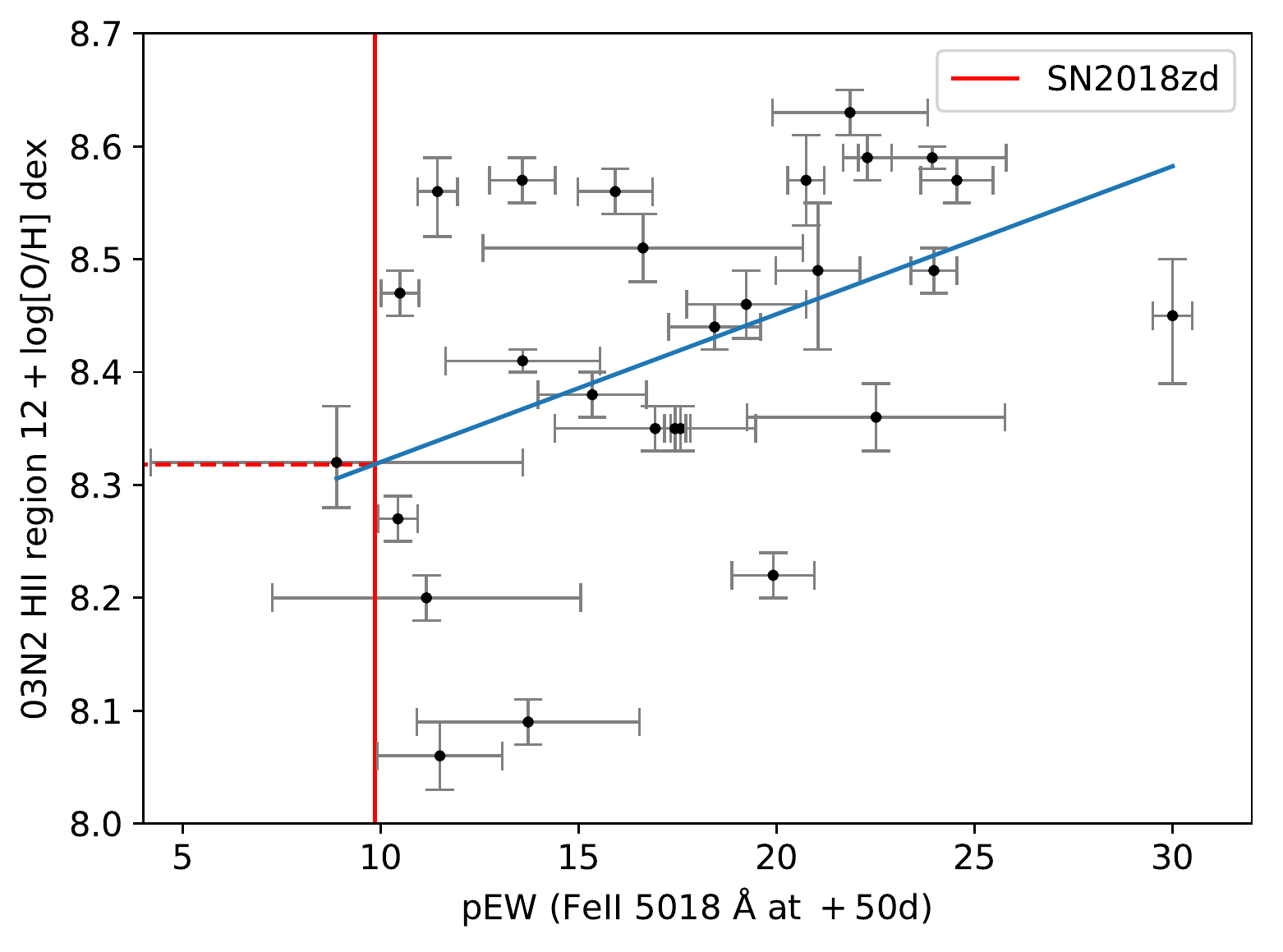}
    \caption{SN~2018zd compared to the correlation found by \citet{Anderson2016_II_metallicity} between the pEW of the Fe~{\sc ii} 5018\AA\ absorption line measured at 50~d and the host H~{\sc ii}-region oxygen abundance. The blue line indicates the mean best fit to the data. The solid red line denotes the measured pEW of SN~2018zd, and the dashed red line indicating the estimated 03N2 H~{\sc ii} region 12+log[O/H] for SN~2018zd. 
    }
    \label{fig:anderson_metallicity_comp}
\end{figure*}


\bsp	
\label{lastpage}
\end{document}